\documentclass{aa} 
\usepackage{graphics} 
\usepackage{graphicx} 
\usepackage[]{pict2e}
\usepackage{color, colortbl}
\usepackage{tikz}
\usepackage{pgfplots}
\usepackage{mathtools,amssymb}
\pgfplotsset{compat=1.16}

\usepackage{txfonts}
\def\ammo{\rm NH_3}
\def\dammo{\rm NH_2D}
\def\ddammo{\rm NHD_2}
\def\ndthree{\rm ND_3}
\def\odammo{\rm oNH_2D}
\def\pdammo{\rm pNH_2D}
\def\oddammo{\rm oNHD_2}
\def\pddammo{\rm pNHD_2}
\def\diaz{\rm N_2H^+}
\def\ddiaz{\rm N_2D^+}
\def\htwo{\rm H_2}
\def\htwodplus{\rm H_2D^+}
\def\dtwohplus{\rm D_2H^+}
\def\dthreeplus{\rm D_3^+}
\def\hthreeplus{\rm H_3^+}
\def\percc{\rm cm^{-3}}

\def\kms{\rm km\,s^{-1}}
\def\ms{\rm m\,s^{-1}}
\def\persqcm{\rm cm^{-2}}
\thicklines
\newcommand*\circled[1]{\tikz[baseline=(char.base)]{
            \node[shape=circle,draw,inner sep=2pt] (char) {#1};}}
\begin{document}

   \title{Nuclear spin ratios of deuterated ammonia in prestellar cores}
   \subtitle{LAsMA observations of H-MM1 and Oph D
   \thanks{This publication is based on data acquired with the Atacama
Pathfinder EXperiment (APEX). During the present observations, APEX was a 
collaboration between the Max-Planck-Institut für Radioastronomie (MPIfR), 
the European Southern Observatory (ESO) and the Onsala Space Observatory. 
Since the beginning of 2023, it is operated by ESO on behalf of the MPIfR.}}

   \author{J. Harju
          \inst{1,2}
          \and
          J. E. Pineda \inst{1}
                    \and
          O. Sipil{\"a} \inst{1}
          \and
          P. Caselli \inst{1}
          \and
          A. Belloche \inst{3}
          \and
          F. Wyrowski \inst{3}
          \and
          W. Riedel\inst{1}
          \and
          E. Redaelli\inst{1}
          \and
          A.I. Vasyunin\inst{4}
          }

 \institute{Max-Planck-Institut f{\"u}r extraterrestrische Physik, Gie{\ss}enbachstra{\ss}e 1, D-85748 Garching, Germany 
   \and
   Department of Physics, P.O. Box 64, FI-00014, University of Helsinki, Finland
   \and
    Max-Planck-Institut f{\"u}r Radioastronomie,
    Auf dem H{\"u}gel 69, D-53121 Bonn, Germany
    \and
    Research Laboratory for Astrochemistry, Ural Federal University, 620002, 19 Mira street, Yekaterinburg, Russia
    }

   \date{Received 3 April 2023; accepted 13 November 2023}

 
  \abstract
   {Molecules containing two or more hydrogen or deuterium atoms have different nuclear spin states which behave as separate chemical species. The relative abundances of these species can give clues to their origin. Formation on grains is believed to yield statistical spin ratios whereas gas-phase reactions are predicted to result in clear deviations from them. This is also true for ammonia and its deuterated forms $\dammo$, $\ddammo$, and $\ndthree$. }
   {Here we aim to determine the ortho/para ratios of $\dammo$ and $\ddammo$ in dense, starless cores, where their formation is supposed to be dominated by gas-phase reactions.}
   {The Large APEX sub-Millimeter Array (LAsMA) multibeam receiver of the    Atacama Pathfinder EXperiment (APEX) telescope was used to observe the    prestellar cores H-MM1 and Oph\,D in Ophiuchus in the ground-state lines of ortho and para $\dammo$ and $\ddammo$. The fractional abundances of these molecules were derived employing  three-dimensional radiative transfer modelling, using different assumptions about the abundance profiles as functions of density. We also ran gas-grain chemistry models with different scenarios concerning proton or deuteron exchanges and chemical desorption from grains {to find out} if one of these models can reproduce the observed spin ratios.}
   {The observationally deduced ortho/para ratios of $\dammo$ and $\ddammo$ are in both cores within 10\% of their statistical values 3 and 2, respectively, and taking $3\,\sigma$ limits, deviations from these of about 20\%  are allowed. {Of the chemistry models tested here, { the model} that assumes proton hop (as opposed to full scrambling) in reactions contributing to ammonia formation, and a constant efficiency of chemical desorption, comes nearest to the observed abundances and spin ratios.}}
   {The nuclear spin ratios derived here are in contrast with spin-state chemistry models that assume full scrambling in proton donation and hydrogen abstraction reactions leading to deuterated ammonia. The efficiency of chemical desorption influences strongly the predicted abundances of $\ammo$, $\dammo$, and $\ddammo$, but has a lesser effect on their ortho/para ratios. For these the proton exchange scenario in the gas is decisive. We suggest that this is because of rapid re-processing of ammonia and related cations by gas-phase ion-molecule reactions.}


   \maketitle
%

\section{Introduction}

Astrochemical models that distinguish between different nuclear spin modifications assume that ion-molecule reactions in interstellar gas proceed via the formation of a relatively long-lived intermediate complex, where the light H and D nuclei can be completely mixed. In this scenario, one of the main deuteration reactions of ammonia in dense gas can be written as $\ammo + {\rm H_2D^+} \rightarrow ({\rm NH_5D^+})^\dagger \rightarrow {\rm NH_3D^+} + \htwo$, where the reaction complex is indicated with a dagger. This reaction is followed by electron recombination of $\rm NH_3D^+$, which can yield $\ammo$ or $\dammo$. The increasing abundances of $\rm H_3^+$, $\rm H_2D^+$, $\rm HD_2^+$, and $\rm D_3^+$ in the interiors of dense, starless cores, and the possibility of mixing between nuclei in gas-phase reactions, leads to a situation where the nuclear spin ratios of ammonia and its deuterated forms deviate from their statistical values. This is caused by the conservation of spin angular momentum and permutation symmetry.  In the models of \cite{2015A&A...581A.122S} and \cite{2018MNRAS.477.4454H}, the ortho-to-para ratio (hereafter o/p) of $\dammo$ approaches 2 or gets slightly below it (instead of the statistical value 3), whereas o/p-$\ddammo$ reaches up to $\sim 3$ (Hily-Blant et al.) or even $4$ (Sipil\"a et al.) at high densities (instead of the statistical 2).

While interesting from a theoretical point of view, this phenomenon also provides a means of determining the origin of species that possess different nuclear spin modifications: hydrogenated species formed on grains should always show statistical spin ratios because they are formed  through H or D atom addition reactions, whereas species formed in the gas phase can have non-statistical spin ratios (e.g., \citealt{1989MNRAS.240P..25B}; \citealt{2015A&A...581A.122S}; \citealt{2017A&A...600A..61H}; \citealt{2018MNRAS.477.4454H}; \citealt{2019A&A...631A..63S}).

The few observational determinations of the o/p ratios of $\dammo$ or  $\ddammo$ in dense cores are consistent with statistical ratios (\citealt{2006A&A...454L..63G}; \citealt{2006ApJ...636..916L}; \citealt{2016MNRAS.457.1535D}; \citealt{2017A&A...600A..61H}; \citealt{2021A&A...649A..21W}). The uncertainties of these estimates are, however, large - typically greater than 30\%, and one cannot confidently exclude the deviations from statistical values predicted by chemistry models.   

Here we attempt to determine the o/p ratios of $\dammo$ and $\ddammo$ in two dense and cold interstellar gas clouds with greater accuracy than achieved previously, using the Large APEX sub-Millimeter Array (LAsMA) of the Atacama Pathfinder EXperiment (APEX). The ground state transitions of the ortho and para modifications of $\dammo$ and $\ddammo$ (hereafter $\odammo$, $\pdammo$, $\oddammo$, and $\pddammo$) at $\lambda \sim0.9$\,mm  were observed with this instrument towards the starless cores H-MM1 and Oph\,D in the region of the dark cloud L1688 in Ophiuchus. The large-scale structure and general properties of the cores in the L1688 region have been discussed in, for example, \cite{2015MNRAS.450.1094P}, \cite{2017ApJ...843...63F}, \cite{2020A&A...640L...6C}, and \cite{2020A&A...638A..74L}. Detailed studies of the structures and the ammonia distributions in the two target cores have been presented in \cite{2020ApJ...895..101H} and \cite{2022AJ....163..294P} (for H-MM1), and in \cite{2011A&A...534A.122R} (for Oph\,D). The two cores can be classified as prestellar, meaning that they are on the verge of collapse, as they have central densities significantly above $10^5\,\percc$, and high levels of deuterium fractionation (e.g., \citealt{2005ApJ...619..379C}; \citealt{2017A&A...600A..61H}). 

In Sect.~\ref{observations} of this paper we describe the LAsMA observations, and in Sect.~\ref{results} we present the data and estimate the abundances and the o/p ratios of $\dammo$ and $\ddammo$ in the two cores. {Spin ratios expected from the proton hop and full scrambling scenarios, and the results of previous spin-state chemistry models are discussed in Sect.~\ref{old_models}. In Sect.~\ref{new_model},} we show results from an updated version of our gas-grain chemistry model. Implications of the observed spin ratios and $\dammo/\ammo$ and $\ddammo/\dammo$ abundance ratios are discussed in Sect.~\ref{discussion}. Finally, in Sect.~\ref{conclusions} we draw our conclusion from this study. 
 
\section{Observations}
\label{observations}

The ground-state rotational lines of o$\dammo$, p$\dammo$, o$\ddammo$, and p$\ddammo$ near  333 and 335\,GHz were observed with the LAsMA array \citep{2008SPIE.7020E..10G} on the APEX telescope \citep{2006A&A...454L..13G} towards the prestellar cores H-MM1 and Oph\,D in Ophiuchus. The observations were done between April 10 and September 28, 2022, under the project number M-0109.F-9503A-2022.  The APEX telescope is located at an altitude of 5100\,m at Llano de Chajnantor, in Chile. 

\begin{figure*}
\unitlength=1mm
\begin{picture}(160,80)(0,0)
\put(85,0){
\begin{picture}(0,0) 
\includegraphics[width=9.5cm,angle=0]{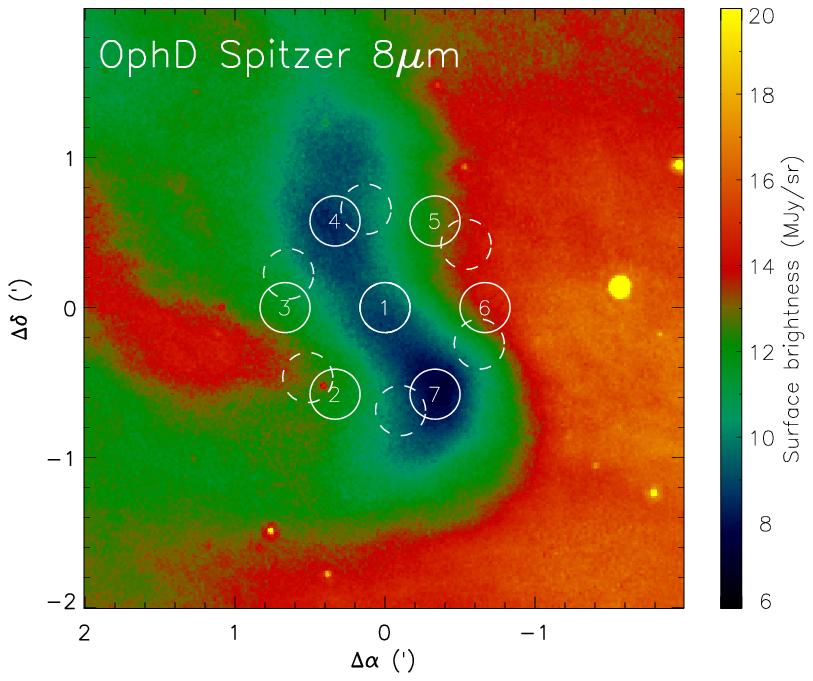}
\end{picture}}
\put(-5,0){
\begin{picture}(0,0) 
\includegraphics[width=9.5cm,angle=0]{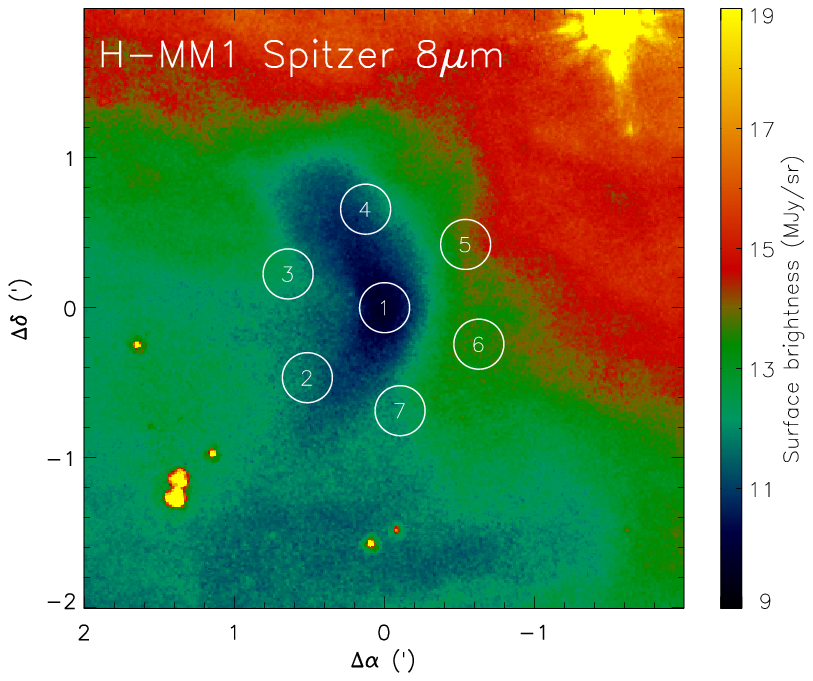}
\end{picture}}
\end{picture}  
   \caption{Footprint of the APEX/LAsMA array  on the $8\,\mu$m surface brightness images of H-MM1 (left) and Oph\,D (right) observed by the {\sl Spitzer} satellite. The dashed and solid circles on the right represent two orientations of the array in observations towards Oph\,D. The positions of the central pixels are RA $16^{\rm h}27^{\rm m}58\fs5$, Dec. $-24\degr33\arcmin40\arcsec$ (J2000.0) for H-MM1,  and RA $16^{\rm h}28^{\rm m}30\fs2$, Dec. $-24\degr18\arcmin42\arcsec$ for Oph\,D.}
   \label{lasma_footprint}
    \end{figure*}

The LAsMA array has 7 pixels separated by approximately $40\arcsec$. Figure~\ref{lasma_footprint} shows the footprint of the instrument superposed on the $8\,\mu$m maps of the cores measured by the InfraRed Array Camera (IRAC) of the Spitzer Space Telescope. The beamsize of APEX is $18\arcsec$ (FWHM) at 335\,GHz. The pattern was unintentionally tilted by $19\fdg1$ during observations in the Spring and Summer 2022. This was corrected during additional observations in September towards Oph D (solid circles on the right). With two tilt angles, the total number of positions observed towards Oph\,D is 13. 

The backend was composed of Fast Fourier Transform Spectrometers (FFTS4G) that covered two 4\,GHz intermediate frequency (IF) bands resolved into 65 536 spectral channels of 61.03\,kHz in width (corresponding to $\sim 55\,\ms$). The lines of ortho and para $\dammo$ and $\ddammo$ were observed in the lower sideband (LSB) of the receiver which was centered at 334.145\,GHz. The frequencies of the observed transitions are listed in Table~\ref{transitions}. Also listed there are the energies of the upper transition level and the Einstein $A$ coefficients of the transitions. The spectroscopic data are adopted from \cite{2021JMoSp.37711431M}. At 10\,K, the optically thin critical densities of the four transitions (estimated using Eq. (4) of \citealt{2015PASP..127..299S}) lie in a narrow range between $6.9\times 10^4\,\percc$ and $7.9\times10^4\,\percc$. 

\begin{figure*}
\centering
\unitlength=1mm
\begin{picture}(160,120)(0,0)
\put(0,60){
\begin{picture}(0,0) 
\includegraphics[width=8cm,angle=0]{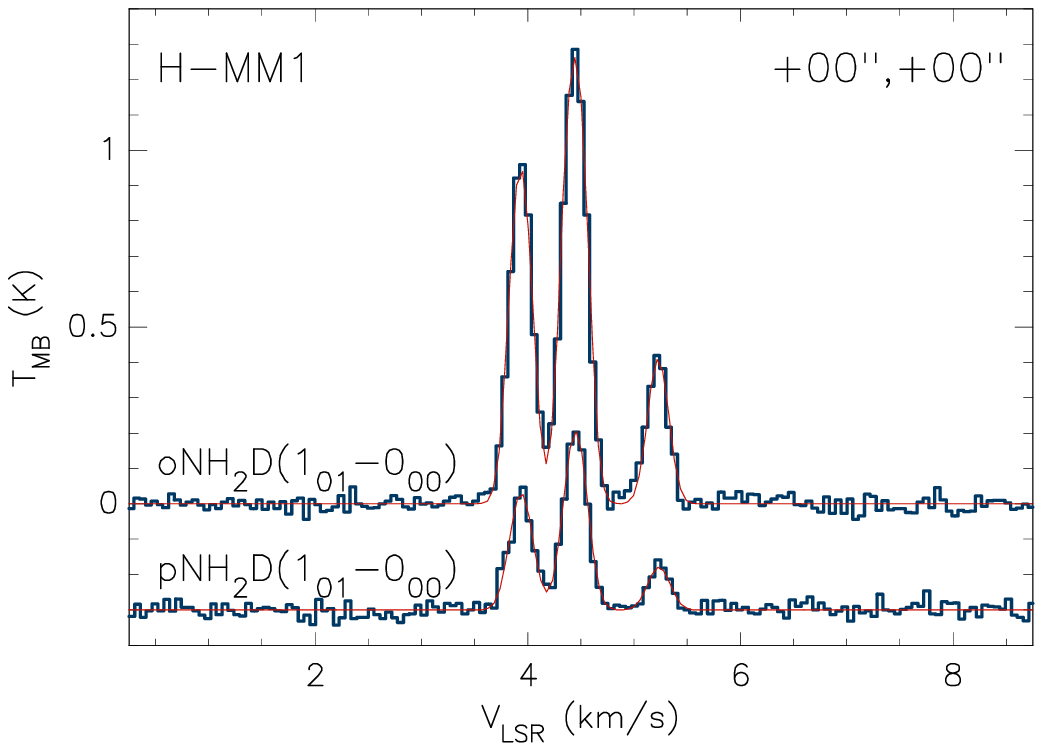}
\end{picture}}
\put(0,0){
\begin{picture}(0,0) 
\includegraphics[width=8cm,angle=0]{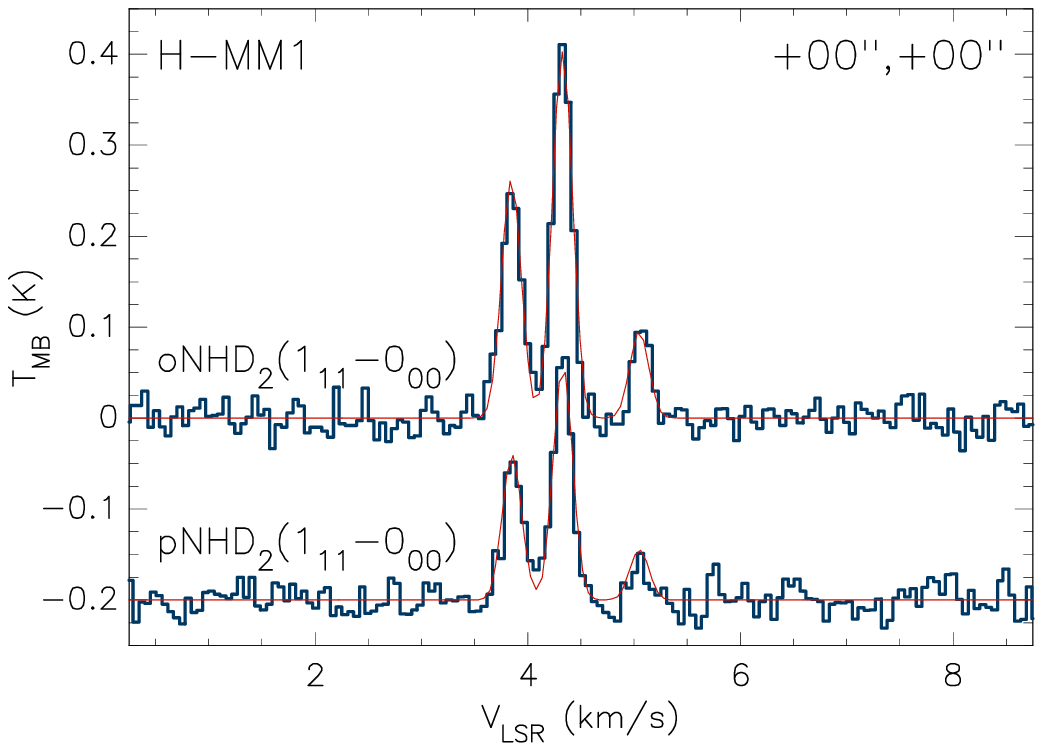}
\end{picture}}
\put(90,60){
\begin{picture}(0,0) 
\includegraphics[width=8cm,angle=0]{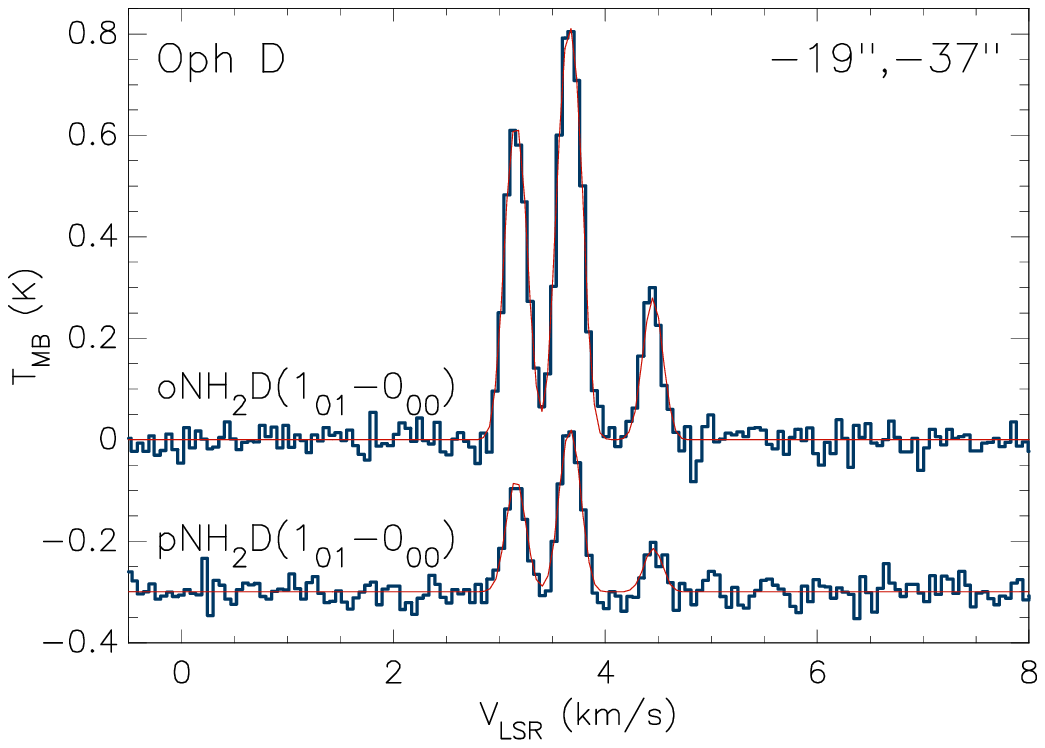}
\end{picture}}
\put(90,0){
\begin{picture}(0,0) 
\includegraphics[width=8cm,angle=0]{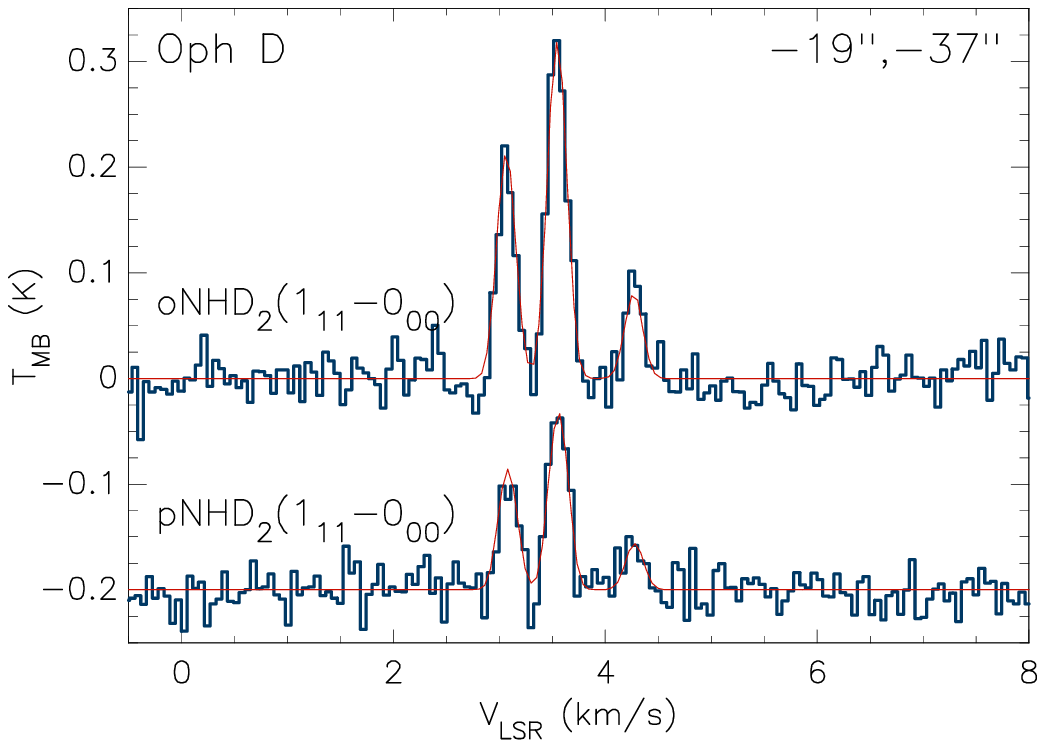}
\end{picture}}
\end{picture}  
\caption[]{$\dammo$ and $\ddammo$ spectra from the peak positions of H-MM1 and Oph\,D (positions \circled{1} and \circled{7} in Fig.~\ref{lasma_footprint}, respectively). Gaussian fits to the hyperfine structure are shown with thin red lines.}
\label{peak_spectra}
\end{figure*}

The $\dammo$ and $\ddammo$ spectra towards the density peaks of the two cores are shown in Fig.~\ref{peak_spectra}. The positions correspond to pixel \circled{1} for H-MM1 and pixel \circled{7} for Oph\,D in Fig.~\ref{lasma_footprint}. The spectra are shown on the main-beam brightness temperature, $T_{\rm MB}$, scale. In the conversion from $T_{\rm A}^*$ scale, we assumed the following main-beam and forward-scattering and spill-over efficiencies: $\eta_{\rm MB}=0.73$, $\eta_{\rm fss}=0.97$. In the course of the measurements, we noticed line intensity changes up to 15\%, which probably are caused by occasional rapid variations in the water vapour content of the atmosphere. We estimate that the absolute calibration is accurate to 20\%. {Because the four lines were observed in the same spectral band, this uncertainty does not concern their relative intensities,  and does not propagate to the derived o/p ratios.}     

\begin{table}
\caption[]{Observed transitions$^*$. }
\begin{tabular}{lllll} \hline
\multicolumn{2}{c}{Transition} & Freq. (MHz) & $E_{\rm u}$ (K) & $A_{\rm ul}$ (s$^{-1}$) \\ \hline
$\odammo$ & $1_{01}^{\rm a}-0_{00}^{\rm a}$ & 332781.803 & 16.6 & $7.81\times10^{-6}$ \\
$\pdammo$ & $1_{01}^{\rm s}-0_{00}^{\rm s}$ & 332822.521 & 16.0 & $7.29\times10^{-6}$ \\
$\oddammo$ & $1_{11}^{\rm s}-0_{00}^{\rm s}$ & 335513.715 & 16.1 & 
$1.29\times10^{-5}$ \\ 
$\pddammo$ & $1_{11}^{\rm a}-0_{00}^{\rm a}$ & 335446.240 & 16.3 &
$1.47\times10^{-5}$ \\ \hline
\noalign{\smallskip}
\end{tabular}
$^*$ The frequencies are adopted from \cite{2021JMoSp.37711431M}
\label{transitions}
\end{table}

\section{Results}
\label{results}

\subsection{LTE analysis of the spectra}
\label{lte_analysis}

The line parameters and column densities derived from Gaussian fits to the hyperfine structure of the lines observed towards the density peaks of H-MM1 and Oph\,D are presented in Table~\ref{line_parameters}. The fits are shown with thin red lines in Fig.~\ref{peak_spectra}. In the column density estimates we have assumed line-of-sight homogeneity and that the excitation temperature, $T_{\rm ex}$, is constant for all rotational transitions of the molecule. Furthermore, the calculation involves the assumption that populations of hyperfine states in a certain rotational level are proportional to their statistical weights. The transitions have three groups which consist of several unresolved components (\citealt{2006A&A...449..855C}; \citealt{2021JMoSp.37711431M}). The linewidth listed in Table~\ref{line_parameters} refers to the intrinsic width of an individual hyperfine component.  One can see in this Table that the $\dammo$ lines are broader than those of $\ddammo$, and the difference in H-MM1 clearly exceeds that expected from thermal broadening, $\sim 5\,{\rm m\,s}^{-1}$ at 10\,K.  The $\ddammo$ abundance is likely to peak at slightly higher densities and deeper in the core as compared to $\dammo$, and the linewidth difference could possibly be understood in terms of lower temperatures and smaller non-thermal velocity dispersion in the interior parts of the cores. The 3D models described below include realistic temperature gradients but assume a constant non-thermal velocity dispersion. 

\begin{figure}[htb]
\unitlength=1mm
\begin{picture}(80,65)(0,0)
\put(-5,0){
\begin{picture}(0,0) 
\includegraphics[width=9cm,angle=0]{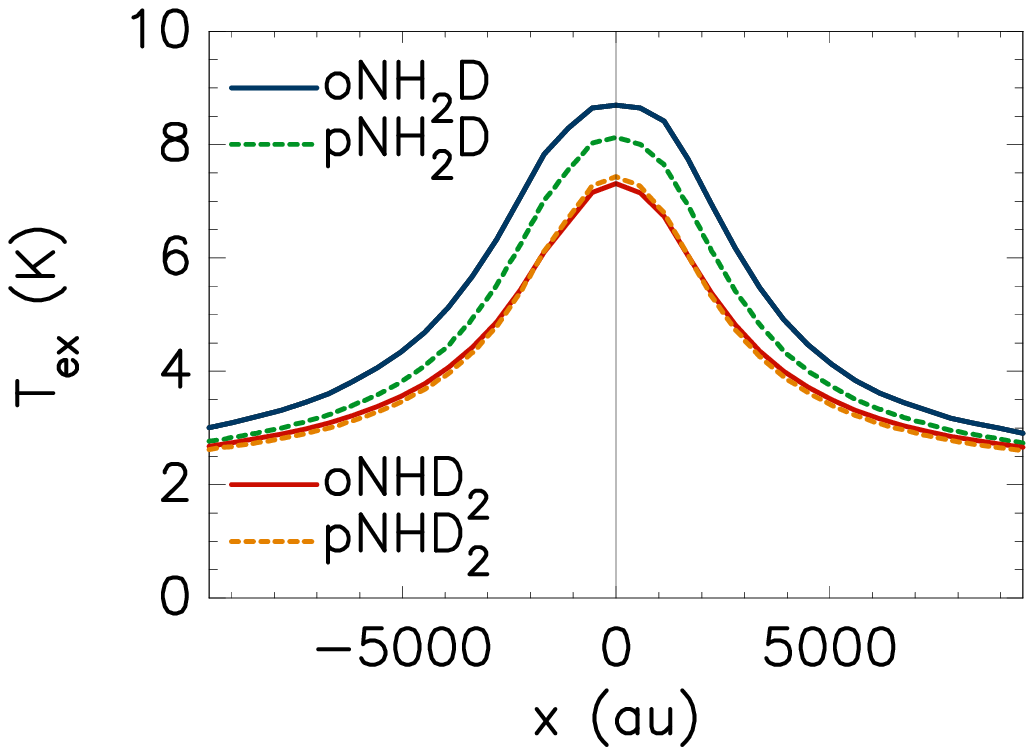}
\end{picture}}
\end{picture}  
   \caption{Excitation temperature profiles of the ground-state lines of
   ortho and para $\dammo$ and $\ddammo$ in a core model resembling H-MM1.} 
   \label{tex_profiles}
\end{figure}

Despite rather high signal-to-noise ratios especially in H-MM1,  the optical thicknesses and the excitation temperatures of the $\pdammo$ and $\pddammo$ lines have large relative uncertainties which propagate to the column densities. The total optical thicknesses (sums of three separate groups) of the $\pdammo$ and $\pddammo$ lines derived towards H-MM1 are below 1, suggesting that one could use for the para lines the optically thin approximation, where the column density is proportional to the integrated intensity. In this method, the $T_{\rm ex}$ must be assumed. We examined the excitation of the observed transitions by applying radiative transfer calculations to the 3D model of H-MM1 used previously by \cite{2022AJ....163..294P}, and described in more detail in Sect.~\ref{rt_modelling}.  Here, the molecules are assumed to have constant fractional abundances, corresponding to the column densities derived from the hyperfine fits described above. The $T_{\rm ex}$ profiles along a line crossing the density peak (with $n(\htwo)\sim1.2\times10^6\,\percc$) are shown in Fig.~\ref{tex_profiles}. According to this calculation, the $T_{\rm ex}$ of p$\dammo$ line is consistently $\sim 10\%$ lower than that of the o$\dammo$ line, whereas for the para- and ortho-$\ddammo$ lines the $T_{\rm ex}$ ratio is 1. The higher $T_{\rm ex}$ of the o$\dammo$ line compared to that of the p$\dammo$ line is likely to be caused by an increased effect of radiative trapping with larger optical thickness (\citealt{2015PASP..127..299S}; \citealt{2021A&A...645A..26R}). In the last column of Table~\ref{line_parameters} we give the $\pdammo$ and $\pddammo$ column densities derived from the integrated intensities assuming that $T_{\rm ex,para}=0.9\times T_{\rm ex,ortho}$ for $\dammo$, and $T_{\rm ex,para}=T_{\rm ex,ortho}$ for $\ddammo$.

Using para column densities from optically thin approximation, the values listed in Table~\ref{line_parameters} give the following o/p-$\dammo$ ratios: $3.0\pm 0.2$ (H-MM1) and $3.5\pm0.6$ (Oph\,D). The corresponding o/p-$\ddammo$ ratios are $2.2\pm0.5$ (H-MM1) and $2.4\pm1.6$ (Oph D). If we use the assumption $T_{\rm ex,para}=T_{\rm ex,ortho}$ also for $\dammo$, the o/p-$\dammo$ ratios increase by $\sim 20\%$ from the values given above. To distinguish between different formation scenarios for these molecules, the o/p ratios should be determined more accurately, which can be achieved via non-LTE modelling as we will discuss below.

Besides the uncertainty involved in the determination of the optical thickness, $\tau$, from only three resolved components, the assumption of a constant $T_{\rm ex}$ along the line of sight in the presence of large density gradients is an additional source of  uncertainty, as demonstrated in Fig.~\ref{tex_profiles}. In what follows, we construct three-dimensional physical models for the H-MM1 and Oph\,D cores, and estimate the fractional abundances of the observed species using radiative transfer calculations. 

\begin{centering}
\begin{table*}
\caption[]{Line parameters and column densities from Gaussian fits to the hyperfine structure.}
\label{line_parameters}
\begin{tabular}{ccccccccc} \hline
Species & Line & $T_{\rm MB}$ & $V_{\rm LSR}$ & $\Delta V$  &  $T_{\rm ex}$ & 
$\tau_{\rm tot}$ & $N$  & $N_{\rm thin}^*$  \\ 
 &  & (K) & $\kms$) & ($\kms$) &  (K) &  & ($10^{14}\persqcm$) &  ($10^{14}\persqcm$)\\ \hline
     \noalign{\smallskip}
        \multicolumn{9}{c}{H-MM1 (0,0)} \\
        \noalign{\smallskip}
        \hline 
        \noalign{\smallskip}
$\odammo$ & $1_{01}^{\rm a}-0_{00}^{\rm a}$ & 
1.285 $\pm$ 0.018 & 4.358 $\pm$ 0.001 & 0.217 $\pm$ 0.002 & 6.95 $\pm$ 0.10 & 2.50 $\pm$ 0.15 & 1.50 $\pm$ 0.11  &  - \\
$\pdammo$ & $1_{01}^{\rm s}-0_{00}^{\rm s}$ & 
0.504 $\pm$ 0.018 & 4.364 $\pm$ 0.003 & 0.237 $\pm$ 0.007 & 6.61 $\pm$ 0.92 & 0.77 $\pm$ 0.35 & 0.50 $\pm$ 0.31 & 0.50 $\pm$ 0.03 \\
$\oddammo$ & $1_{11}^{\rm s}-0_{00}^{\rm s}$ & 
0.411 $\pm$ 0.015 & 4.325 $\pm$ 0.002 & 0.201 $\pm$ 0.005 & 6.02 $\pm$ 0.23 & 0.80 $\pm$ 0.12 & 0.28 $\pm$ 0.06 & - \\
$\pddammo$ & $1_{11}^{\rm a}-0_{00}^{\rm a}$ & 
0.266 $\pm$ 0.014 & 4.330 $\pm$ 0.003 & 0.203 $\pm$ 0.008 & 6.76$\pm$ 0.43 & 0.33 $\pm$ 0.07 & 0.12 $\pm$ 0.03 & 0.13$\pm$0.02 \\
\hline
     \noalign{\smallskip}
        \multicolumn{9}{c}{Oph\,D  ($-19\arcsec,-37\arcsec$)} \\
        \noalign{\smallskip}
        \hline \noalign{\smallskip}
$\odammo$ & $1_{01}^{\rm a}-0_{00}^{\rm a}$ & 
0.805 $\pm$ 0.021 & 3.581 $\pm$ 0.002 & 0.205 $\pm$ 0.004 & 5.83 $\pm$ 0.11 & 2.85 $\pm$ 0.29 & 1.32 $\pm$ 0.16 & -\\
$\pdammo$ & $1_{01}^{\rm s}-0_{00}^{\rm s}$ &  
0.315 $\pm$ 0.020 & 3.583 $\pm$ 0.004 & 0.195 $\pm$ 0.011 & 5.07 $\pm$ 0.59 & 1.29 $\pm$ 0.68 & 0.53 $\pm$ 0.38 & 0.38$\pm$0.04 \\
$\oddammo$ & $1_{11}^{\rm s}-0_{00}^{\rm s}$ & 
0.320 $\pm$ 0.017 & 3.546 $\pm$ 0.003 & 0.189 $\pm$ 0.008 & 5.14 $\pm$ 0.54 & 1.16 $\pm$ 0.43 & 0.32 $\pm$ 0.18 & - \\
$\pddammo$ & $1_{11}^{\rm a}-0_{00}^{\rm a}$ & 
0.163 $\pm$ 0.016 & 3.558 $\pm$ 0.006 & 0.190 $\pm$ 0.016 & 4.26 $\pm$ 0.56 & 1.32 $\pm$ 1.02 & 0.28 $\pm$ 0.28 & 0.13$\pm$0.04\\
\hline
\noalign{\smallskip}
\end{tabular}
$^*$ Column density using optically thin approximation with $T_{\rm ex}$ estimated from the ortho line (see text). 
\end{table*}
\end{centering}

\subsection{3D radiative transfer modelling}
\label{rt_modelling}

We derived the fractional abundances of the observed molecules by comparing simulated spectra with the observations. Physical models for H-MM1 and Oph\,D were constructed using the $8\,\mu$m surface brightness maps shown in Fig.~\ref{lasma_footprint}. Both cores appear as absorption features in the maps, and this was  used to derive high-resolution $\htwo$ column density maps. The method is described in Appendix A of \cite{2020ApJ...895..101H} and in \cite{2022AJ....163..294P}. The density structures of the cores were estimated by fitting a Plummer-type function to the cross-sectional $\htwo$ column density profiles in different positions along the ridges of the cores. The method is adopted from \cite{2011A&A...529L...6A}. We assumed that the density distribution has circular symmetry in the plane perpendicular to the spine of the core. The latter is defined by a spline fit through local $N(\htwo)$ peaks. A more detailed description of the method is presented in Appendix D of \cite{2022AJ....163..294P}. 

The $\htwo$ column densities of the models of H-MM1 and Oph\,D are comparable to those that can be obtained from the high-resolution $N(\htwo)$ map of L1688 derived from multi-wavelength {\sl Herschel} data \citep{2020A&A...638A..74L}, which has an effective resolution of $18\farcs2$.  When the model column density distributions are smoothed to the same resolution, the density peaks have the following $\htwo$ column densities (values from the {\sl Herschel} map of \cite{2020A&A...638A..74L} are given in brackets): H-MM1 $(0,0)$ $N(\htwo)=3.7\times10^{22}\,\persqcm$ ($4.4\times10^{22}\,\persqcm$); Oph\,D $(-19\arcsec,-37\arcsec)$ $N(\htwo)=3.2\times10^{22}\,\persqcm$ ($2.2\times10^{22}\,\persqcm$); Oph\,D $(+20\arcsec,+35\arcsec)$ $N(\htwo)=2.2\times10^{22}\,\persqcm$ ($2.2\times10^{22}\,\persqcm$). The fact that the total $N(\htwo)$ column density derived from {\sl Herschel} is {\sl lower} than that of the core model at Oph\,D ($-19\arcsec,-37\arcsec$) can be understood by the relatively high dust colour temperature ($T_{\rm C}=12.9$\,K) derived from the far-infrared data. The mass-averaged physical temperature is probably lower towards this position (see below). The high-resolution $N(\htwo)$ and $T_{\rm C}$ maps of L1688 from {\sl Herschel} are accessible via {http://gouldbelt-herschel.cea.fr}. 

\begin{figure*}
\centering
\unitlength=1.0mm
\begin{picture}(160,160)(0,0)
\put(0,80){
\begin{picture}(0,0) 
\includegraphics[width=8cm,angle=0]{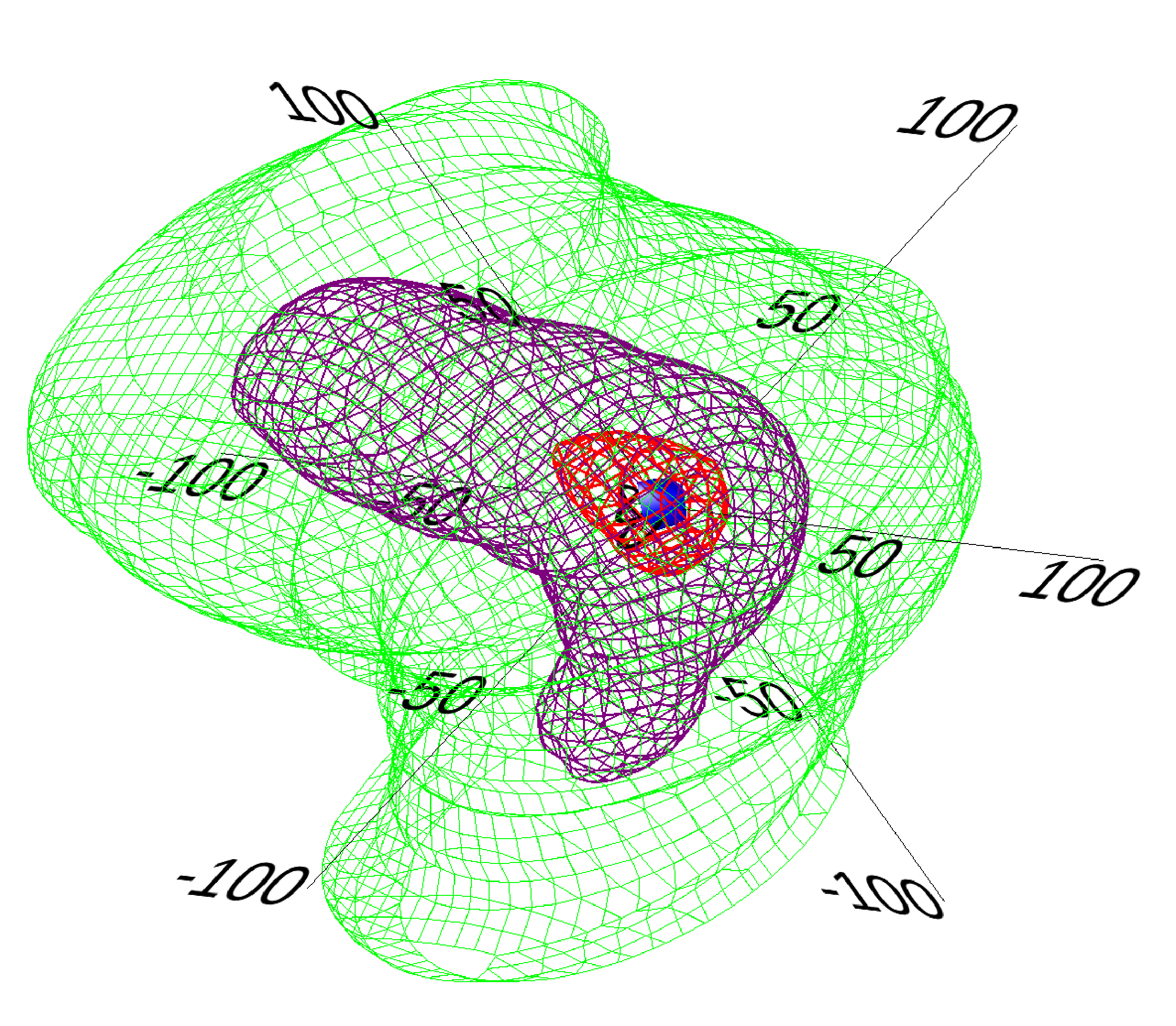}
\end{picture}}
\put(90,80){
\begin{picture}(0,0) 
\includegraphics[width=8cm,angle=0]{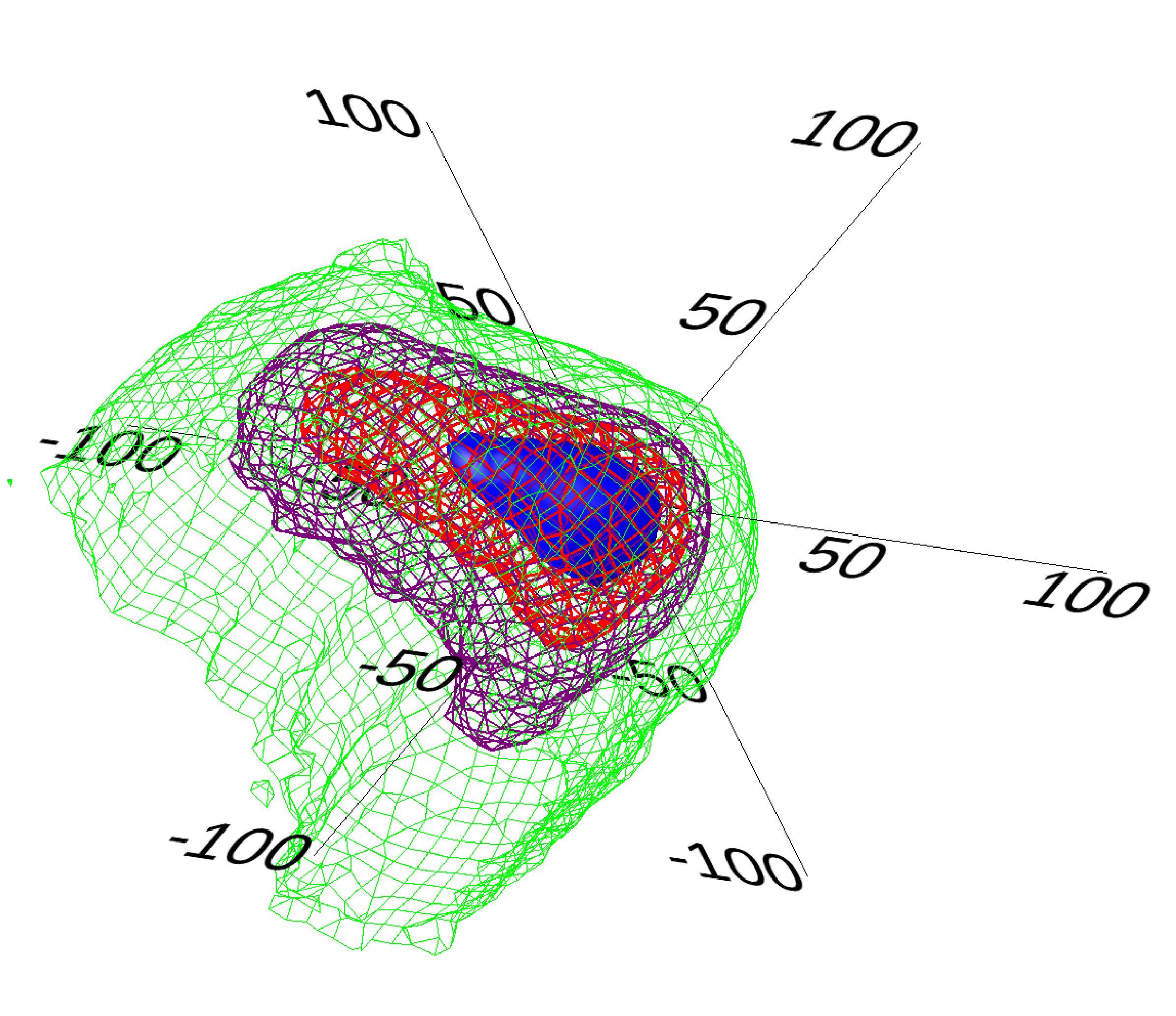}
\end{picture}}
\put(0,0){
\begin{picture}(0,0) 
\includegraphics[width=8.5cm,angle=0]{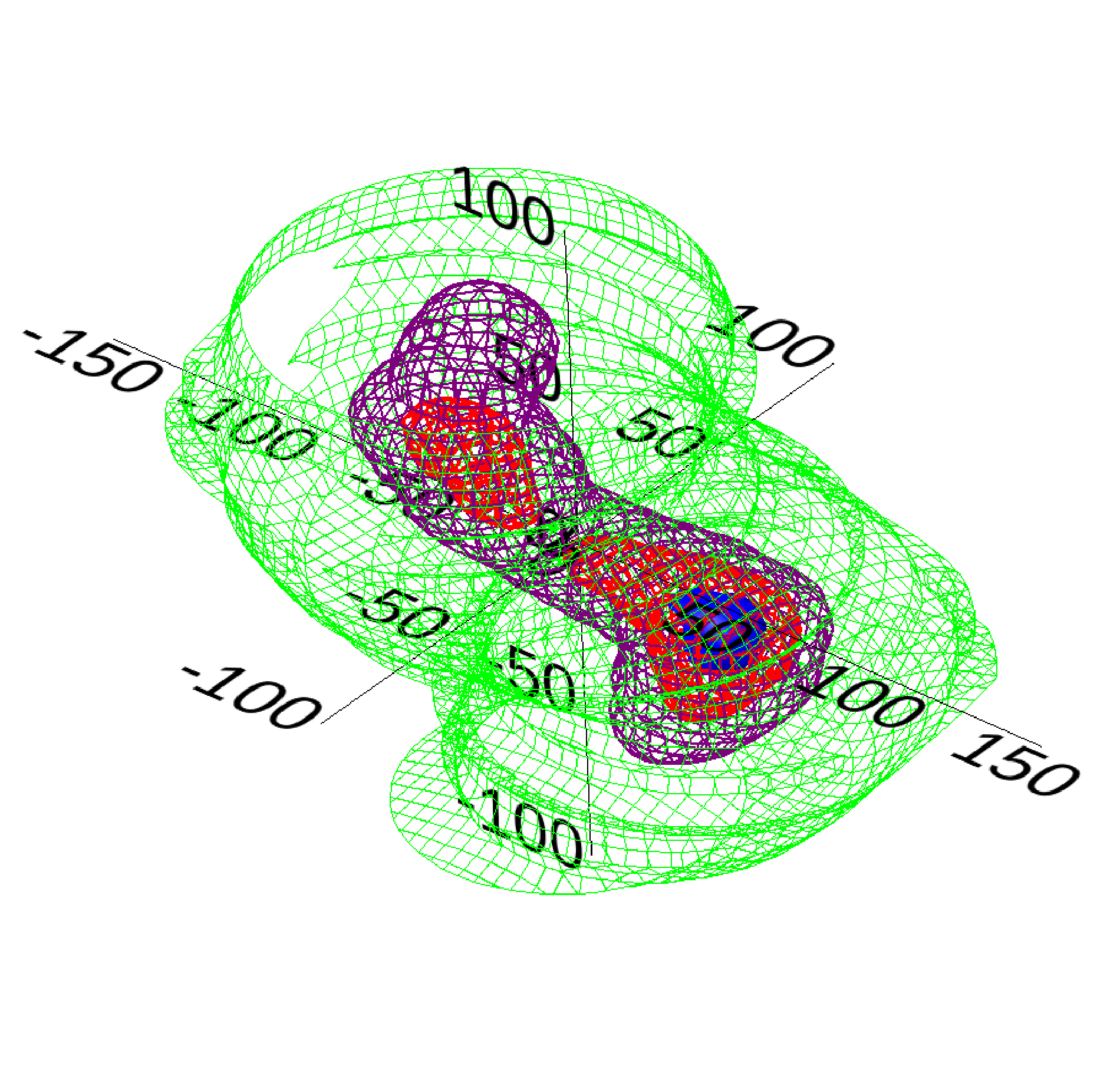}
\end{picture}}
\put(90,0){
\begin{picture}(0,0) 
\includegraphics[width=8cm,angle=0]{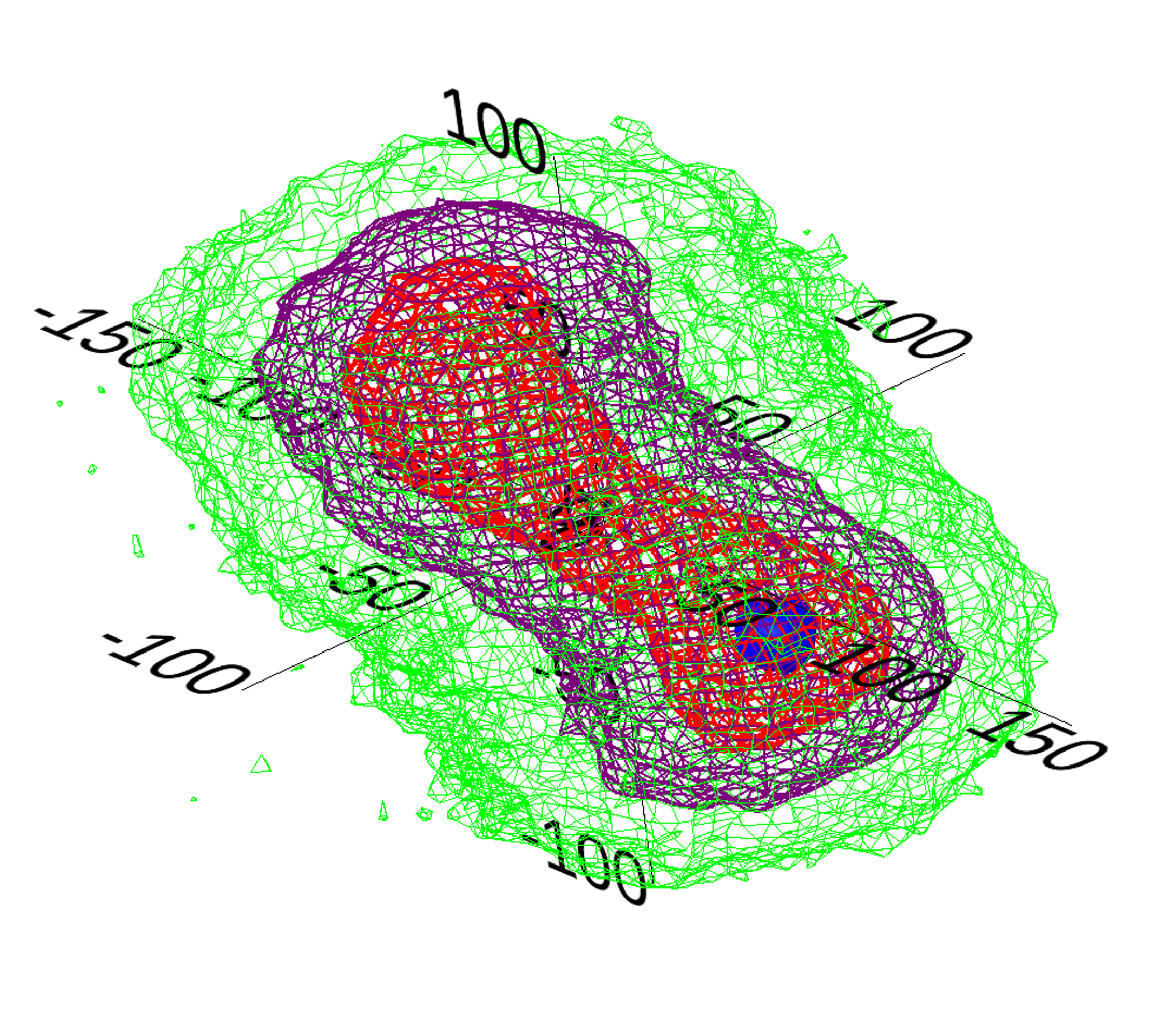}
\end{picture}}
\put(0,100){\bf H-MM1}
\put(0,15){\bf Oph D}
\put(0,150){\bf Density}
\put(90,150){\bf Temperature}
\put(170,110){\bf X}
\put(159,140){\bf Y}
\put(124,140){\bf Z}
\put(73,142){\bf Y}
\put(68,55){\bf Y}
\put(162,47){\bf Y}
\put(88,20){\bf X}
\end{picture}
\caption[]{Density and dust temperature models for H-MM1 and Oph D. {\bf Top: H-MM1}~ The isodensity surfaces (left) are $10^4\,\percc$ (green), $10^5\,\percc$ (purple), $5\times10^5\,\percc$ (red), and $10^6\,\percc$ (blue). The isotemperature $T_{\rm dust}$ surfaces (right) are 14\,K (green), 12\,K, (purple), 11\,K (red), and 10\,K (blue). {\bf Bottom: Oph\,D}~ The isodensity surfaces (left) are $10^4\,\percc$ (green), $10^5\,\percc$ (purple), $2\times10^5\,\percc$ (red), and $4\times10^5\,\percc$ (blue). The isotemperature $T_{\rm dust}$ surfaces (right) are 11\,K (green), 10\,K (purple), 9\,K (red), and 8\,K (blue). The positive X-axis points to the observer, and the Y-axis points to the equatorial west. The scale is in arcseconds. In both cores, the gas kinetic temperature, $T_{\rm kin}$,  is assumed to be equal to $T_{\rm dust}$ in the inner parts, and settle to 11\,K in the outer parts.}   
\label{3D_models}
\end{figure*}

The gas temperature was assumed to be equal to the dust temperature, $T_{\rm dust}$, in the inner parts of the cores. Based on the $T_{\rm kin}$ distribution in H-MM1 derived from $\ammo$ by \cite{2022AJ....163..294P} (see their Figs. 5 and 11) we assumed, however, that the gas temperature does not rise above 11\,K in the outer parts of the cores owing to molecular line cooling. The dust temperature was computed using the continuum radiative transfer program CRT \citep{2005A&A...440..531J}. In the case of H-MM1, the external radiation field was assumed to consist of an isotropic component and a point source located on the western side of the core. The point source represents the subgiant binary HD 147889 located about 1.2\,pc west of H-MM1. The density model of H-MM1 and the assumptions about the external radiation field are the same as those used in \cite{2022AJ....163..294P}. For Oph\,D, which is known to be cooler than H-MM1 ($\sim6-7$\,K in the densest parts; \citealt{2008A&A...482..535H}; \citealt{2011A&A...534A.122R}), the external radiation field was assumed isotropic and weaker by a factor of two than that around H-MM1. The spectrum of the unattenuated interstellar radiation field (ISRF) was taken from \cite{1994ASPC...58..355B}. We used the dust opacity data from \cite{1994A&A...291..943O} for unprocessed dust grains with thin ice coatings\footnote{hera.ph1.uni-koeln.de/~ossk/Jena/tables.html}. The density and dust temperature models of H-MM1 and Oph\,D are shown in Fig.~\ref{3D_models}. The assumed levelling off of the gas temperature at 11\,K only affects the model of H-MM1, where $T_{\rm dust}$ rises to $\sim 14$\,K in the outer parts.

The fractional abundances were determined by applying minimum chi-square estimation. We ran three sets of radiative transfer calculations with different assumptions about the abundance distributions: (a) one assuming that the fractional $\dammo$ and $\ddammo$ abundances are constant throughout the core,  (b) another assuming that the abundances are constant in the outer parts but start to decrease when the $\htwo$ density exceeds $2\times10^5 \,\percc$, and (c) a third one assuming that the abundances are constant in the density range $5\times10^4\,\percc < n(\htwo) < 2\times 10^5\,\percc$ but decrease both towards the outer edge and the core centre. In models (b) and (c), the fractional abundances in the central parts were assumed to be inversely proportional to the density, that is, $X\propto n^{-1}$, owing to freezing onto the grains. This tendency was recently found in H-MM1 for $\ammo$ \citep{2022AJ....163..294P}, and it is possible that also the deuterated forms of ammonia behave that way (see, e.g., \citealt{2022ApJ...929...13C}). In model (c), the abundance in the outer part was assumed to be, somewhat arbitrarily, directly proportional to the density, $X \propto n$. At the core edge, the abundances of $\dammo$ and $\ddammo$ are expected to increase towards higher densities as the production of these molecules is favoured by CO depletion. This behaviour is also predicted by our chemistry models (see Fig.~\ref{X_radii_dens} below). 

We varied the undepleted baseline abundances, and calculated from the 3D model described above spectral cubes that were smoothed to the $20\arcsec$ resolution of APEX. Thereafter spectra corresponding to the pixels of LAsMA were extracted. The simulations were done using the LOC radiative transfer program \citep{2020A&A...644A.151J}. We compared the simulated and observed spectra in the brightest positions where the signal-to-noise ratios of the p$\dammo$ spectra exceed 10. For H-MM1, we included positions $(0,0)$ and $(+8\arcsec,+39\arcsec$), and for Oph\,D, we used positions $(0,0)$, $(+20\arcsec,+35\arcsec)$, and $(-19\arcsec,-37\arcsec)$ for fitting the fractional abundances.  

For each test value of the fractional abundance and in each position included, we evaluated the statistic
$$
\chi^2 = \sum_i \frac{(T_{{\rm obs},i}-T_{{\rm mod},i})^2}{\sigma^2} \; ,
$$
where $T_{{\rm obs},i}$ is the observed intensity and $T_{{\rm  mod},i}$ is the computed intensity in channel $i$, and $\sigma$ is the RMS noise of the observed spectrum. The summation is over approximately 100 channels including the three hyperfine groups. The sum of the obtained $\chi^2$ values over two or three positions shows a parabolic distribution as a function of the fractional abundance. The $x$-coordinate of the minimum corresponds to the best-fit fractional abundance. We fitted a parabola to this distribution, and denote the fitted function with $\tilde{\chi}^2$. The best-fit values, $\bar{X}$, and their  $1\sigma$ errors, $\sigma_X$, 
were calculated from 
$$
\bar{X} = \int\,X P_{\tilde{\chi}}(X) dX \; , \hspace{1cm} 
\sigma^2_X = \int (X-\bar{X})^2 P_{\tilde{\chi}}(X) dX \; ,
$$
where $P_{\tilde{\chi}} \propto e^{\tilde{\chi}^2/2}$ is the normalised probability distribution. The method is based on the assumption that the error distribution of the spectra is Gaussian \citep{2010arXiv1009.2755A}. We tested the error estimates of the $X$ values using Monte Carlo. To this end, we generated a few thousand realisations of the spectra by adding normally distributed noise ($\sigma$ = RMS) to each channel, and calculated the 'best-fit' $X$ values by minimum chi-squared estimation. The distribution of these $X$ values is Gaussian and the range containing $68\,\%$ of the values agrees with $\bar{X} \pm \sigma_X$ derived above. The resulting {maximum and average} fractional abundances and their $1\,\sigma$ uncertainties are presented in Table~\ref{chi2_fracs}. {By 'maximum' abundance we mean the undepleted, baseline abundance, and by 'average' abundance we mean the column density ratio $N(\rm  mol)/N(\htwo)$ in the direction of the density peak of the core model. The average abundances are used for comparison with the predictions of the chemistry model in Sect.~\ref{new_model}.} 


The assumption of molecular depletion at high (or low) densities results naturally in higher baseline (undepleted) abundances {and lower average abundances towards the density peak compared to} the assumption of a constant $X$. However, the relative abundances of the four species are approximately the same for all three $X$ profiles. As the main interest of this study is in the o/p ratios, we do not attempt to optimise the depletion power laws or the density thresholds. The simulated spectra using the $X$=constant model are shown in Appendix A together with the observed spectra. 

One can see from Table~\ref{chi2_fracs} that the best-fit o/p-$\dammo$ and o/p-$\ddammo$ ratios derived towards both cores are close to their statistical values 3 and 2, respectively, and that these ratios do not depend much on the abundance profile assumed. For H-MM1, the $3\,\sigma$ limits are within 12\,\% from the statistical ratios. For Oph D, both o/p-$\dammo$ and o/p-$\ddammo$ are higher than those derived in H-MM1, and the $3\,\sigma$ limits allow deviations up to 15\,\% and 24\,\% in excess of the statistical values 3 and 2, respectively.   

The fractionation ratio $\ddammo/\dammo$ is lower in H-MM1 (0.23-0.25) than in Oph\,D (0.27-0.29), where an exceptionally high $\ddiaz/\diaz$ column density ratio, $0.44\pm0.08$, has been derived previously by \cite{2005ApJ...619..379C}. The baseline fractional $\dammo$ abundance of $\sim 9\times10^{-9}$ (ortho+para) obtained here for H-MM1 assuming molecular depletion, together with the (undepleted) fractional $\ammo$ abundance $2\times10^{-8}$ (ortho+para) derived by \cite{2022AJ....163..294P}, imply an $\dammo/\ammo$ abundance ratio of 0.45.  The o/p ratios  of $\dammo$ and $\ddammo$, and the abundance ratios $\dammo/\ammo$ and $\ddammo/\dammo$ derived in H-MM1 are consistent with the previous estimates by \cite{2017A&A...600A..61H}. The noise level of the present 0.8\,mm spectra is a factor of two lower than those used in the earlier study, but perhaps a more important difference is that the analysis is now based on several positions and a realistic 3D model is used for the core. 

\begin{table*}
      \caption[]{Fractional abundances, $X$, of the observed species in H-MM1 and Oph\,D from minimum chi-squared estimation, together with the $\ddammo/\dammo$ ratios and the o/p ratios of $\dammo$ and $\ddammo$ derived from these.}
         \label{chi2_fracs}
         \begin{centering}
        \begin{tabular}{l|lll|llllll|llllll}\hline
        \noalign{\smallskip}
        \multicolumn{16}{c}{H-MM1} \\
        \noalign{\smallskip}
        \hline
                    & \multicolumn{3}{c}{(a)$^\dagger$} &
                       \multicolumn{6}{|c|}{(b)$^\dagger$} & 
                       \multicolumn{6}{c}{(c)$\dagger$} \\
                       & \multicolumn{3}{|c}{} & 
                       \multicolumn{3}{|c}{maximum} &
                       \multicolumn{3}{c|}{average} & 
                       \multicolumn{3}{c}{maximum} & 
                       \multicolumn{3}{c}{average}  \\
                            & \multicolumn{3}{c}{($\times10^{-9}$)} &
                       \multicolumn{3}{|c}{($\times10^{-9}$)} &  
                        \multicolumn{3}{c|}{($\times10^{-9}$)} &
                                   \multicolumn{3}{c}{($\times10^{-9}$)} &  
                        \multicolumn{3}{c}{($\times10^{-9}$)}   \\             
        $X(\odammo)^\ddagger$ & 4.520 &$\pm$& 0.028 & 6.617 &$\pm$& 0.042 & 2.581 &$\pm$& 0.017 & 6.720 &$\pm$& 0.043 & 2.471 &$\pm$& 0.016\\
        $X(\pdammo)$ & 1.469 &$\pm$& 0.021 & 2.374 &$\pm$& 0.032 & 0.926 &$\pm$& 0.013 & 2.426 &$\pm$& 0.035 & 0.892 &$\pm$& 0.016\\
        $X(\oddammo)$ & 0.925 &$\pm$& 0.014 & 1.477 &$\pm$& 0.022 & 0.576 &$\pm$& 0.009 & 1.491 &$\pm$& 0.023 & 0.548 &$\pm$& 0.008 \\
        $X(\pddammo)$ & 0.455 &$\pm$& 0.012 & 0.734 &$\pm$& 0.018 & 0.286 &$\pm$& 0.007 &  0.745 &$\pm$& 0.019 & 0.274 &$\pm$& 0.007 \\
   
        $X(\dammo)$ & 5.989 &$\pm$& 0.035 & 8.991 &$\pm$& 0.053 & 3.507 &$\pm$& 0.021 &  9.146 &$\pm$& 0.055 & 3.363 &$\pm$& 0.022 \\
        $X(\ddammo)$ & 1.380 &$\pm$& 0.018 & 2.211 &$\pm$& 0.029 & 0.862 &$\pm$& 0.011 &   2.236 &$\pm$& 0.030 & 0.822 &$\pm$& 0.011 \\ \hline
        \noalign{\smallskip}
        $\ddammo/\dammo$ & 0.230 &$\pm$& 0.003 & 0.246 &$\pm$& 0.004 & 0.246 &$\pm$& 0.003  & 0.244 &$\pm$& 0.004 & 0.244 &$\pm$& 0.007 \\
          o/p$\dammo$ & 3.077 &$\pm$& 0.048 & 2.787 &$\pm$& 0.042 & 2.788 &$\pm$& 0.042 &  2.770  &$\pm$& 0.043 & 2.770 &$\pm$& 0.052 \\
         o/p$\ddammo$ & 2.033 &$\pm$& 0.061 & 2.011 &$\pm$& 0.058 & 2.012 &$\pm$& 0.058 &  2.002 &$\pm$& 0.060 & 2.001 &$\pm$& 0.083 \\      
        \hline
        \noalign{\smallskip}
        \multicolumn{16}{c}{Oph D} \\ \hline
        \noalign{\smallskip}
                     & \multicolumn{3}{c}{(a)} &
                       \multicolumn{6}{|c|}{(b)} & 
                       \multicolumn{6}{c}{(c)} \\
                       & \multicolumn{3}{|c}{} & 
                       \multicolumn{3}{|c}{maximum} &
                       \multicolumn{3}{c|}{average} & 
                       \multicolumn{3}{c}{maximum} & 
                       \multicolumn{3}{c}{average}  \\
                            & \multicolumn{3}{c}{($\times10^{-9}$)} &
                       \multicolumn{3}{|c}{($\times10^{-9}$)} &  
                        \multicolumn{3}{c|}{($\times10^{-9}$)} &
                                   \multicolumn{3}{c}{($\times10^{-9}$)} &  
                        \multicolumn{3}{c}{($\times10^{-9}$)}   \\  
        $X(\odammo)^\ddagger$ & 4.640 &$\pm$& 0.045 & 5.345 &$\pm$& 0.052 & 3.170 &$\pm$& 0.031 & 5.535 &$\pm$& 0.052 & 3.113 &$\pm$& 0.029\\
        $X(\pdammo)$ & 1.469 &$\pm$& 0.026 & 1.649 &$\pm$& 0.030 & 0.978 &$\pm$& 0.018 &    1.710 &$\pm$& 0.032 & 0.962 &$\pm$& 0.018\\
        $X(\oddammo)$ & 1.122 &$\pm$& 0.019 & 1.271 &$\pm$& 0.022 & 0.754 &$\pm$& 0.013 &   1.434 &$\pm$& 0.026 & 0.806 &$\pm$& 0.015\\
        $X(\pddammo)$ & 0.502 &$\pm$& 0.015 & 0.580 &$\pm$& 0.018 & 0.344 &$\pm$& 0.011 &    0.651 &$\pm$& 0.021 & 0.366 &$\pm$& 0.012\\

        $X(\dammo)$ & 6.109 &$\pm$& 0.052 & 6.994 &$\pm$& 0.061 & 4.148 &$\pm$& 0.036 &
        7.245 &$\pm$& 0.061 & 4.075 &$\pm$& 0.034\\
        $X(\ddammo)$ & 1.624 &$\pm$& 0.025 & 1.851 &$\pm$& 0.028 & 1.098  &$\pm$& 0.017 &
        2.085 &$\pm$& 0.034 & 1.173 &$\pm$& 0.019 \\ \hline
        \noalign{\smallskip}
        $\ddammo/\dammo$ & 0.266 &$\pm$& 0.005 & 0.265 &$\pm$& 0.005 &0.265 &$\pm$& 0.005 &
        0.288 &$\pm$& 0.005 & 0.288 &$\pm$& 0.005 \\
         o/p$\dammo$ & 3.159 &$\pm$& 0.064 & 3.241 &$\pm$& 0.068 & 3.242 &$\pm$& 0.067 & 
         3.237 &$\pm$& 0.067 & 3.237 &$\pm$& 0.068\\
         o/p$\ddammo$ & 2.235 &$\pm$& 0.077 & 2.191 &$\pm$& 0.076 & 2.191 &$\pm$& 0.078 &
         2.201 &$\pm$& 0.081 & 2.203 &$\pm$& 0.120 \\       
        \hline
 \noalign{\smallskip}
        \end{tabular}

\end{centering}
$^\dagger${\footnotesize The letters indicate three different assumptions about the abundance distributions: (a) $X$ is constant throughout the core; (b) $X$ decreases in the densest parts with $n(\htwo)>2\times10^5\,\percc$, following the power law $n^{-1}$; (c) $X$ increases proportional to $n(\htwo)$ at densities below $5\times10^4\,\percc$,  and decreases again as in model (b) above $2\times10^5\,\percc$. For models including depletion, 'maximum' means the undepleted fractional abundance and 'average' means the column density ratio $N({\rm mol})/N(\htwo)$ through the density peak of the core.}

$\ddagger${\footnotesize The $X$ values are given in units of $10^{-9}$. }         
  
\end{table*}


\section{Spin ratios from the proton hop and full scrambling scenarios and the predictions of previous chemistry models}
\label{old_models}

Deuterium enrichment of $\ammo$ in dense interstellar gas is supposed to occur primarily through deuteron donation from $\htwodplus$, $\dtwohplus$, or $\dthreeplus$ to ammonia.  {Deuterium can also enter the ammonia formation sequence in reactions of other nitrogen hydrides, such as NH, with the cations $\htwodplus$, $\dtwohplus$, and $\dthreeplus$.} {These exothermic reactions} can be thought to proceed via a direct {proton or deuteron hop (hereafter PH)} or via the formation of an intermediate complex where H and D nuclei can be exchanged. In the latter mechanism, called full scrambling {(hereafter FS)}, the nuclear spin state of the resulting cation, for example ${\rm NH_3D^+}$, depends on the initial spin states of both reactants according to the selection rules that take into account the conservation of the spin and symmetry (e.g., \citealt{2013JPCA..117.9800R}; \citealt{2013ApJ...770L...2F}; \citealt{2015A&A...581A.122S}; \citealt{2018MNRAS.477.4454H}). The spin states of the $\htwodplus$, $\dtwohplus$, and $\dthreeplus$ ions, in turn, are determined by spin-state conversions in the $\hthreeplus + \htwo$ reaction system and its deuterated variants. The state-to-state and ground-state-to-species rate coefficients of this system were derived by \cite{2009JChPh.130p4302H}, and these have been used in spin-state chemistry models ever since (e.g., \citealt{2009A&A...494..623P}; \citealt{2010A&A...509A..98S}; \citealt{2014A&A...562A..83L}; \citealt{2018MNRAS.477.4454H}).

To illustrate how the reaction details affect the nuclear spin ratios, we { discuss} the outcome of a basic sequence leading to $\dammo$, assuming that this follows either the PH or the FS scenario. The sequence is initiated by $\ammo + \htwodplus$ and leads to $\dammo$ through dissociative electron recombination of ${\rm NH_3D^+}$. Each reaction step has several   possible products, but here we concentrate on the branch ending with   $\dammo$. { The spin-separated sequences according to the PH and FS scenarios are shown in Fig.~\ref{branching_diagram}.}

{In the PH mechanism, $\htwodplus$ simply donates the deuteron to   ammonia, and the spin symmetry of the hydrogen complex does not change. The H$_3$ complex of $E$ (``para'') symmetry dissociates to form ${\rm o\htwo + H}$ and ${\rm p\htwo + H}$ at equal probabilities, whereas the $A_1$ (``ortho'') symmetry dissociates to form ${\rm o\htwo + H}$ (if not $\rm H+H+H$). Assuming equal amounts of o$\ammo$ and p$\ammo$, this sequence produces three times more o$\dammo$ than p$\dammo$. In the FS scenario, the first step forms the reaction complex $\rm (NH_5D^+)^\dagger$. With reactants $\rm o\ammo$ and $\rm p\htwodplus$, the resulting H$_5$ complex is   of $G_1$ symmetry, which gives ${\rm H_3}(A_1)$ ("ortho") and ${\rm H_3}(E)$ ("para") at equal probabilities when dissociating (\citealt{2009JChPh.130p4302H}, Table III;   \citealt{2018MNRAS.477.4454H}, Table C2). The H$_5$ complex formed in the reaction $\rm p\ammo + p\htwodplus$ has $H_1$ symmetry with strong   preference to ${\rm H_3}(E)$ in dissociation. Assuming again equal amounts of o$\ammo$ and p$\ammo$, the FS sequence results in an o/p-$\dammo$ ratio of $27/13\approx 2.1$.

  The difference between the PH and FS scenarios is similar for other routes to $\dammo$. For example, according to the PH model, the sequence ${\rm NHD^+ \stackrel{\htwo}{\rightarrow} NH_2D^+ \stackrel{\htwo}{\rightarrow} NH_3D^+ \stackrel{e^-}{\rightarrow} \dammo}$ produces again o$\dammo$ and p$\dammo$ in the statistical ratio 3:1, whereas the same sequence in the FS model results in an o/p-$\dammo$ ratio of $7/5=1.4$, assuming that the reacting $\htwo$ is always in the more common para form. Finally, the formation of $\ddammo$ involves the combination of two deuterium nuclei. If this happens one by one, as in the PH model, the resulting o/p-$\ddammo$  ratio is always 2. In the FS model, the outcome depends on the deuteron donor. For example, in a reaction between a singly deuterated species and o$\dtwohplus$, the o/p ratio of the $\rm D_2$ complex formed is 13/5=2.6. The branching ratios of the reactions discussed here can be verified by inspecting the nuclear spin symmetry induction and subduction matrices for systems with multiple H and D atoms presented in \cite{2009JChPh.130p4302H},  \cite{2015A&A...581A.122S}, and  \cite{2018MNRAS.477.4454H}.}

{The fact that the FS assumption in proton-donation reactions leads to deviations of the o/p ratios of $\dammo$ and $\ddammo$ from their high-temperature statistical values was noted by 
\cite{2015A&A...581A.122S},  \cite{2016MNRAS.457.1535D}, and  \cite{2018MNRAS.477.4454H} 
(see Table~5 of \citealt{2018MNRAS.477.4454H}).} This disagreed with observations which suggested statistical ratios, albeit with large uncertainties. \cite{2016MNRAS.457.1535D} pointed out that the observed o/p ratios of $\dammo$ and $\ddammo$ are consistent with formation on grains, while \cite{2017A&A...600A..61H} argued that  statistical spin ratios could be explained by gas-phase reactions if they can be characterised as single-particle hops, in analogy with H and D atom additions on grains. The latter suggestion was examined by \cite{2019A&A...631A..63S} who considered the effects of {PH vs. FS} in proton-donation reactions, such as $\ammo + \dtwohplus$, and in the hydrogen abstraction chain leading to the ammonium ion, ${\rm N^+ \stackrel{\htwo}{\rightarrow} NH^+ \stackrel{\htwo}{\rightarrow} NH_2^+ \stackrel{\htwo}{\rightarrow} NH_3^+ \stackrel{\htwo}{\rightarrow} NH_4^+}$, and its deuterated variants. It turned out that replacing the {FS with PH} in proton-donation reactions brought the predicted o/p ratios a little closer to the statistical values but the difference was still greater than 30\%. On the other hand, by extending the {PH} scenario also to H abstraction reactions, the o/p ratios agreed with those observed previously towards H-MM1, but the modelled fractionation ratios $\dammo/\ammo$ and $\ddammo/\dammo$ exceeded those derived from the observations.

\section{Gas-grain chemistry model with dynamic chemical desorption}   
\label{new_model}

One way to explain the discrepancy between the observed and modelled spin ratios is that the desorption of ammonia from grain surfaces is in reality more efficient than what has been considered in the previous simulations. Ammonia does not desorb thermally in low-temperature conditions owing to its high binding energy on ice; usually values around $4000-5000$\,K are assumed for the $\ammo$ binding energy on water ice (\citealt{Minissale22};  \citealt{2023ApJ...944..142F}). Although recent studies indicate that we should rather speak of a broad distribution of binding energies on amorphous water ice (\citealt{2020A&A...643A.155G}; \citealt{2022ESC.....6.1514T}), the desorption of ammonia most likely occurs through non-thermal mechanisms.

There are several non-thermal desorption mechanisms that are typically considered in astrochemical models, namely photodesorption, cosmic-ray induced desorption, and chemical desorption (also known as reactive desorption). The first two mechanisms do not release ammonia into the gas phase efficiently enough to cause significant ``contamination'' of the gas-phase ammonia reservoir. Chemical desorption (CD) on the other hand has been shown to have a very large impact on ammonia abundance in starless core conditions \citep{Sipila19a}, assuming that the desorption efficiency is constant at 1\,\% \citep{Garrod07}. Hence, out of the three options mentioned here, CD is the most promising candidate for pushing the gas-phase ammonia spin ratios toward statistical values. However, the constant efficiency assumption for CD was found to lead to gas-phase ammonia abundances much higher than what is indicated by observations for volume densities around $10^4\,\percc$ \citep{Sipila19a}.

{ It is reasonable to posit that} in reality the efficiency of CD is in fact not universally constant, but would instead depend not only on the desorbing molecule but also on the type of surface on which the desorption is taking place. An alternative description of CD exploring this scenario has been presented by \citet{Minissale16a}, and applied to a chemical model by \citet{Vasyunin17}. An important conclusion in these works is that the CD efficiency tends to increase when switching from a water ice surface to a CO ice surface, which in a starless core would correspond to the transition from outer regions to the dense inner core where the freeze-out timescale is short and the grains are covered by a multitude of species, and not only water. Motivated by these findings, { we investigate here} whether the dynamic nature of the CD process {influences the spin ratios of deuterated ammonia} -- we recall that the observed ammonia lines originate in volume densities approaching $10^5\,\percc$ where the ices on the grains are made up of a mixture of molecules.

 We use for the simulations the chemical model described in \citet{2019A&A...631A..63S}, and adopt the full scrambling and proton hop chemical networks presented therein. In the proton hop model, proton donation and abstraction reactions proceed via the exchange of one proton (or deuteron) between the reactants, while in the full scrambling model multiple atom exchanges are possible. Also, for the sake of simplicity, we employ a one-dimensional physical model for H-MM1 recently presented in \citeauthor{2022AJ....163..294P}\,(\citeyear{2022AJ....163..294P}; their Fig. 11). The adoption of a one-dimensional spherically symmetric model means that we can with reasonable accuracy compare the simulation results to the observations obtained toward the dust peak, but we cannot attempt to reproduce the morphology of the deuterated ammonia distributions beyond a distance of a few tens of arcseconds from the core center.
 
{ We consider two different approaches to simulating CD: one where the CD efficiency is set to a constant value of 1\,\% \citep{Garrod07}, and another where the CD efficiency changes dynamically. The latter is an updated version of the model of \citet{Vasyunin17} -- in the present work, the energy released in a chemical desorption event is partitioned among the reaction products according to their total degrees of freedom. In the dynamic model, the CD efficiency of a given species depends on the ice composition, and hence on the location in the core and on the simulation time, as well as the properties (mass, binding energy) of the species itself. At late simulation times, however, we obtain nearly constant CD efficiencies as a function of radius. For example for $\rm NH_2D$, the dynamic model predicts a CD efficiency of 4.1\% at the center of the core, and 4.9\% at the edge, at a simulation time of $5\times10^5 \, \rm yr$. A detailed account of the new dynamic CD model will be given in \cite{2023arXiv231008389R}. 

We employ the same initial abundances as in \citet{2019A&A...631A..63S}, not recounted here for brevity, and most of the simulation parameters are also the same as in that paper. There are two notable differences: here we assume an external visual extinction $A_{\rm V} = 3 \, \rm mag$ following \citet{2022AJ....163..294P} (2 mag in \citealt{2019A&A...631A..63S}), and we use the new description for cosmic-ray induced desorption from \citet{Sipila21}.
}

\begin{figure}
\unitlength=1mm
\begin{picture}(80,65)(0,0)
\put(0,0){
\begin{picture}(0,0) 
\includegraphics[width=9cm,angle=0]{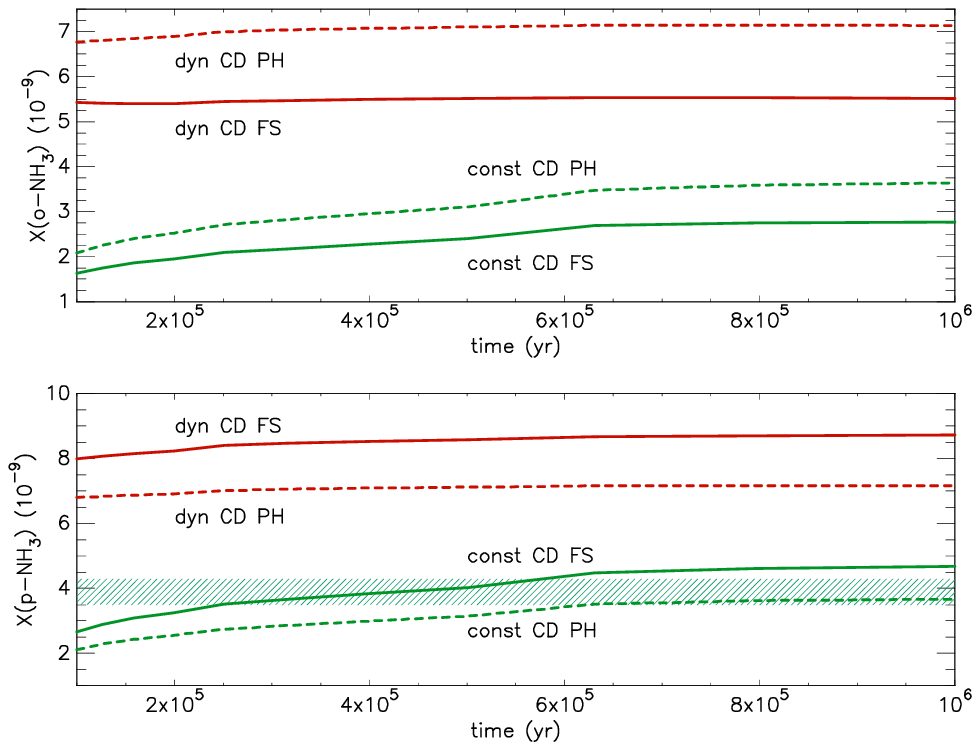}
\end{picture}}
\end{picture}  
  \caption{Evolution of the average fractional ortho- and para-$\ammo$ abundances according to different varieties of our chemistry model: "const CD" means chemical desorption (CD) with a constant (1\,\%) efficiency, "dyn CD" means changing efficiency of CD depending on the composition of the grain surface, "FS" means full scrambling of protons in proton donation and hydrogen abstraction reactions, and "PH"  means proton hop in these reactions (see text). The averages are {column density ratios $N(\rm mol)/N(\htwo)$} taken through the density maximum of the core model used in chemical simulations. The shaded area in the lower panel shows the range of similarly taken average of the fractional p$\ammo$ abundance through the 3D model of H-MM1 based on the results of \citet{2022AJ....163..294P}.}
\label{XNH3_times}
\end{figure}

\begin{figure}
\unitlength=1mm
\begin{picture}(80,65)(0,0)
\put(0,0){
\begin{picture}(0,0) 
\includegraphics[width=9cm,angle=0]{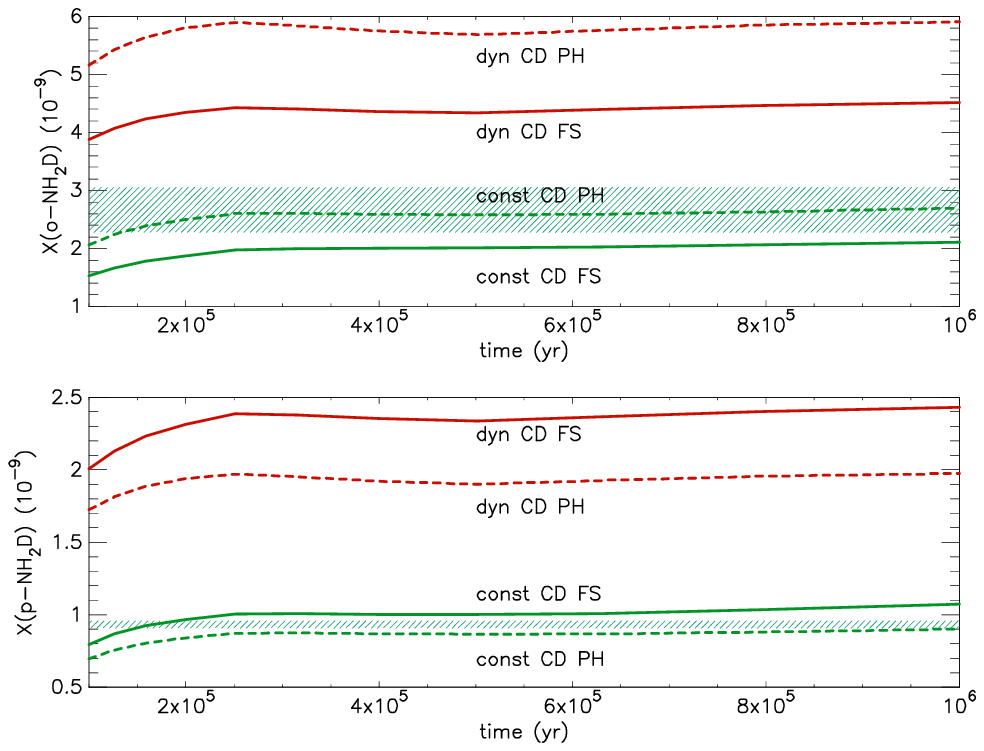}
\end{picture}}
\end{picture}  
   \caption{Same as Fig.~\ref{XNH3_times} but for ortho- and para-$\dammo$. The shaded areas show the ranges of averages through the 3D model with the abundance profiles (b) and (c) used in the fitting of the observed spectra towards H-MM1 {and Oph\,D} (see Table~\ref{chi2_fracs}).}
   \label{XNH2D_times}
\end{figure}

\begin{figure}
\unitlength=1mm
\begin{picture}(80,65)(0,0)
\put(0,0){
\begin{picture}(0,0) 
\includegraphics[width=9cm,angle=0]{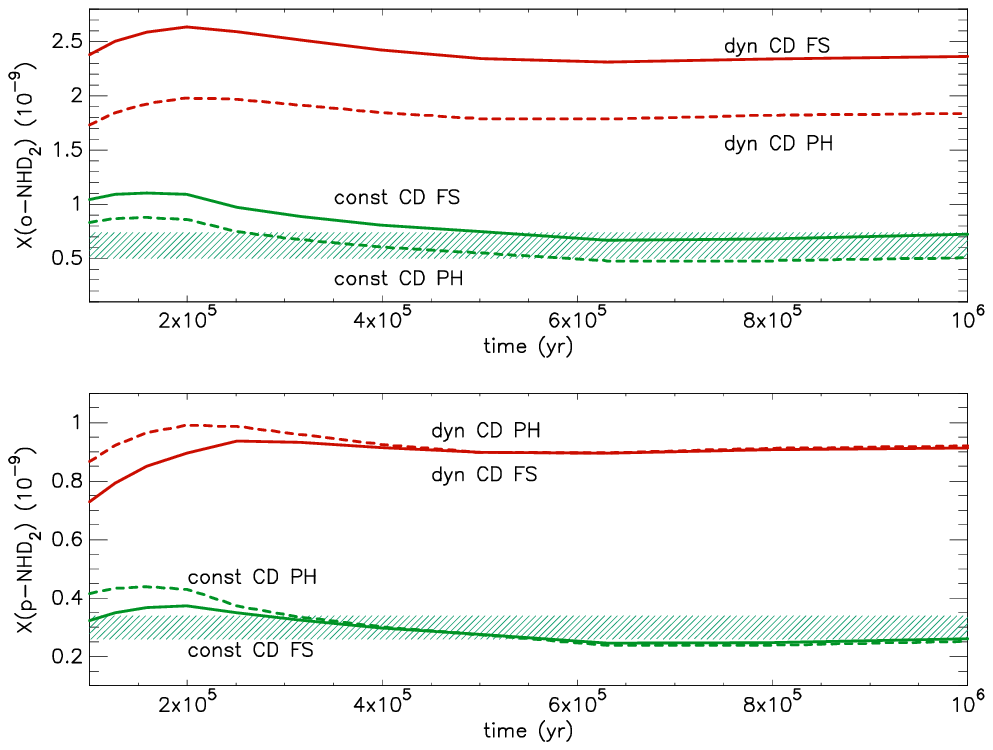}
\end{picture}}
\end{picture}  
   \caption{Same as Fig.~\ref{XNH2D_times} but for ortho- and para-$\ddammo$.}
   \label{XNHD2_times}
\end{figure}

\begin{figure}
\unitlength=1mm
\begin{picture}(80,65)(0,0)
\put(0,0){
\begin{picture}(0,0) 
\includegraphics[width=9cm,angle=0]{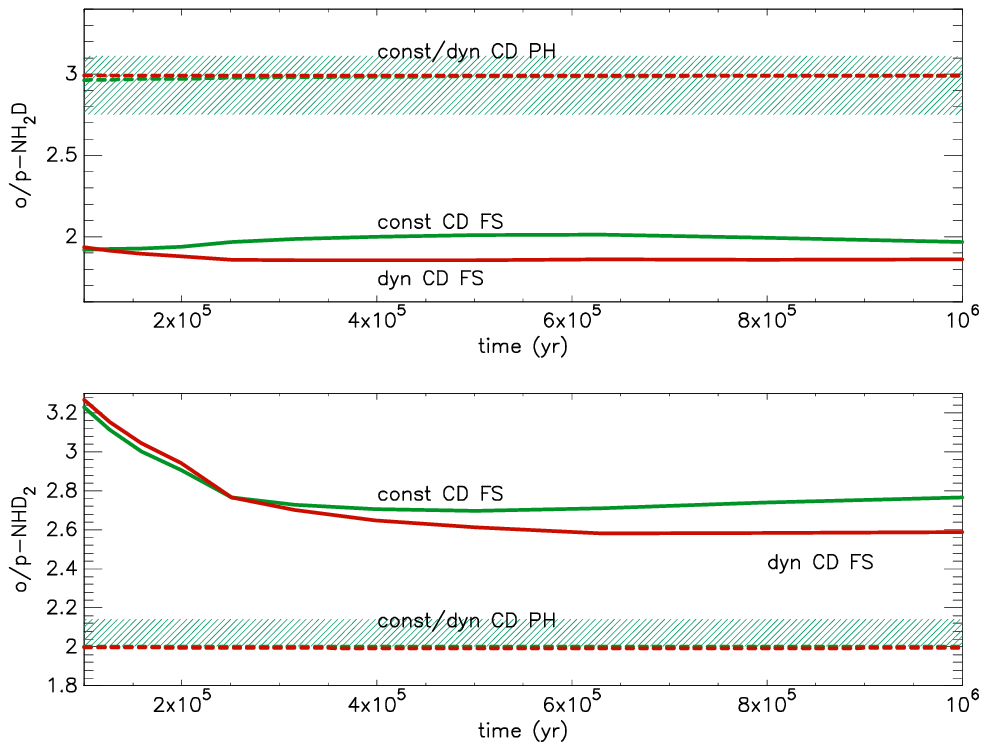}
\end{picture}}
\end{picture}  
   \caption{Evolution of the average o/p-$\dammo$ and o/p-$\ddammo$ ratios according to our chemistry model. The labelling is the same as in Fig.~\ref{XNH3_times}. The ranges derived from observations using the 3D {models for H-MM1 and Oph\,D} with abundance profiles (b) and (c) (see Table~\ref{chi2_fracs}) are indicated 
   with shaded areas.}
   \label{op_times}
\end{figure}

\begin{figure}
\unitlength=1mm
\begin{picture}(80,65)(0,0)
\put(0,0){
\begin{picture}(0,0) 
\includegraphics[width=9cm,angle=0]{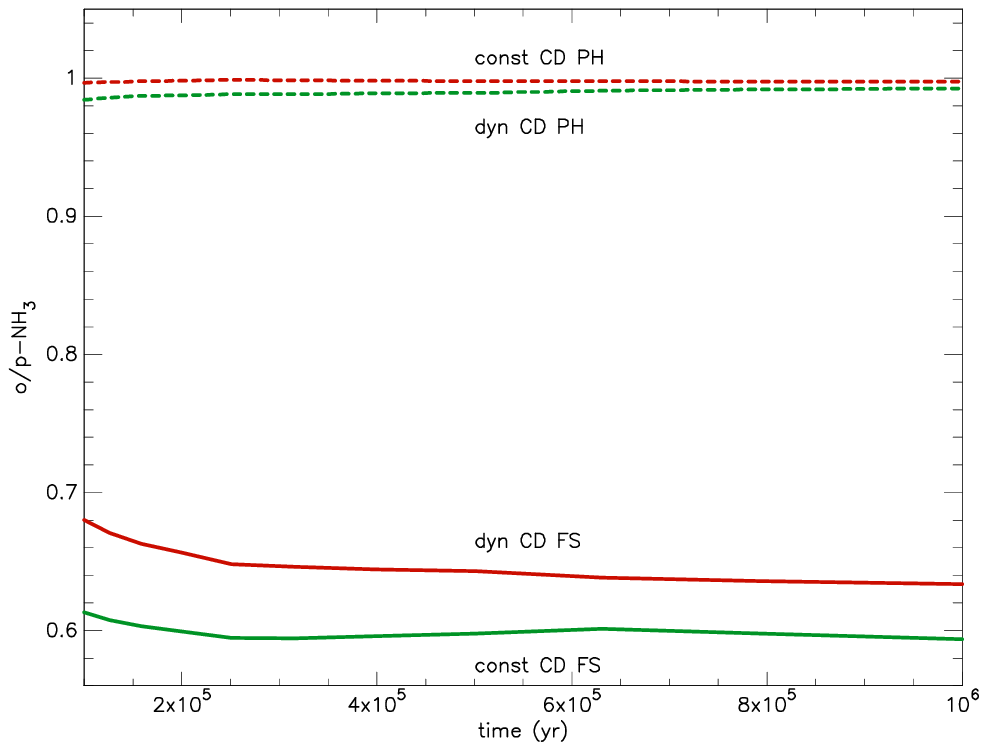}
\end{picture}}
\end{picture}  
   \caption{Same as Fig.~\ref{op_times} but for $\ammo$. There is no observational estimate of the o/p-$\ammo$ ratio towards H-MM1 {or Oph\,D}.}
   \label{nh3_op_times}
\end{figure}

\begin{figure}
\unitlength=1mm
\begin{picture}(80,65)(0,0)
\put(0,0){
\begin{picture}(0,0) 
\includegraphics[width=9cm,angle=0]{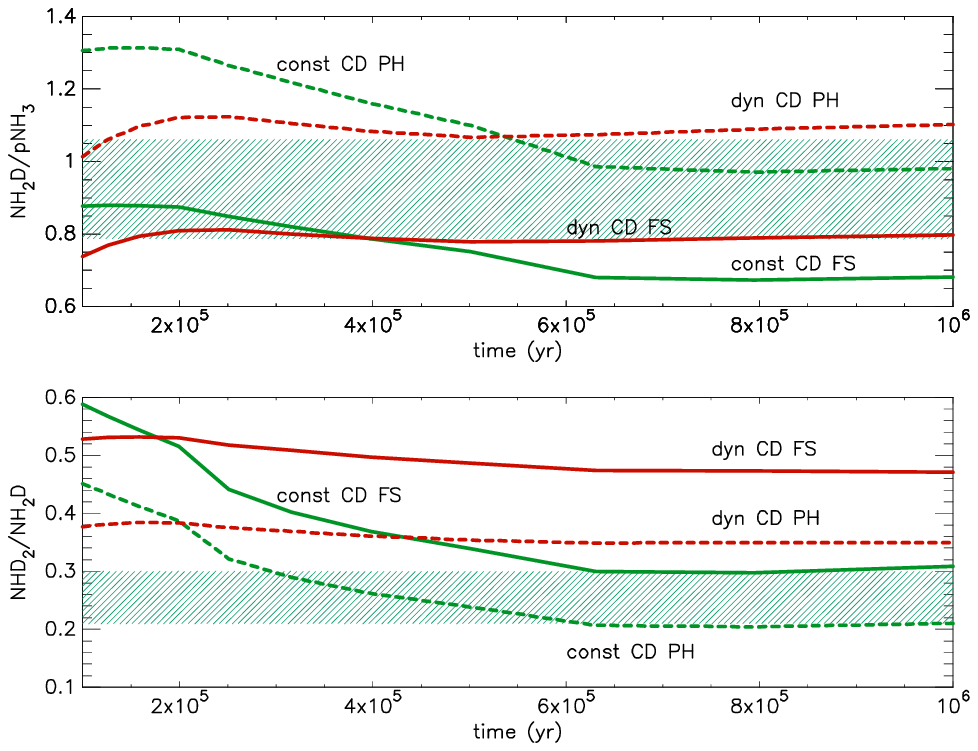}
\end{picture}}
\end{picture}  
   \caption{{ Evolution of the average deuterium fractionation ratios $\dammo/{\rm p}\ammo$ and $\ddammo/\dammo$ according to the chemistry model.} The labelling is the same as in Fig.~\ref{XNH3_times}. {In the top panel, the shaded area indicates { the ratio derived from observations towards H-MM1, where the fractional p$\ammo$ abundance has been determined by \citet{2022AJ....163..294P}.} In the bottom panel, the "observed" ratios include both H-MM1 and Oph\,D.}}
   \label{Dfrac_times}
\end{figure}

 In Figs.~\ref{XNH3_times} to \ref{Dfrac_times}, we compare { the average fractional abundances, that is, the column density ratios $N({\rm mol})/N(\htwo)$ towards the density peaks of the cores, and similarly derived average o/p and deuterium fractionation ratios} predicted by the chemical models to those derived by fitting spectra observed towards H-MM1 and Oph\,D. The { fractional $\dammo$ and $\ddammo$ abundances relative to $\htwo$ and the abundance ratios} derived from observations are column density ratios through the density peak of the 3D model with abundance profiles (b) and (c) provided in Table~\ref{chi2_fracs}. { The ranges indicated with hatched areas correspond to the inverse-variance weighted averages and the standard deviations of the values derived towards H-MM1 and Oph\,D.} For the observed para-$\ammo$ ({ hereafter p$\ammo$}) abundance, we adopt the result from \citet{2022AJ....163..294P} that the fractional (ortho+para) $\ammo$ abundance is $2\times10^{-8}$ ($\pm 10\%$) in the outer core, but decreases with the density when $n(\htwo)>2\times10^5\,\percc$. \citet{2022AJ....163..294P} could only derive the p$\ammo$ abundance, and assumed o/p-$\ammo=1$.  

One can see from Figs.~\ref{XNH3_times}, \ref{XNH2D_times}, and \ref{XNHD2_times} that the two CD scenarios result in large differences in the average $\ammo$, $\dammo$, and $\ddammo$ abundances. The fractional p$\ammo$, $\dammo$, and $\ddammo$ abundances predicted by the constant CD models are close to those derived from the observations, whereas the dynamic CD models seem to produce too much of these molecules. {On the other hand,  the assumed gas-phase reaction mechanism - either PH or FS - has a minor effect on the abundances, and for most species the predicted abundances from the two models are contained within { or lie close to} the range of observationally derived values. The only exception is o$\dammo$, for which the model assuming constant CD with PH seems to give a better agreement than the FS model.} 

{It is evident from Fig.~\ref{op_times} that both CD models assuming PH agree with the o/p ratios of $\dammo$ and $\ddammo$ derived from the observations, whereas { the} FS models under-predict o/p-$\dammo$ ratio by approximately $30\%$ and over-predict the o/p-$\ddammo$ ratio by a similar amount.  Including results for both H-MM1 and Oph\,D, the observationally derived ratios are { approximately o/p-$\dammo\sim 2.9 \pm 0.2$ and o/p-$\ddammo\sim 2.1 \pm 0.1$}.  As expected, the PH models predict statistical spin ratios. In the FS models, the o/p-$\dammo$ ratios settle around 1.9-2.0, and the o/p-$\ddammo$ ratios are found in the range 2.6-2.8 at late times.}

{ According to Fig.\ref{Dfrac_times}, the $\ddammo/\dammo$ fractionation ratio is reproduced by the constant CD {PH} model at simulation times around $5\times10^5$\,yr}, and also the constant CD FS model approaches the observed values at late times of the simulation. { The constant CD PH model over-predicts the $\dammo/{\rm p}\ammo$ ratio except at late times ($>5.5\times10^5$\,yr), whereas the prediction of the constant CD FS model agrees with the observed ratio only at early times ($<4\times10^5$\,yr). The dynamic FS models seem to work better for this ratio.} 

\begin{table}
\caption[]{Relative deviations, $|q_{\rm mod}-q_{\rm obs}|/q_{\rm obs}$, of the predicted abundance ratios from the observed values.} 

\begin{centering}
\begin{tabular}{lcccc} \hline
\noalign{\smallskip}
quantity & \multicolumn{4}{c}{relative deviation$^\dagger$} \\
 & \multicolumn{2}{c}{constant CD} & \multicolumn{2}{c}{dynamic CD} \\ 
  & FS & PH & FS & PH  \\ \hline
$\dammo/{\rm p}\ammo$ & 0.19 & 0.19 & 0.16 & 0.16 \\
$\ddammo/\dammo$ & 0.33 & 0.06 & 0.91 & 0.39 \\ 
o/p-$\dammo$     & 0.29 & 0.05 & 0.35 & 0.05 \\
o/p-$\ddammo$    & 0.31 & 0.03 & 0.27 & 0.03 \\ 
X($\dammo$)      & 0.16 & 0.04 & 0.85 & 1.11 \\
X($\ddammo$)     & 0.11 & 0.10 & 2.53 & 1.92 \\ \hline
mean             & 0.23 & 0.08 & 0.84 & 0.61 \\
\noalign{\smallskip}
\end{tabular}

$^\dagger${\footnotesize The deviations are given for the simulation time $5\times10^5$\,yr.}
\end{centering}

\label{deviations}
\end{table}

{To summarize, the models assuming constant CD (with PH or FS) come closer to the observed abundances than the dynamic CD models. The models assuming PH in the gas-phase reactions (combined with constant or dynamic CD) can reproduce the observed spin ratios, whereas the FS models predict too low o/p-$\dammo$ ratios and too { high o/p-$\ddammo$ ratios}. Furthermore, the $\ddammo/\dammo$ ratio { can be reproduced by the constant CD PH model at a certain time interval ($\sim 3-6\times10^5$\,yr), but this model over-predicts the $\dammo/{\rm p}\ammo$ ratio before  the simulation time $\sim 5.5\times10^5$\,yr, reaching the observed ratio only later. Both deuterium fractionation ratios predicted by the constant CD PH model are within the observed range only during a short period before $6\times10^5$\,yr.} 

{ In an attempt to quantify} the overall agreement or disagreement of the four models with the observations, we present in Table~\ref{deviations} the relative deviations of the { predicted $\dammo/{\rm p}\ammo$ and $\ddammo/\dammo$ ratios, o/p ratios, and fractional $\dammo$ and $\ddammo$ abundances from the values derived from observations}. The relative deviation of the quantity $q$ is defined by $\Delta q \equiv |q_{\rm mod}-q_{\rm obs}|/q_{\rm obs}$. The modelled values are taken { at the time $5\times10^5$\,yr. The observed values are variance-weighted averages of those derived towards the centres of H-MM1 and Oph\,D}. One can see that both PH models give small relative deviations for the o/p ratios, but the constant CD PH model provides the smallest mean deviation when all quantities are included. As discussed above, the dynamic CD models over-predict the fractional abundances, and this causes the large relative deviations in the last two rows of { Table~\ref{deviations}} for these models. { As can be expected from Figs.~\ref{XNH2D_times}, \ref{XNHD2_times}, \ref{op_times}, and \ref{Dfrac_times}, the relative deviations change with time, especially at early times ($<3\times 10^5$\,yr), except for the o/p ratios predicted by the PH models which are nearly constant. The constant CD PH model gives the lowest mean deviation at all times. We note that in general, the abundances of $\dammo$ and $\ddammo$, and thus the deuterium fractionation ratios, depend on the physical conditions (temperature and density), whereas the spin ratios are directly related to the chemical reaction mechanisms.}}

Spectra calculated from the 3D cloud model for H-MM1 with abundances interpolated from the 1D chemical model for three simulation times are shown in Figs.~\ref{constCD_FS_spectra} to \ref{dynCD_PH_spectra} in Appendix\,\ref{appendix:simulated_spectra_pyRate}. The interpolation uses fractional abundances as functions of density in the 1D model. The spectra observed towards the centre of H-MM1 are superposed on the simulated spectra.  The abundances from the constant CD model with {FS} can approximately reproduce the p$\dammo$ and p$\ddammo$ lines, but predict too weak o$\dammo$ lines and too strong o$\ddammo$ lines (Fig.~\ref{constCD_FS_spectra}). {The constant CD model with {PH} (Fig.~\ref{constCD_PH_spectra}) reaches rather a good agreement in all four spectra at late times.} As expected from Figs.~\ref{XNH2D_times} and \ref{XNHD2_times}, abundances from the dynamic CD models over-predict the line intensities of all the observed lines, and the difference is particularly stark for the $\ddammo$ lines. The simulated spectra, which depend on the beam-averaged column densities towards the centre of the core model, {support the idea conveyed by Table~\ref{deviations} that the constant CD model with PH agrees better with the observations than the other three chemical models tested here.}

\section{Discussion}
\label{discussion}

The observations and {their} analysis presented here indicate that the o/p ratios of $\dammo$ and $\ddammo$ in two cold, dense cores deviate at most 20\%  from their high-temperature statistical values o/p-$\dammo=3$ and o/p-$\ddammo=2$. The result conforms with previous observational determinations, with a higher accuracy than previously attained. Statistical spin ratios do not agree with {previously published} astrochemical models that assume complete scrambling of light nuclei in gas-phase reactions. These typically predict an o/p-$\dammo$ ratio of $\sim2$ or slightly below (\citealt{2015A&A...581A.122S}; \citealt{2018MNRAS.477.4454H}). For o/p-$\ddammo$, the predictions diverge depending on some differences in complex-forming reactions. In the model of \cite{2018MNRAS.477.4454H}, o/p-$\dammo$ remains below 3, whereas it approaches 4 in the original spin-state chemistry model of \cite{2015A&A...581A.122S}; see Sect. 3.3.4 and Fig.~6 of \cite{2018MNRAS.477.4454H}.

Using chemical models that include both gas-phase and grain-surface reactions, and two different chemical desorption scenarios, we tested two suggestions that have been put forward to explain the discrepancy between observations and models: 1) gas-phase ammonia has a substantial contribution from grains where H and D addition reactions result in statistical spin ratios \citep{2016MNRAS.457.1535D}, and 2) gas-phase reactions that form ammonia and its deuterated isotopologues occur through proton or atom hops instead of particle exchange in a reaction complex (\citealt{2017A&A...600A..61H}; \citealt{2019A&A...631A..63S}). According to nuclear spin selection rules, reactions where particles are added one by one should produce species in proportion to their nuclear spin statistical weights.

It turns out that while the adopted chemical desorption mechanism (constant or dynamic) affects strongly the average abundances of ammonia and its deuterated isotopologues (Figs.~\ref{XNH3_times}, \ref{XNH2D_times}, and \ref{XNHD2_times}), the effect of this mechanism on their spin ratios is less marked (Figs.~\ref{op_times} and \ref{nh3_op_times}). {Switching from constant to dynamic CD increases the fractional $\ammo$, $\ddammo$, and $\ddammo$ abundances by a factor of 2-3. In the FS scenario, dynamic CD
decreases the o/p-$\dammo$ and o/p-$\ddammo$ ratios by $\sim 10\,\%$, and increases the o/p-$\ammo$ ratio by a similar amount, but has hardly any effect on the spin ratios predicted by the PH model.} 
{The poor agreement of the dynamic CD model with the observed fractional abundances suggests that desorption is not as efficient as assumed there, rendering the constant CD model the more realistic of the two.} 

{ An input of ammonia} from grains seems plausible in view of the fact that the fractional abundance of $\ammo$ residing on grains is probably several orders of magnitude higher than that in the gas in starless cores (e.g., \citealt{2022ESC.....6.1514T}). The recent  quantum mechanical simulations of \cite{2023ApJ...944..142F} indicate that direct chemical desorption of ammonia from water ice is unlikely, because the reaction energy of hydrogenation is efficiently absorbed by the ice surface. However, the situation may be different in CO-rich ices, which are supposed to be characteristic of dense, starless cores, where also deuterated ammonia is found. The dynamic CD model discussed above is based { on these ideas, but it seems to over-predict the desorption efficiency from CO-rich ices}. 

{ Our modelling results suggest that the observed statistical o/p ratios of $\dammo$ and $\ddammo$ are primarily determined by gas phase reactions and are not an effect of vigorous desorption from grains. Firstly, the spin ratios depend strongly on the assumed gas-phase proton exchange scenario (FS or PH) also in the dynamic CD model with very efficient desorption. Secondly, the predicted deuterium fractionation ratios in the grain mantles ($\dammo/\ammo < 0.1$ and $\ddammo/\dammo\sim0.1$) are lower than those observed in starless dense cores, including the targets of the present study.  The observed fractionation ratios agree with the predictions for gas-phase formation. Inspection of the rates of the principal reactions producing and destroying ammonia in a given simulation cell at a certain time shows that desorption and accretion typically account for a few percent of the production and destruction of ammonia at densities ($\sim 10^5\,\percc$) where we believe the observed lines to originate, while the formation through ${\rm NH_4^+ + e^-}$ and destruction through $\ammo + {\rm H_3^+}$ (and isotopologues), and to a lesser extent through charge exchange between $\ammo$ and ${\rm H^+}$, dominate (see also Fig.\,2 in \citealt{2015A&A...581A.122S}). This means that, according to our simulations, ammonia and its deuterated forms are rapidly re-processed by ion-molecule reactions in the gas phase, even though desorption from grain surfaces may have a signiﬁcant contribution to their total abundances. By re-processing we mean constant conversion of nitrogen hydrides to cations through charge or proton exchange reactions, and their return to neutral species in recombination reactions with electrons, for example,  $\ammo\stackrel{\rm H^+}{\rightarrow} NH_3^+ \stackrel{\htwo}{\rightarrow} NH_4^+ \mathrel{\substack{{\rm e^-}\\ \rightleftarrows \\ {\rm H_3^+}}} \ammo$.}

\begin{figure}
\unitlength=1mm
\begin{picture}(80,70)(0,0)
\put(0,-2){
\begin{picture}(0,0) 
\includegraphics[width=9cm,angle=0]{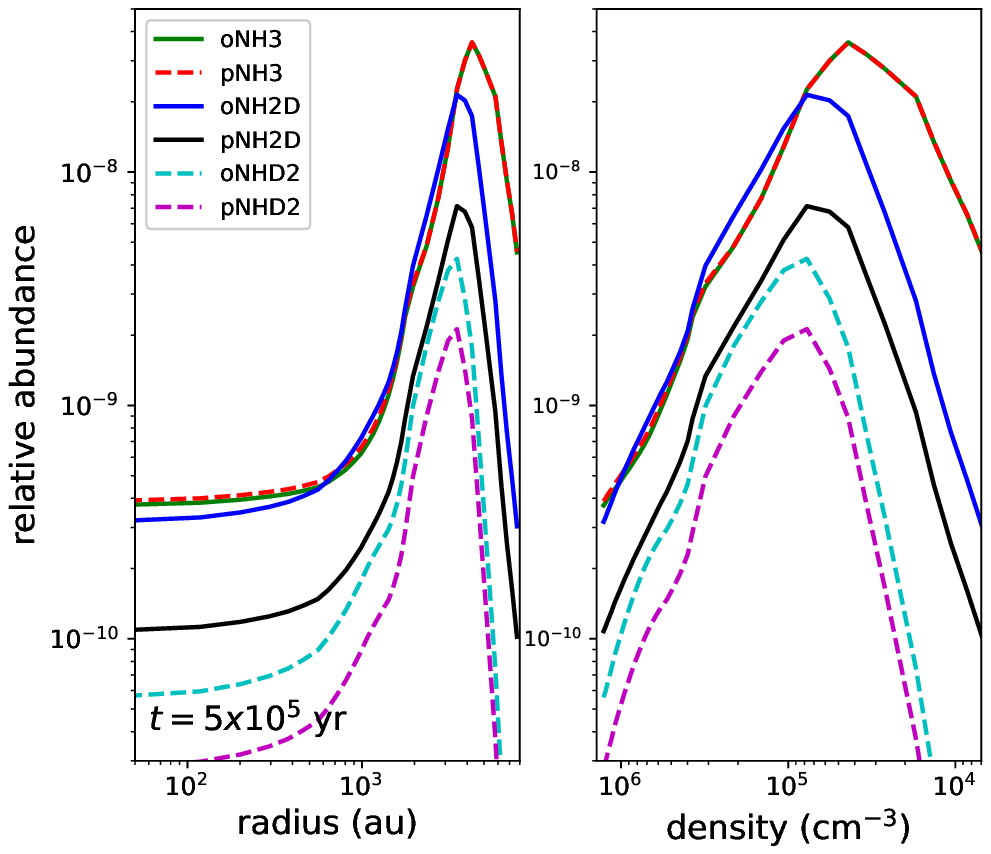}
\end{picture}}
\end{picture}  
   \caption{Fractional abundances of the ortho and para modifications of $\ammo$, $\dammo$, and $\ddammo$ as functions of radial distance from the centre (left) and density (right) in the 1D cloud model used for chemical calculations. The chemistry model assumes constant CD with {PH. The abundances are taken { at the simulation time $5\times10^5$\,yr.}}}
\label{X_radii_dens}
\end{figure}

Besides the fact that many uncertainties are involved in chemical modelling, also observational determination of molecular abundances is non-trivial.  There, we are dealing with changing abundances and excitation conditions along the line of sight. To demonstrate this, we show in Fig.~\ref{X_radii_dens} the fractional abundances of $\ammo$, $\dammo$, and $\ddammo$ as functions of the radial distance from the core centre and the density, predicted by the constant CD model with {PH} at a certain time of the simulation. When comparing the best-fit abundances derived using the 3D model of H-MM1 to the model predictions, we used 1) {column density ratios $N(\rm mol)/N(\htwo)$} along a line going through the core centre (Figs.~\ref{XNH3_times} to \ref{Dfrac_times} {and relative deviations given in Table~\ref{deviations}}), and 2) simulated and observed spectra towards the core centre, sampling a cylindrical region including the densest parts of the core (Figs.~\ref{constCD_FS_spectra} to \ref{dynCD_PH_spectra}). In both comparisons, the constant CD model with {PH} gave the best overall agreement with the observations. The interpretation of the four spectral lines used here is facilitated by the fact that they lie close in frequency, meaning that they can be observed in the same spectral band and have similar critical densities.

The {PH} scenario was previously tested by \cite{2019A&A...631A..63S} with a core model resembling H-MM1.  They found that {PH} reactions favour the singly deuterated $\dammo$ at the cost of $\ddammo$ and $\ndthree$, but also that all the predicted deuterium fractionation ratios, $\dammo/\ammo$, $\ddammo/\dammo$, and $\ndthree/\ddammo$, were higher than observed in H-MM1 when the ammonia production only involved proton hop reactions. Here, we do not find large discrepancies between the modelled and { observed $\dammo/{\rm p}\ammo$ and} $\ddammo/\dammo$ ratios when using the constant CD {PH} model.  The main difference between the present and the former modelling efforts is that \cite{2019A&A...631A..63S} did not include CD. The replenishment of gas by ammonia formed on grains can diminish the deuterium fractionation as the $\dammo/\ammo$ and $\ddammo/\dammo$ ratios on grains are predicted to be substantially lower than those in the gas. For example, in all four models examined here, the $\dammo/\ammo$ ratio on grains always stays below 0.1.

\section{Conclusions}
\label{conclusions}

We have determined the o/p ratios of $\dammo$ and $\ddammo$ in two prestellar cores by observing the ground-state rotational lines of these molecules with the LAsMA multi-beam receiver on APEX. By simulating line emission from realistic 3D models of the cores, and using different assumptions regarding the abundance distributions, we could establish that the o/p ratios deviate at most 20\% from their statistical values, o/p-$\dammo=3$, o/p-$\ddammo=2$.

We also ran chemical models to test if the observed statistical nuclear spin ratios can be explained by enhanced chemical desorption (CD) in the core, employing two different scenarios for proton/deuteron exchanges in gas-phase chemical reactions.  An increase in the efficiency of CD in the core interior can be envisioned because the grain surface composition is expected to change from ${\rm H_2O}$-dominated to CO-rich ice. However, models assuming constant CD, with 1\% of exothermic reactions leading to desorption, can reproduce the observed abundances better than the so called dynamic CD models with changing efficiency. { Despite diverging predictions for the fractional abundances, both constant and dynamic CD models assuming single-particle hop in proton donation and hydrogen abstraction reactions  gave the better agreement with the observed o/p ratios than the corresponding models with full scrambling.}

The efficiency of CD naturally affects the abundances of $\ammo$ and its deuterated isotopologues in the gas, but according to our modelling results the effect of the proton exchange scenario in gas-phase reactions is more important to the spin ratios. { This is because the accretion and desorption processes are outcompeted by rapid conversion of gas-phase nitrogen hydrides to cations through charge or proton exchange reactions, and the return of cations to neutral species in recombination with electrons.}

\begin{acknowledgements}
We thank {the anonymous referee for helpful comments,} and Mika Juvela for his advice on error analysis and on the use of the LOC radiative transfer program. We also thank the Max Planck Society for financial support. 
\end{acknowledgements}

\bibliographystyle{aa} 

   \bibliography{bibliography.bib} 

\begin{thebibliography}{47}
\expandafter\ifx\csname natexlab\endcsname\relax\def\natexlab#1{#1}\fi

\bibitem[{{Andrae}(2010)}]{2010arXiv1009.2755A}
{Andrae}, R. 2010, arXiv e-prints, arXiv:1009.2755

\bibitem[{{Arzoumanian} {et~al.}(2011){Arzoumanian}, {Andr{\'e}}, {Didelon},
  {K{\"o}nyves}, {Schneider}, {Men'shchikov}, {Sousbie}, {Zavagno}, {Bontemps},
  {di Francesco}, {Griffin}, {Hennemann}, {Hill}, {Kirk}, {Martin}, {Minier},
  {Molinari}, {Motte}, {Peretto}, {Pezzuto}, {Spinoglio}, {Ward-Thompson},
  {White}, \& {Wilson}}]{2011A&A...529L...6A}
{Arzoumanian}, D., {Andr{\'e}}, P., {Didelon}, P., {et~al.} 2011, \aap, 529, L6

\bibitem[{{Black}(1994)}]{1994ASPC...58..355B}
{Black}, J.~H. 1994, in Astronomical Society of the Pacific Conference Series,
  Vol.~58, The First Symposium on the Infrared Cirrus and Diffuse Interstellar
  Clouds, ed. R.~M. {Cutri} \& W.~B. {Latter}, 355

\bibitem[{{Brown} \& {Millar}(1989)}]{1989MNRAS.240P..25B}
{Brown}, P.~D. \& {Millar}, T.~J. 1989, \mnras, 240, 25P

\bibitem[{{Caselli} {et~al.}(2022){Caselli}, {Pineda}, {Sipil{\"a}}, {Zhao},
  {Redaelli}, {Spezzano}, {Maureira}, {Alves}, {Bizzocchi}, {Bourke},
  {Chac{\'o}n-Tanarro}, {Friesen}, {Galli}, {Harju}, {Jim{\'e}nez-Serra},
  {Keto}, {Li}, {Padovani}, {Schmiedeke}, {Tafalla}, \&
  {Vastel}}]{2022ApJ...929...13C}
{Caselli}, P., {Pineda}, J.~E., {Sipil{\"a}}, O., {et~al.} 2022, \apj, 929, 13

\bibitem[{{Choudhury} {et~al.}(2020){Choudhury}, {Pineda}, {Caselli},
  {Ginsburg}, {Offner}, {Rosolowsky}, {Friesen}, {Alves}, {Chac{\'o}n-Tanarro},
  {Punanova}, {Redaelli}, {Kirk}, {Myers}, {Martin}, {Shirley}, {Chun-Yuan
  Chen}, {Goodman}, \& {Di Francesco}}]{2020A&A...640L...6C}
{Choudhury}, S., {Pineda}, J.~E., {Caselli}, P., {et~al.} 2020, \aap, 640, L6

\bibitem[{{Coudert} \& {Roueff}(2006)}]{2006A&A...449..855C}
{Coudert}, L.~H. \& {Roueff}, E. 2006, \aap, 449, 855

\bibitem[{{Crapsi} {et~al.}(2005){Crapsi}, {Caselli}, {Walmsley}, {Myers},
  {Tafalla}, {Lee}, \& {Bourke}}]{2005ApJ...619..379C}
{Crapsi}, A., {Caselli}, P., {Walmsley}, C.~M., {et~al.} 2005, \apj, 619, 379

\bibitem[{{Daniel} {et~al.}(2016){Daniel}, {Rist}, {Faure}, {Roueff},
  {G{\'e}rin}, {Lis}, {Hily-Blant}, {Bacmann}, \&
  {Wiesenfeld}}]{2016MNRAS.457.1535D}
{Daniel}, F., {Rist}, C., {Faure}, A., {et~al.} 2016, \mnras, 457, 1535

\bibitem[{{Faure} {et~al.}(2013){Faure}, {Hily-Blant}, {Le Gal}, {Rist}, \&
  {Pineau des For{\^e}ts}}]{2013ApJ...770L...2F}
{Faure}, A., {Hily-Blant}, P., {Le Gal}, R., {Rist}, C., \& {Pineau des
  For{\^e}ts}, G. 2013, \apjl, 770, L2

\bibitem[{{Ferrero} {et~al.}(2023){Ferrero}, {Pantaleone}, {Ceccarelli},
  {Ugliengo}, {Sodupe}, \& {Rimola}}]{2023ApJ...944..142F}
{Ferrero}, S., {Pantaleone}, S., {Ceccarelli}, C., {et~al.} 2023, \apj, 944,
  142

\bibitem[{{Friesen} {et~al.}(2017){Friesen}, {Pineda}, {co-PIs}, {Rosolowsky},
  {Alves}, {Chac{\'o}n-Tanarro}, {How-Huan Chen}, {Chun-Yuan Chen}, {Di
  Francesco}, {Keown}, {Kirk}, {Punanova}, {Seo}, {Shirley}, {Ginsburg},
  {Hall}, {Offner}, {Singh}, {Arce}, {Caselli}, {Goodman}, {Martin}, {Matzner},
  {Myers}, {Redaelli}, \& {GAS Collaboration}}]{2017ApJ...843...63F}
{Friesen}, R.~K., {Pineda}, J.~E., {co-PIs}, {et~al.} 2017, \apj, 843, 63

\bibitem[{{Garrod} {et~al.}(2007){Garrod}, {Wakelam}, \& {Herbst}}]{Garrod07}
{Garrod}, R.~T., {Wakelam}, V., \& {Herbst}, E. 2007, \aap, 467, 1103

\bibitem[{{Gerin} {et~al.}(2006){Gerin}, {Lis}, {Philipp}, {G{\"u}sten},
  {Roueff}, \& {Reveret}}]{2006A&A...454L..63G}
{Gerin}, M., {Lis}, D.~C., {Philipp}, S., {et~al.} 2006, \aap, 454, L63

\bibitem[{{Grassi} {et~al.}(2020){Grassi}, {Bovino}, {Caselli}, {Bovolenta},
  {Vogt-Geisse}, \& {Ercolano}}]{2020A&A...643A.155G}
{Grassi}, T., {Bovino}, S., {Caselli}, P., {et~al.} 2020, \aap, 643, A155

\bibitem[{{G{\"u}sten} {et~al.}(2008){G{\"u}sten}, {Baryshev}, {Bell},
  {Belloche}, {Graf}, {Hafok}, {Heyminck}, {Hochg{\"u}rtel}, {Honingh},
  {Jacobs}, {Kasemann}, {Klein}, {Klein}, {Korn}, {Kr{\"a}mer}, {Leinz},
  {Lundgren}, {Menten}, {Meyer}, {Muders}, {Pacek}, {Rabanus}, {Sch{\"a}fer},
  {Schilke}, {Schneider}, {Stutzki}, {Wieching}, {Wunsch}, \&
  {Wyrowski}}]{2008SPIE.7020E..10G}
{G{\"u}sten}, R., {Baryshev}, A., {Bell}, A., {et~al.} 2008, in Society of
  Photo-Optical Instrumentation Engineers (SPIE) Conference Series, Vol. 7020,
  Millimeter and Submillimeter Detectors and Instrumentation for Astronomy IV,
  ed. W.~D. {Duncan}, W.~S. {Holland}, S.~{Withington}, \& J.~{Zmuidzinas},
  702010

\bibitem[{{G{\"u}sten} {et~al.}(2006){G{\"u}sten}, {Nyman}, {Schilke},
  {Menten}, {Cesarsky}, \& {Booth}}]{2006A&A...454L..13G}
{G{\"u}sten}, R., {Nyman}, L.~{\r{A}}., {Schilke}, P., {et~al.} 2006, \aap,
  454, L13

\bibitem[{{Harju} {et~al.}(2017){Harju}, {Daniel}, {Sipil{\"a}}, {Caselli},
  {Pineda}, {Friesen}, {Punanova}, {G{\"u}sten}, {Wiesenfeld}, {Myers},
  {Faure}, {Hily-Blant}, {Rist}, {Rosolowsky}, {Schlemmer}, \&
  {Shirley}}]{2017A&A...600A..61H}
{Harju}, J., {Daniel}, F., {Sipil{\"a}}, O., {et~al.} 2017, \aap, 600, A61

\bibitem[{{Harju} {et~al.}(2008){Harju}, {Juvela}, {Schlemmer}, {Haikala},
  {Lehtinen}, \& {Mattila}}]{2008A&A...482..535H}
{Harju}, J., {Juvela}, M., {Schlemmer}, S., {et~al.} 2008, \aap, 482, 535

\bibitem[{{Harju} {et~al.}(2020){Harju}, {Pineda}, {Vasyunin}, {Caselli},
  {Offner}, {Goodman}, {Juvela}, {Sipil{\"a}}, {Faure}, {Le Gal}, {Hily-Blant},
  {Alves}, {Bizzocchi}, {Burkert}, {Chen}, {Friesen}, {G{\"u}sten}, {Myers},
  {Punanova}, {Rist}, {Rosolowsky}, {Schlemmer}, {Shirley}, {Spezzano},
  {Vastel}, \& {Wiesenfeld}}]{2020ApJ...895..101H}
{Harju}, J., {Pineda}, J.~E., {Vasyunin}, A.~I., {et~al.} 2020, \apj, 895, 101

\bibitem[{{Hily-Blant} {et~al.}(2018){Hily-Blant}, {Faure}, {Rist}, {Pineau des
  For{\^e}ts}, \& {Flower}}]{2018MNRAS.477.4454H}
{Hily-Blant}, P., {Faure}, A., {Rist}, C., {Pineau des For{\^e}ts}, G., \&
  {Flower}, D.~R. 2018, \mnras, 477, 4454

\bibitem[{{Hugo} {et~al.}(2009){Hugo}, {Asvany}, \&
  {Schlemmer}}]{2009JChPh.130p4302H}
{Hugo}, E., {Asvany}, O., \& {Schlemmer}, S. 2009, \jcp, 130, 164302

\bibitem[{{Juvela}(2005)}]{2005A&A...440..531J}
{Juvela}, M. 2005, \aap, 440, 531

\bibitem[{{Juvela}(2020)}]{2020A&A...644A.151J}
{Juvela}, M. 2020, \aap, 644, A151

\bibitem[{{Ladjelate} {et~al.}(2020){Ladjelate}, {Andr{\'e}}, {K{\"o}nyves},
  {Ward-Thompson}, {Men'shchikov}, {Bracco}, {Palmeirim}, {Roy}, {Shimajiri},
  {Kirk}, {Arzoumanian}, {Benedettini}, {Di Francesco}, {Fiorellino},
  {Schneider}, {Pezzuto}, {Motte}, \& {Herschel Gould Belt Survey
  Team}}]{2020A&A...638A..74L}
{Ladjelate}, B., {Andr{\'e}}, P., {K{\"o}nyves}, V., {et~al.} 2020, \aap, 638,
  A74

\bibitem[{{Le Gal} {et~al.}(2014){Le Gal}, {Hily-Blant}, {Faure}, {Pineau des
  For{\^e}ts}, {Rist}, \& {Maret}}]{2014A&A...562A..83L}
{Le Gal}, R., {Hily-Blant}, P., {Faure}, A., {et~al.} 2014, \aap, 562, A83

\bibitem[{{Lis} {et~al.}(2006){Lis}, {Gerin}, {Roueff}, {Vastel}, \&
  {Phillips}}]{2006ApJ...636..916L}
{Lis}, D.~C., {Gerin}, M., {Roueff}, E., {Vastel}, C., \& {Phillips}, T.~G.
  2006, \apj, 636, 916

\bibitem[{{Melosso} {et~al.}(2021){Melosso}, {Bizzocchi}, {Dore}, {Kisiel},
  {Jiang}, {Spezzano}, {Caselli}, {Gauss}, \&
  {Puzzarini}}]{2021JMoSp.37711431M}
{Melosso}, M., {Bizzocchi}, L., {Dore}, L., {et~al.} 2021, Journal of Molecular
  Spectroscopy, 377, 111431

\bibitem[{Minissale {et~al.}(2022)Minissale, Aikawa, Bergin, Bertin, Brown,
  Cazaux, Charnley, Coutens, Cuppen, Guzman, Linnartz, McCoustra, Rimola,
  Schrauwen, Toubin, Ugliengo, Watanabe, Wakelam, \& Dulieu}]{Minissale22}
Minissale, M., Aikawa, Y., Bergin, E., {et~al.} 2022, ACS Earth and Space
  Chemistry, 6, 597

\bibitem[{{Minissale} {et~al.}(2016){Minissale}, {Dulieu}, {Cazaux}, \&
  {Hocuk}}]{Minissale16a}
{Minissale}, M., {Dulieu}, F., {Cazaux}, S., \& {Hocuk}, S. 2016, \aap, 585,
  A24

\bibitem[{{Ossenkopf} \& {Henning}(1994)}]{1994A&A...291..943O}
{Ossenkopf}, V. \& {Henning}, T. 1994, \aap, 291, 943

\bibitem[{{Pagani} {et~al.}(2009){Pagani}, {Vastel}, {Hugo}, {Kokoouline},
  {Greene}, {Bacmann}, {Bayet}, {Ceccarelli}, {Peng}, \&
  {Schlemmer}}]{2009A&A...494..623P}
{Pagani}, L., {Vastel}, C., {Hugo}, E., {et~al.} 2009, \aap, 494, 623

\bibitem[{{Pattle} {et~al.}(2015){Pattle}, {Ward-Thompson}, {Kirk}, {White},
  {Drabek-Maunder}, {Buckle}, {Beaulieu}, {Berry}, {Broekhoven-Fiene},
  {Currie}, {Fich}, {Hatchell}, {Kirk}, {Jenness}, {Johnstone}, {Mottram},
  {Nutter}, {Pineda}, {Quinn}, {Salji}, {Tisi}, {Walker-Smith}, {di Francesco},
  {Hogerheijde}, {Andr{\'e}}, {Bastien}, {Bresnahan}, {Butner}, {Chen},
  {Chrysostomou}, {Coude}, {Davis}, {Duarte-Cabral}, {Fiege}, {Friberg},
  {Friesen}, {Fuller}, {Graves}, {Greaves}, {Gregson}, {Griffin}, {Holland},
  {Joncas}, {Knee}, {K{\"o}nyves}, {Mairs}, {Marsh}, {Matthews},
  {Moriarty-Schieven}, {Rawlings}, {Richer}, {Robertson}, {Rosolowsky},
  {Rumble}, {Sadavoy}, {Spinoglio}, {Thomas}, {Tothill}, {Viti}, {Wouterloot},
  {Yates}, \& {Zhu}}]{2015MNRAS.450.1094P}
{Pattle}, K., {Ward-Thompson}, D., {Kirk}, J.~M., {et~al.} 2015, \mnras, 450,
  1094

\bibitem[{{Pineda} {et~al.}(2022){Pineda}, {Harju}, {Caselli}, {Sipil{\"a}},
  {Juvela}, {Vastel}, {Rosolowsky}, {Burkert}, {Friesen}, {Shirley},
  {Maureira}, {Choudhury}, {Segura-Cox}, {G{\"u}sten}, {Punanova}, {Bizzocchi},
  \& {Goodman}}]{2022AJ....163..294P}
{Pineda}, J.~E., {Harju}, J., {Caselli}, P., {et~al.} 2022, \aj, 163, 294

\bibitem[{{Riedel} {et~al.}(2023){Riedel}, {Sipil{\"a}}, {Redaelli}, {Caselli},
  {Vasyunin}, {Dulieu}, \& {Watanabe}}]{2023arXiv231008389R}
{Riedel}, W., {Sipil{\"a}}, O., {Redaelli}, E., {et~al.} 2023, A\&A {\sl in
  press}, arXiv:2310.08389

\bibitem[{{Rist} {et~al.}(2013){Rist}, {Faure}, {Hily-Blant}, \& {Le
  Gal}}]{2013JPCA..117.9800R}
{Rist}, C., {Faure}, A., {Hily-Blant}, P., \& {Le Gal}, R. 2013, Journal of
  Physical Chemistry A, 117, 9800

\bibitem[{{Roueff} {et~al.}(2021){Roueff}, {Gerin}, {Gratier}, {Levrier},
  {Pety}, {Gaudel}, {Goicoechea}, {Orkisz}, {de Souza Magalhaes}, {Vono},
  {Bardeau}, {Bron}, {Chanussot}, {Chainais}, {Guzman}, {Hughes},
  {Kainulainen}, {Languignon}, {Le Bourlot}, {Le Petit}, {Liszt}, {Marchal},
  {Miville-Desch{\^e}nes}, {Peretto}, {Roueff}, \&
  {Sievers}}]{2021A&A...645A..26R}
{Roueff}, A., {Gerin}, M., {Gratier}, P., {et~al.} 2021, \aap, 645, A26

\bibitem[{{Ruoskanen} {et~al.}(2011){Ruoskanen}, {Harju}, {Juvela},
  {Miettinen}, {Liljestr{\"o}m}, {V{\"a}is{\"a}l{\"a}}, {Lunttila}, \&
  {Kontinen}}]{2011A&A...534A.122R}
{Ruoskanen}, J., {Harju}, J., {Juvela}, M., {et~al.} 2011, \aap, 534, A122

\bibitem[{{Shirley}(2015)}]{2015PASP..127..299S}
{Shirley}, Y.~L. 2015, \pasp, 127, 299

\bibitem[{{Sipil{\"a}} {et~al.}(2019{\natexlab{a}}){Sipil{\"a}}, {Caselli}, \&
  {Harju}}]{2019A&A...631A..63S}
{Sipil{\"a}}, O., {Caselli}, P., \& {Harju}, J. 2019{\natexlab{a}}, \aap, 631,
  A63

\bibitem[{{Sipil{\"a}} {et~al.}(2019{\natexlab{b}}){Sipil{\"a}}, {Caselli},
  {Redaelli}, {Juvela}, \& {Bizzocchi}}]{Sipila19a}
{Sipil{\"a}}, O., {Caselli}, P., {Redaelli}, E., {Juvela}, M., \& {Bizzocchi},
  L. 2019{\natexlab{b}}, \mnras, 487, 1269

\bibitem[{{Sipil{\"a}} {et~al.}(2015){Sipil{\"a}}, {Harju}, {Caselli}, \&
  {Schlemmer}}]{2015A&A...581A.122S}
{Sipil{\"a}}, O., {Harju}, J., {Caselli}, P., \& {Schlemmer}, S. 2015, \aap,
  581, A122

\bibitem[{{Sipil{\"a}} {et~al.}(2010){Sipil{\"a}}, {Hugo}, {Harju}, {Asvany},
  {Juvela}, \& {Schlemmer}}]{2010A&A...509A..98S}
{Sipil{\"a}}, O., {Hugo}, E., {Harju}, J., {et~al.} 2010, \aap, 509, A98

\bibitem[{Sipil{\"a} {et~al.}(2021)Sipil{\"a}, Silsbee, \& Caselli}]{Sipila21}
Sipil{\"a}, O., Silsbee, K., \& Caselli, P. 2021, \apj, 922, 126

\bibitem[{{Tinacci} {et~al.}(2022){Tinacci}, {Germain}, {Pantaleone},
  {Ferrero}, {Ceccarelli}, \& {Ugliengo}}]{2022ESC.....6.1514T}
{Tinacci}, L., {Germain}, A., {Pantaleone}, S., {et~al.} 2022, ACS Earth and
  Space Chemistry, 6, 1514

\bibitem[{{Vasyunin} {et~al.}(2017){Vasyunin}, {Caselli}, {Dulieu}, \&
  {Jim{\'e}nez-Serra}}]{Vasyunin17}
{Vasyunin}, A.~I., {Caselli}, P., {Dulieu}, F., \& {Jim{\'e}nez-Serra}, I.
  2017, \apj, 842, 33

\bibitem[{{Wienen} {et~al.}(2021){Wienen}, {Wyrowski}, {Walmsley}, {Csengeri},
  {Pillai}, {Giannetti}, \& {Menten}}]{2021A&A...649A..21W}
{Wienen}, M., {Wyrowski}, F., {Walmsley}, C.~M., {et~al.} 2021, \aap, 649, A21

\end{thebibliography}

%
%


\begin{appendix}

\section{Observed and simulated spectra}

\begin{figure*}
\centering
\unitlength=1.0mm
\begin{picture}(160,110)(0,0)
\put(60,40){
\begin{picture}(0,0) 
\includegraphics[width=3.75cm,angle=0]{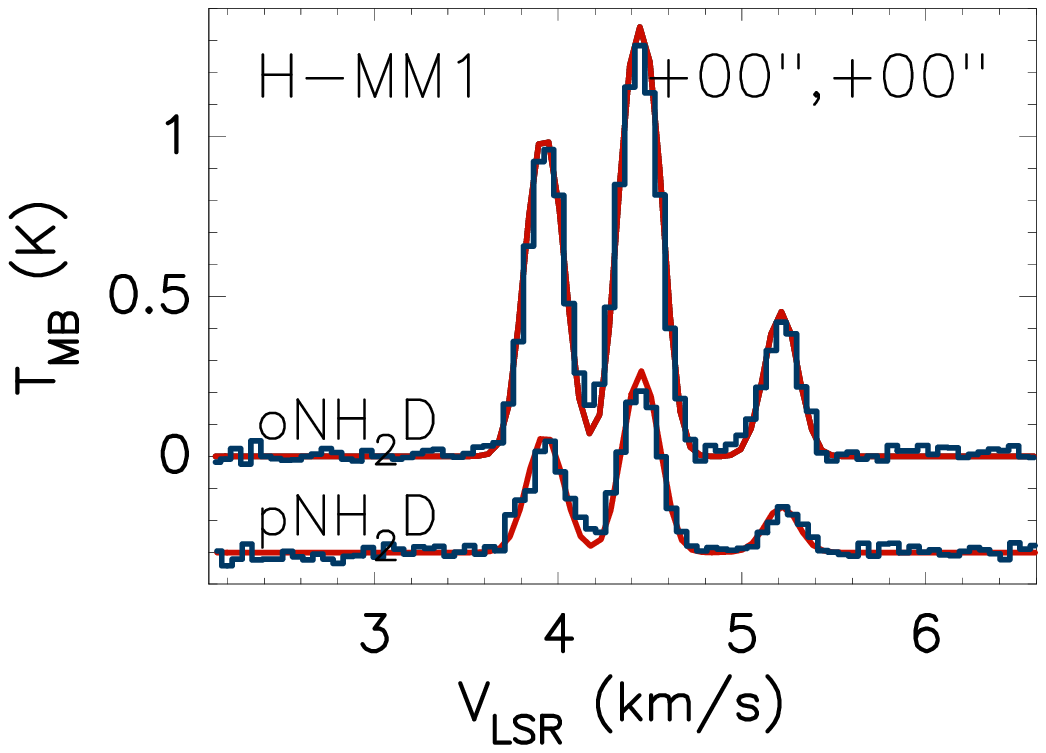}
\end{picture}}
\put(29,12){
\begin{picture}(0,0) 
\includegraphics[width=3.75cm,angle=0]{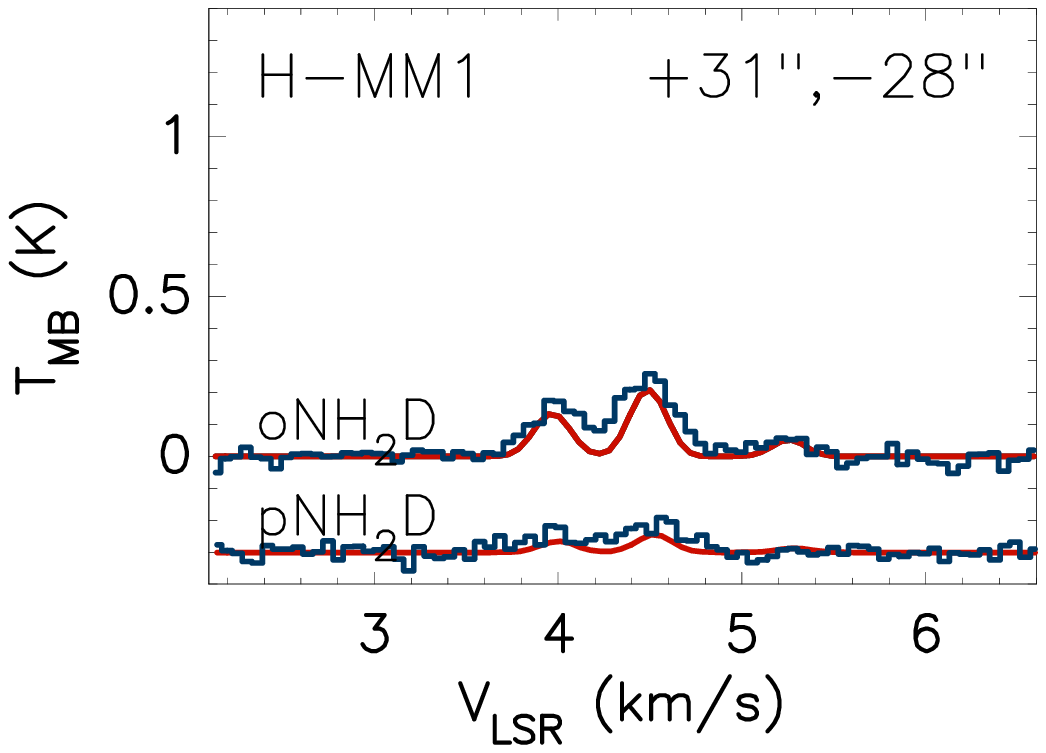}
\end{picture}}
\put(25,54){
\begin{picture}(0,0) 
\includegraphics[width=3.75cm,angle=0]{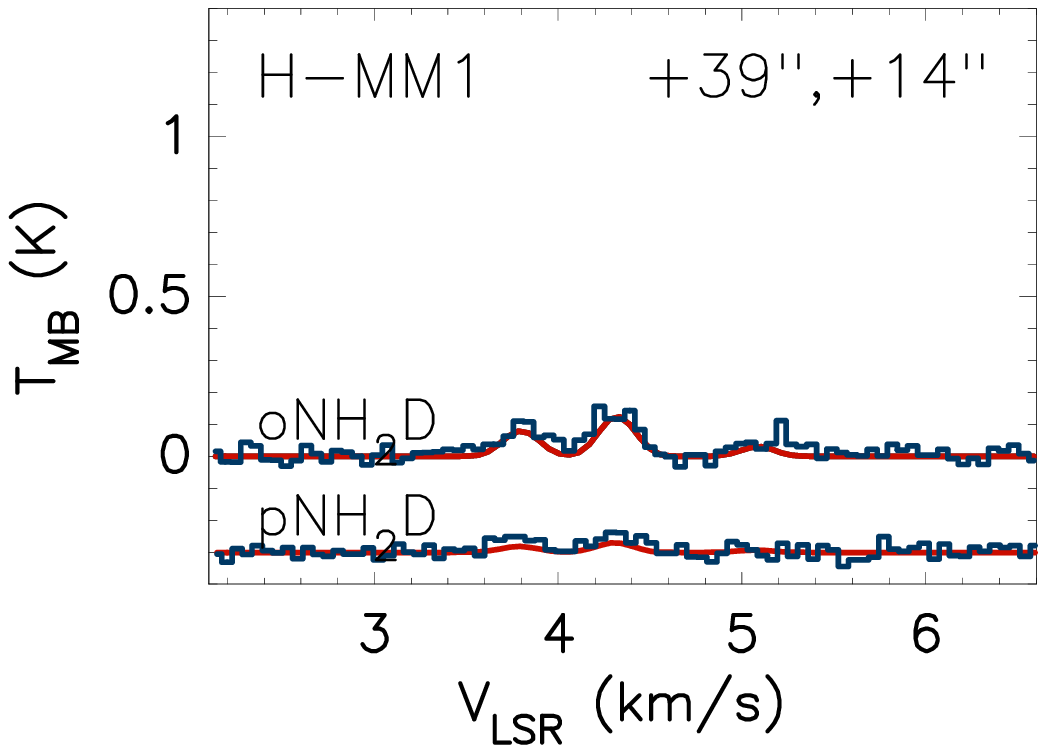}
\end{picture}}
\put(52,79){
\begin{picture}(0,0) 
\includegraphics[width=3.75cm,angle=0]{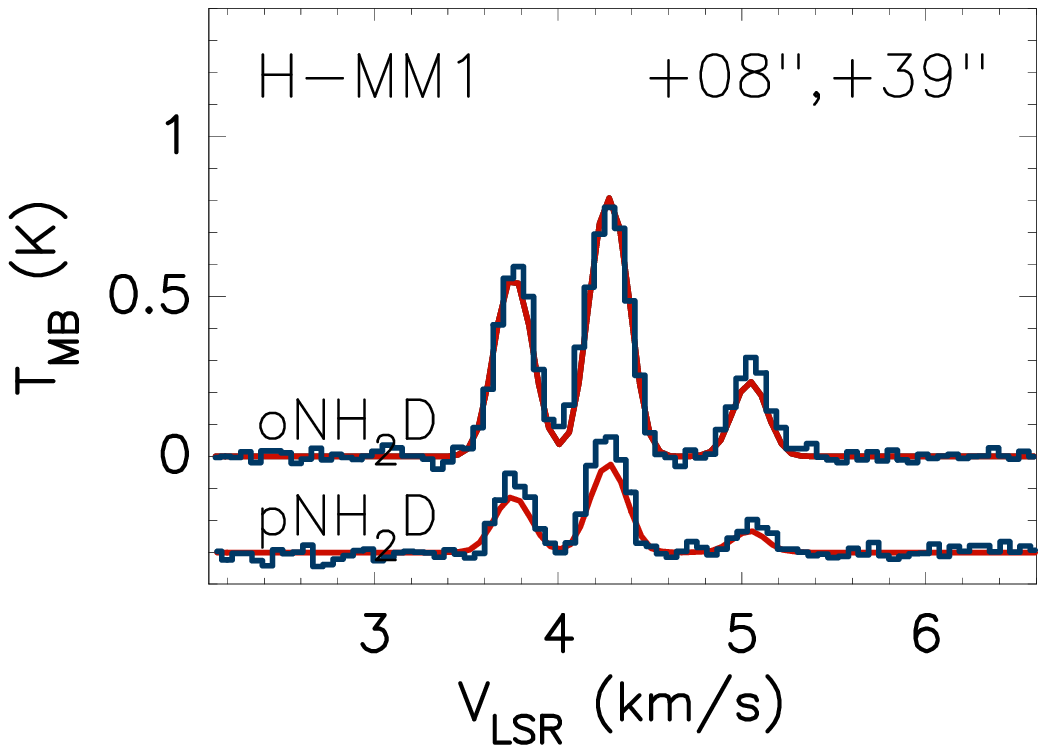}
\end{picture}}
\put(92,65){
\begin{picture}(0,0) 
\includegraphics[width=3.75cm,angle=0]{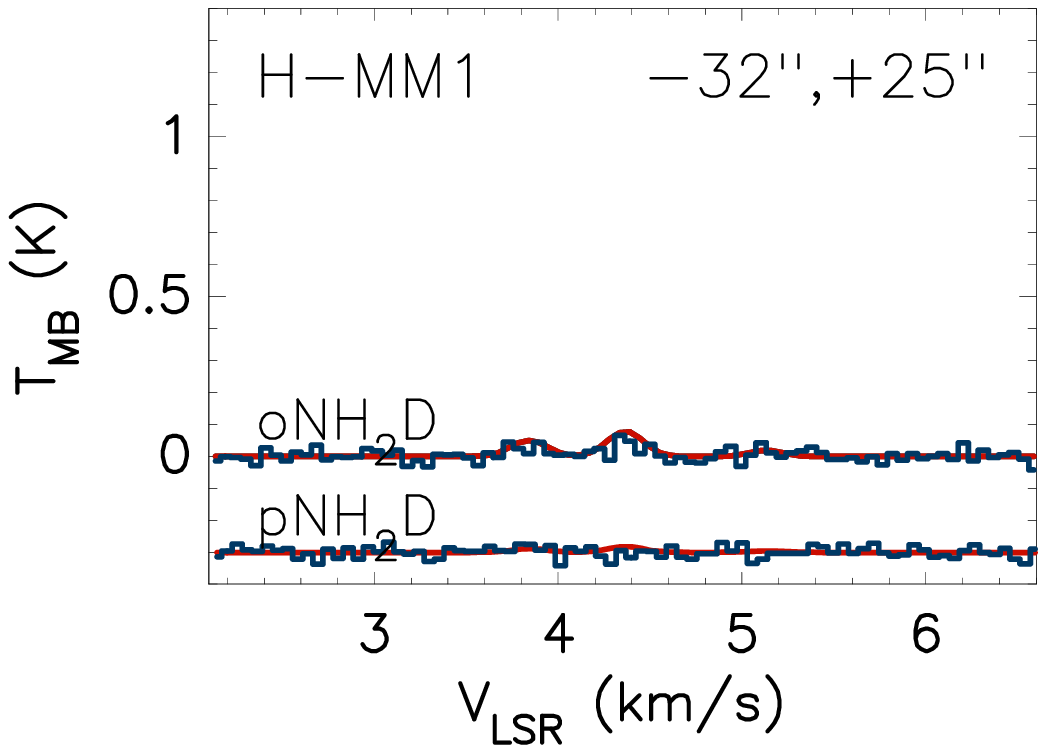}
\end{picture}}
\put(98,25){
\begin{picture}(0,0) 
\includegraphics[width=3.75cm,angle=0]{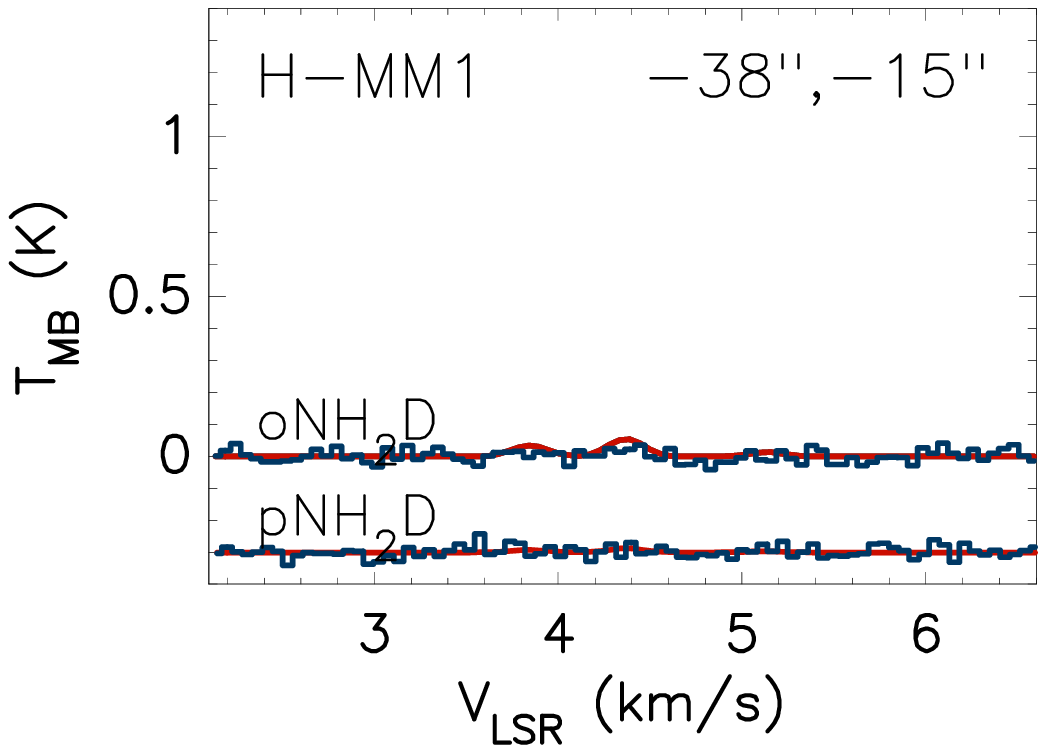}
\end{picture}}
\put(66,-1){
\begin{picture}(0,0) 
\includegraphics[width=3.75cm,angle=0]{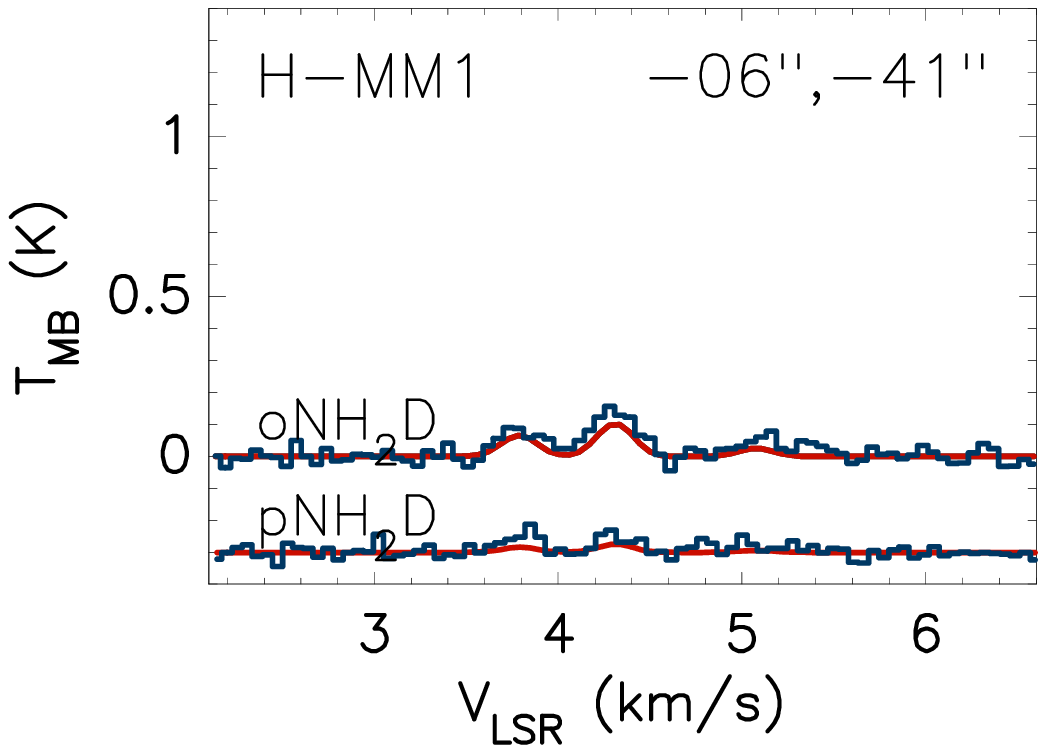}
\end{picture}}
\put(0,100){\large \bf H-MM1 \hspace{0.5cm} $\dammo$}
\end{picture}

\caption[]{$\dammo$ spectra obtained with the LAsMA array towards H-MM1. 
Simulated spectra assuming constant fractional abundances in the 3D core model
are shown with red. The assumed ortho- and para-$\dammo$ abundances are those
given in the left column of Table~\ref{chi2_fracs}, that is, $X(\odammo)=4.520
\times10^{-9}$,  $X(\pdammo)=1.469\times10^{-9}$.}
\label{hmm1_best_fit_nh2d_spectra}
\end{figure*}

\begin{figure*}
\centering
\unitlength=1.0mm
\begin{picture}(160,110)(0,0)
\put(60,40){
\begin{picture}(0,0) 
\includegraphics[width=3.75cm,angle=0]{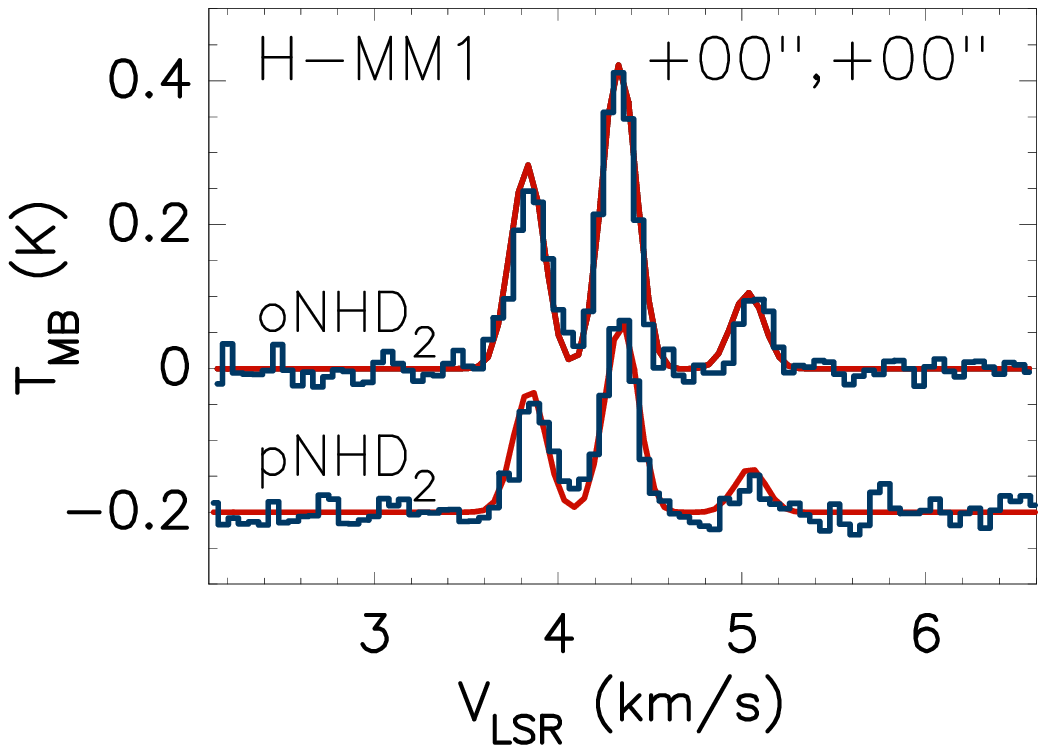}
\end{picture}}
\put(29,12){
\begin{picture}(0,0) 
\includegraphics[width=3.75cm,angle=0]{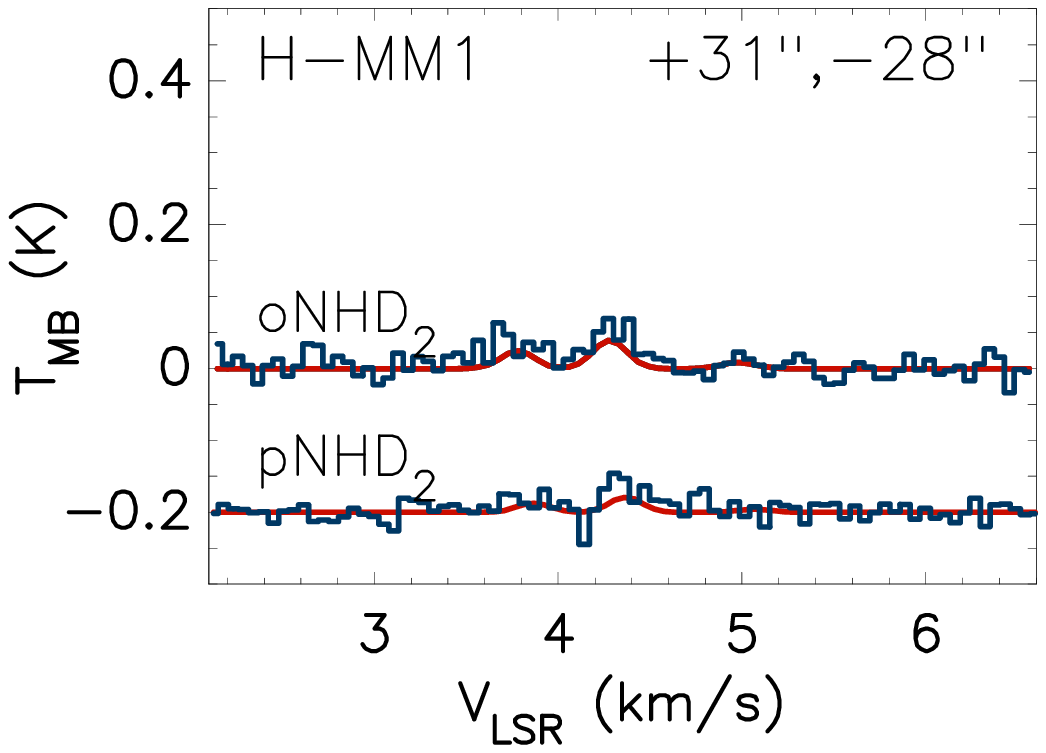}
\end{picture}}
\put(25,54){
\begin{picture}(0,0) 
\includegraphics[width=3.75cm,angle=0]{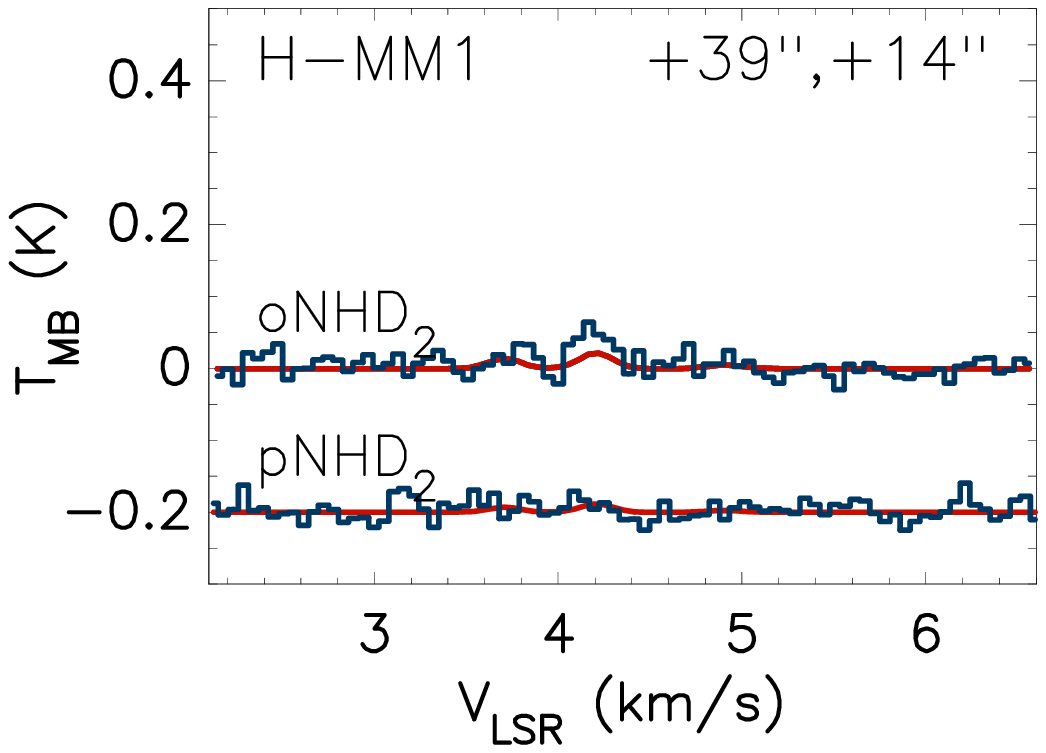}
\end{picture}}
\put(52,79){
\begin{picture}(0,0) 
\includegraphics[width=3.75cm,angle=0]{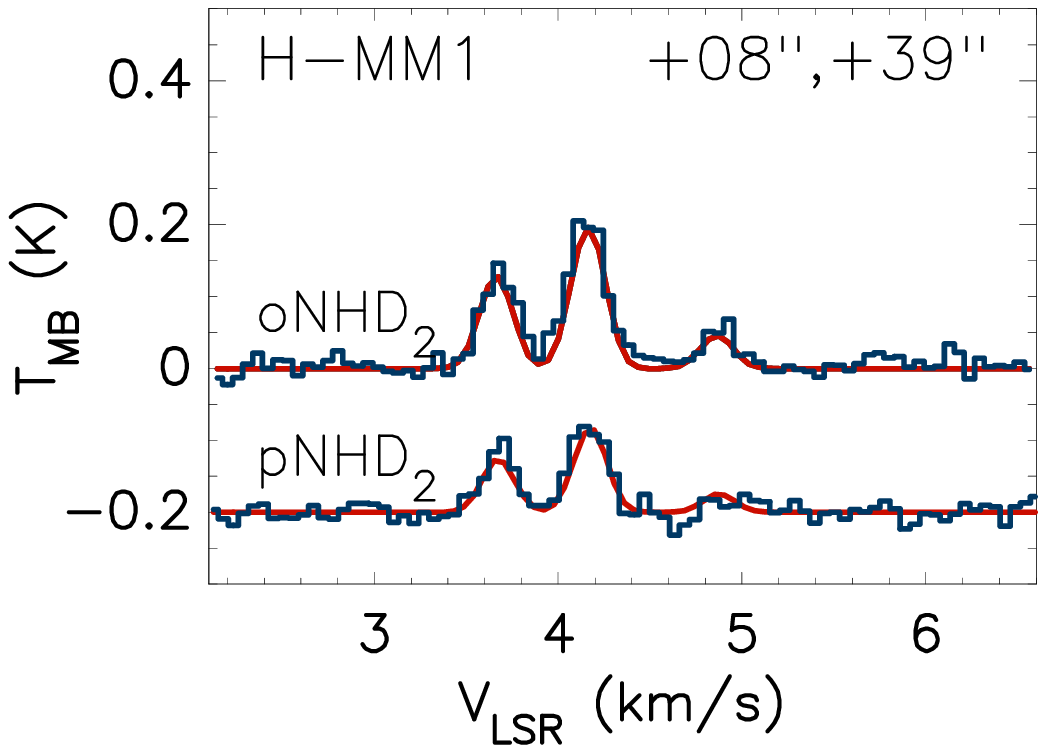}
\end{picture}}
\put(92,65){
\begin{picture}(0,0) 
\includegraphics[width=3.75cm,angle=0]{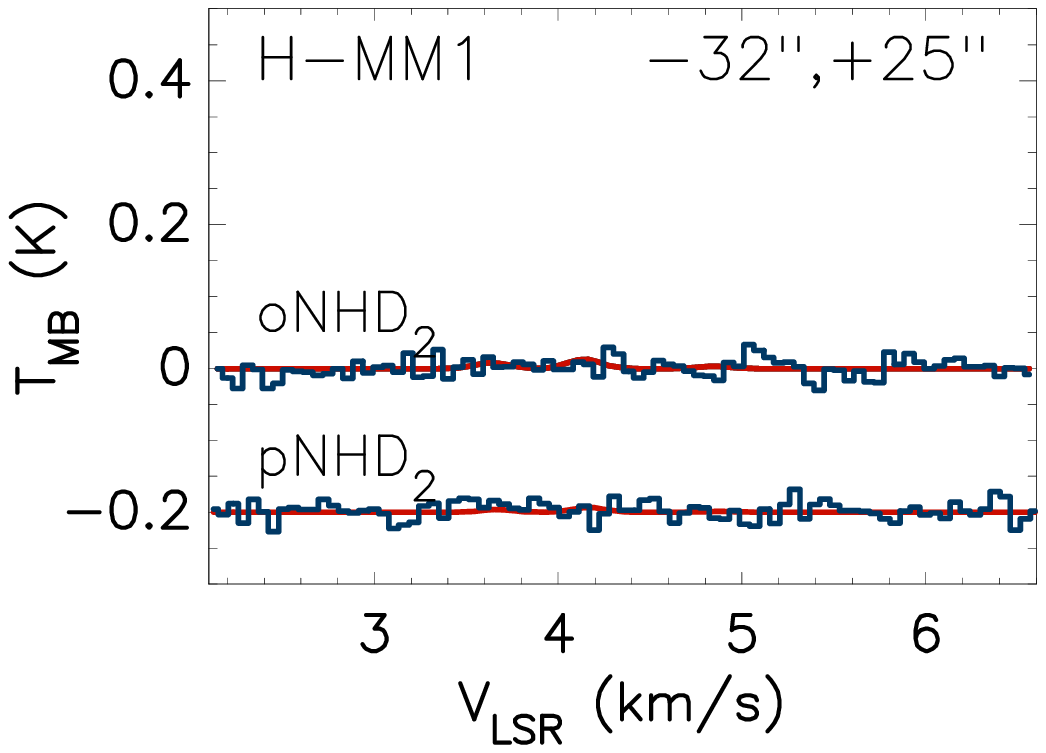}
\end{picture}}
\put(98,25){
\begin{picture}(0,0) 
\includegraphics[width=3.75cm,angle=0]{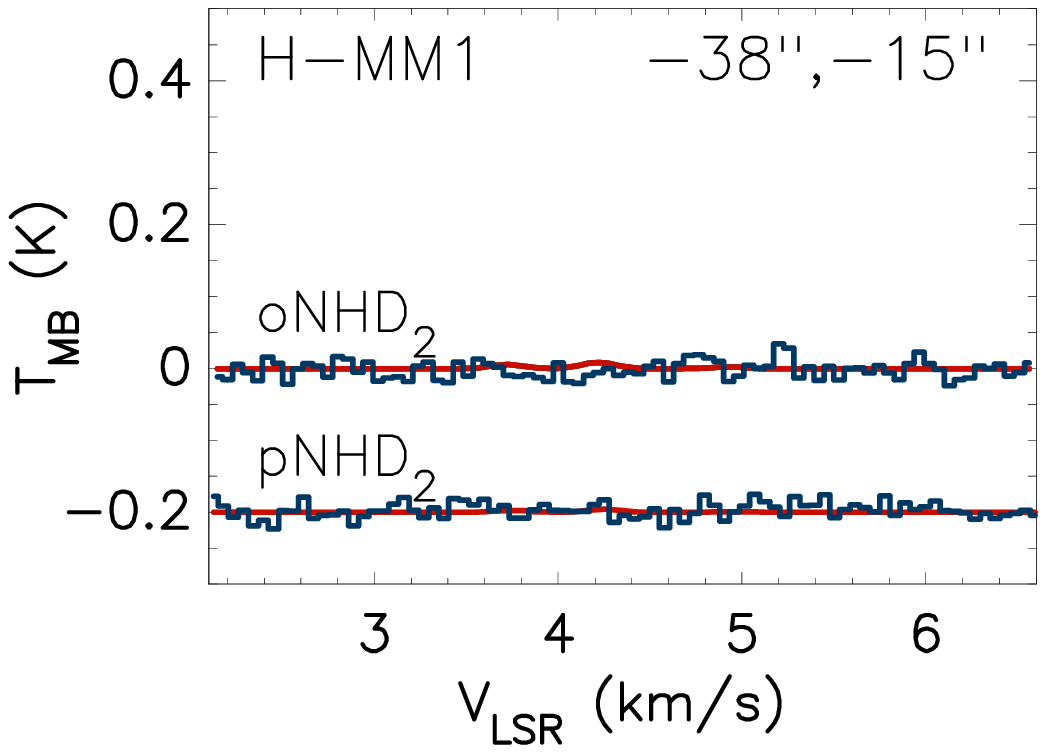}
\end{picture}}
\put(66,-1){
\begin{picture}(0,0) 
\includegraphics[width=3.75cm,angle=0]{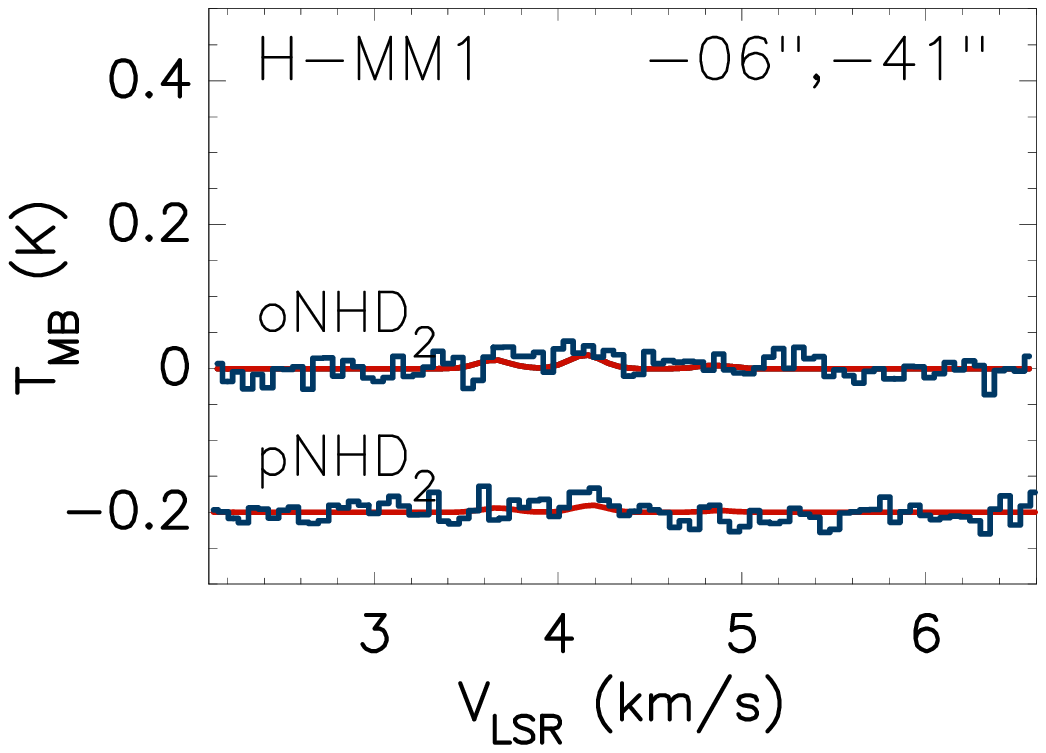}
\end{picture}}
\put(0,100){\large \bf H-MM1 \hspace{0.5cm} $\ddammo$}
\end{picture}

\caption[]{Observed and simulated $\ddammo$ spectra towards H-MM1. 
The model abundances $X(\oddammo)=9.25\times10^{-10}$, 
$X(\pddammo)=4.55\times10^{-10}$.}
\label{hmm1_best_fit_nhd2_spectra}
\end{figure*}

\begin{figure*}
\centering
\unitlength=1.0mm
\begin{picture}(160,110)(0,0)
\put(60,40){
\begin{picture}(0,0) 
\includegraphics[width=3.75cm,angle=0]{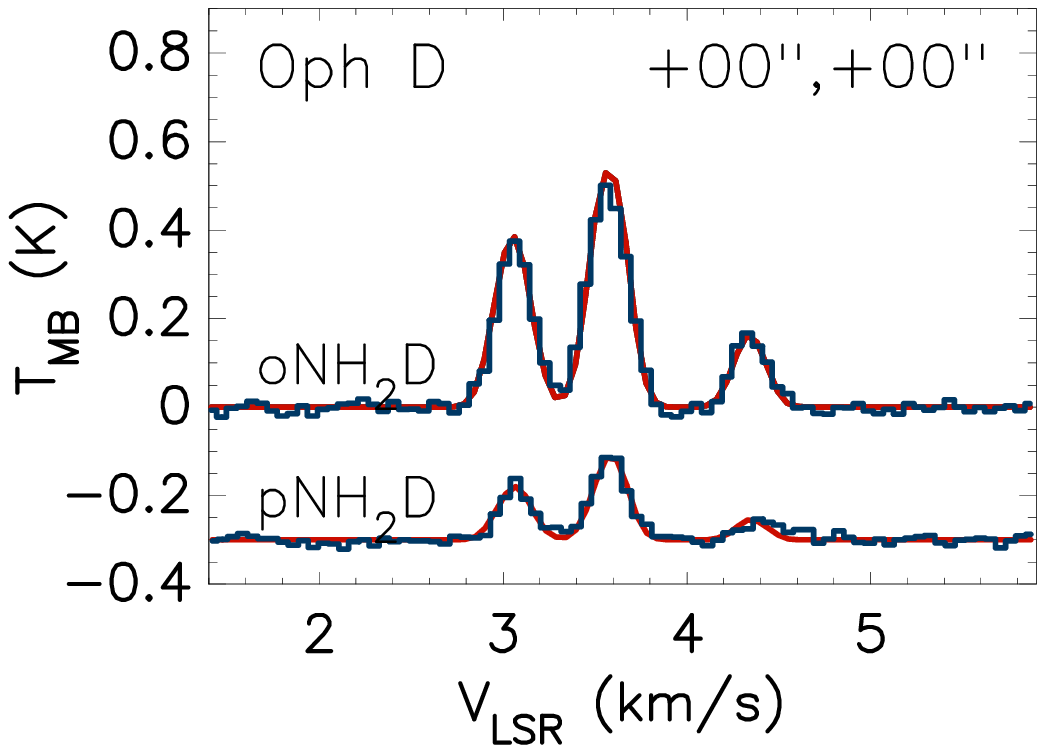}
\end{picture}}
\put(29,12){
\begin{picture}(0,0) 
\includegraphics[width=3.75cm,angle=0]{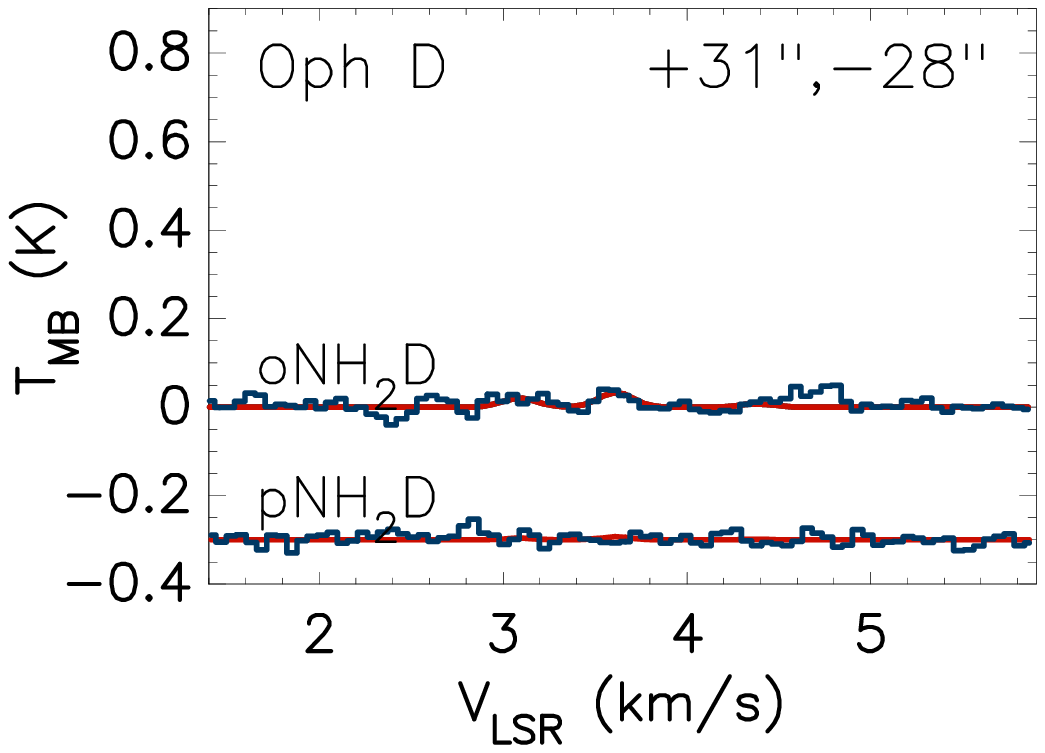}
\end{picture}}
\put(21,54){
\begin{picture}(0,0) 
\includegraphics[width=3.75cm,angle=0]{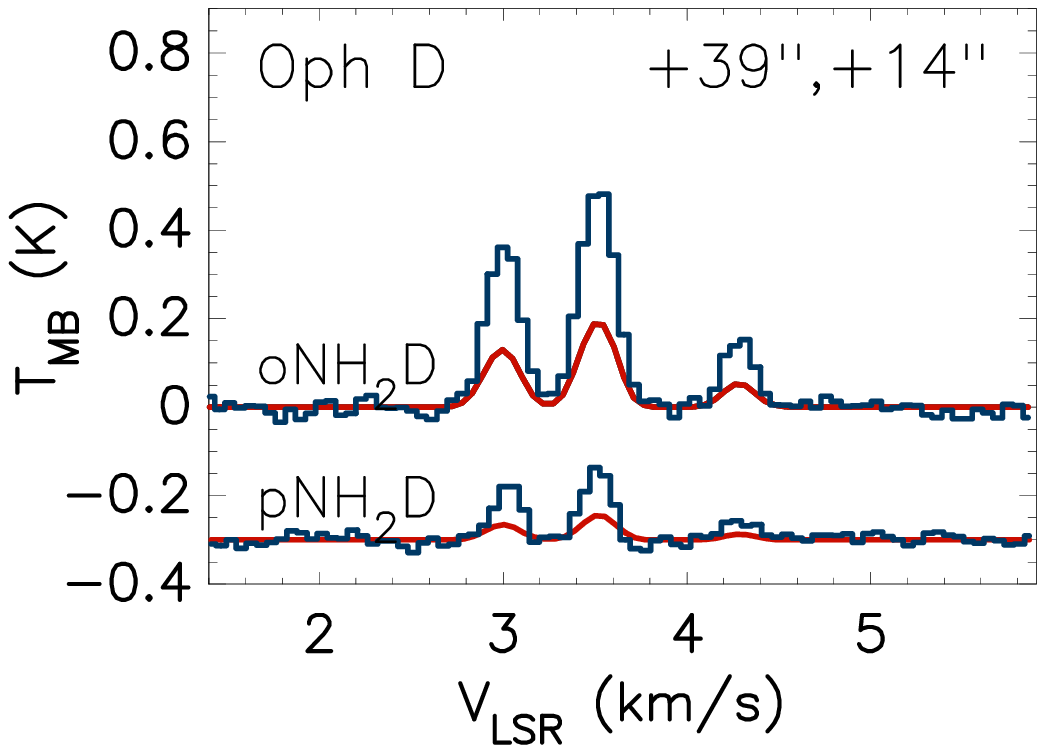}
\end{picture}}
\put(52,79){
\begin{picture}(0,0) 
\includegraphics[width=3.75cm,angle=0]{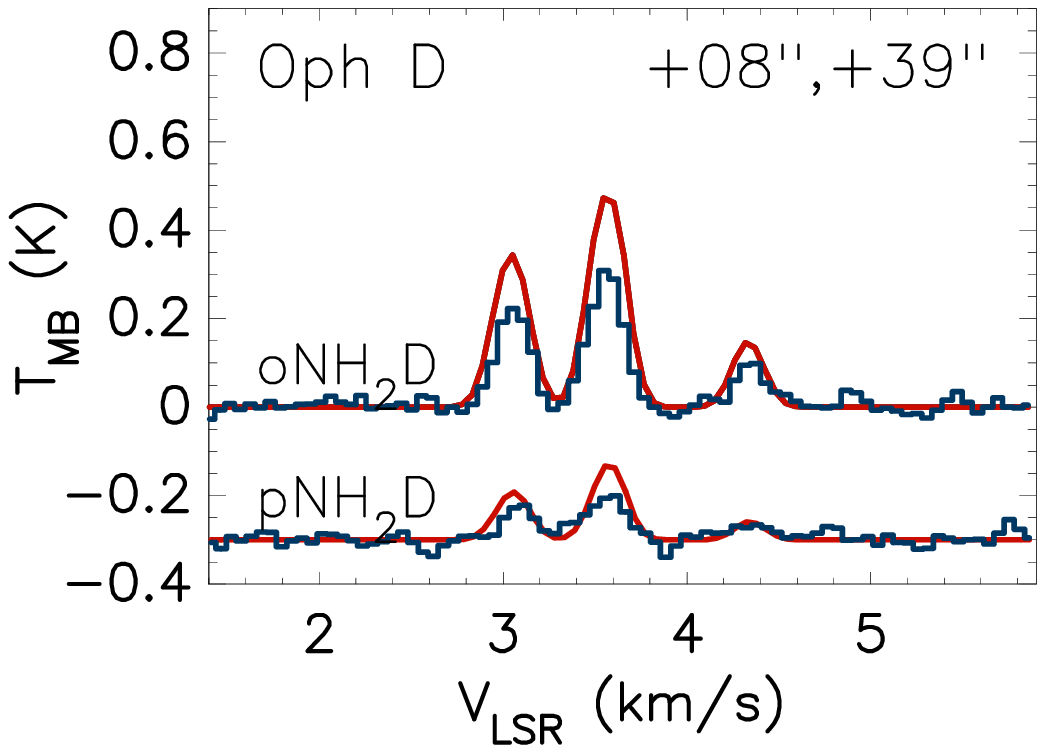}
\end{picture}}
\put(92,65){
\begin{picture}(0,0) 
\includegraphics[width=3.75cm,angle=0]{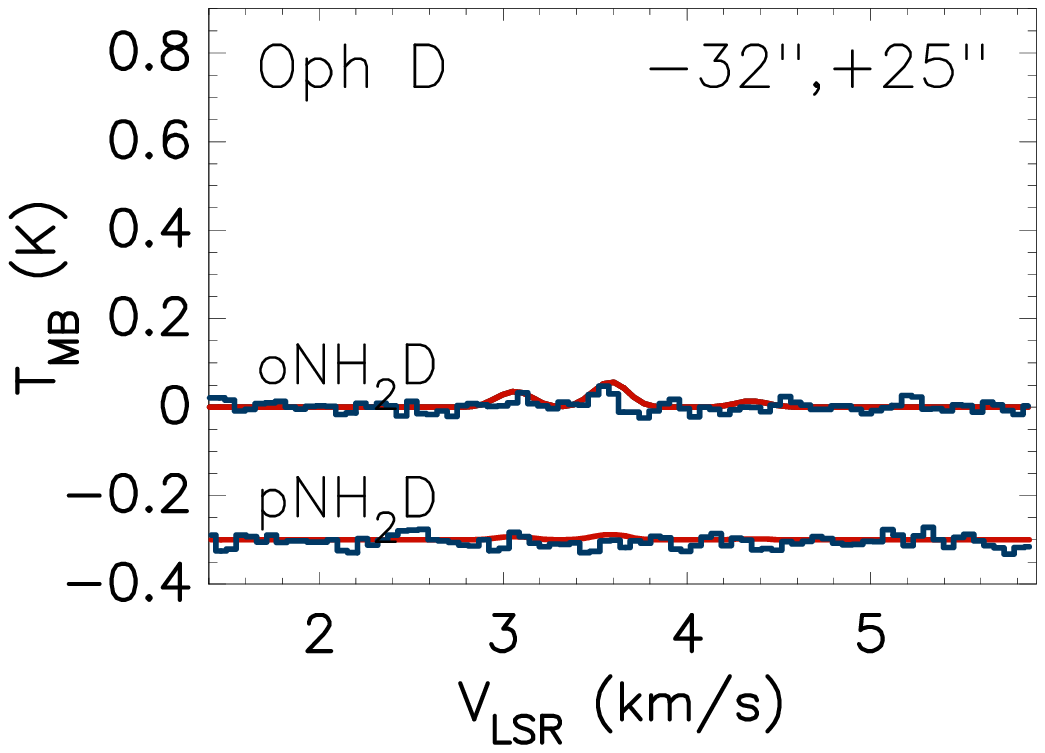}
\end{picture}}
\put(98,25){
\begin{picture}(0,0) 
\includegraphics[width=3.75cm,angle=0]{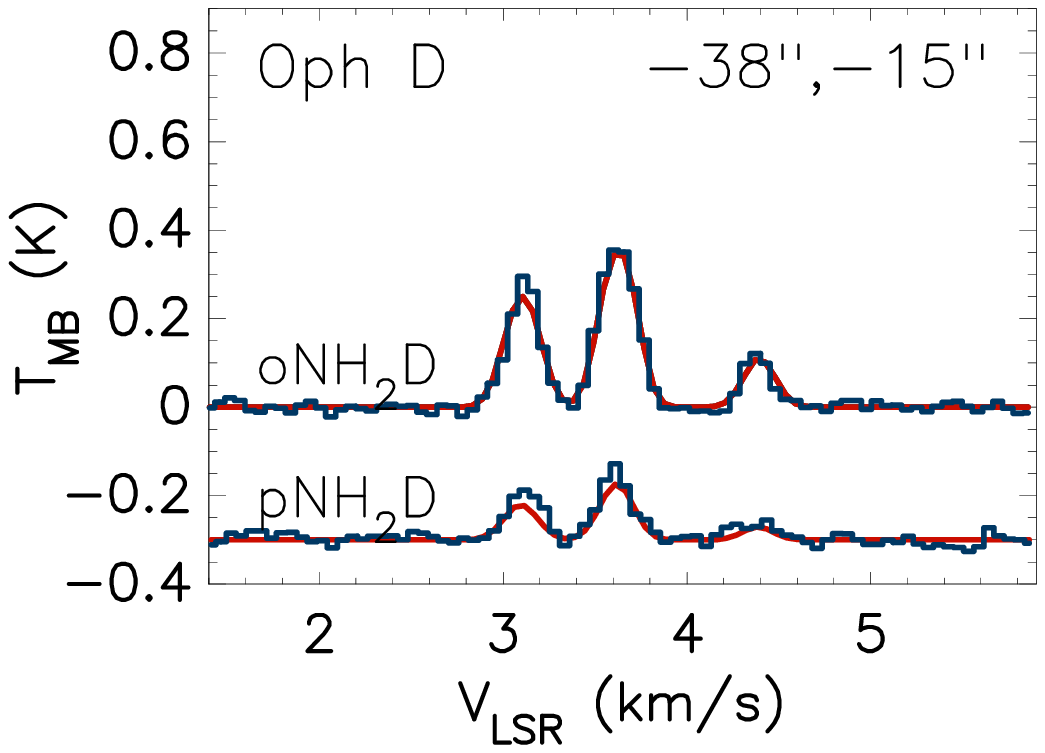}
\end{picture}}
\put(66,-1){
\begin{picture}(0,0) 
\includegraphics[width=3.75cm,angle=0]{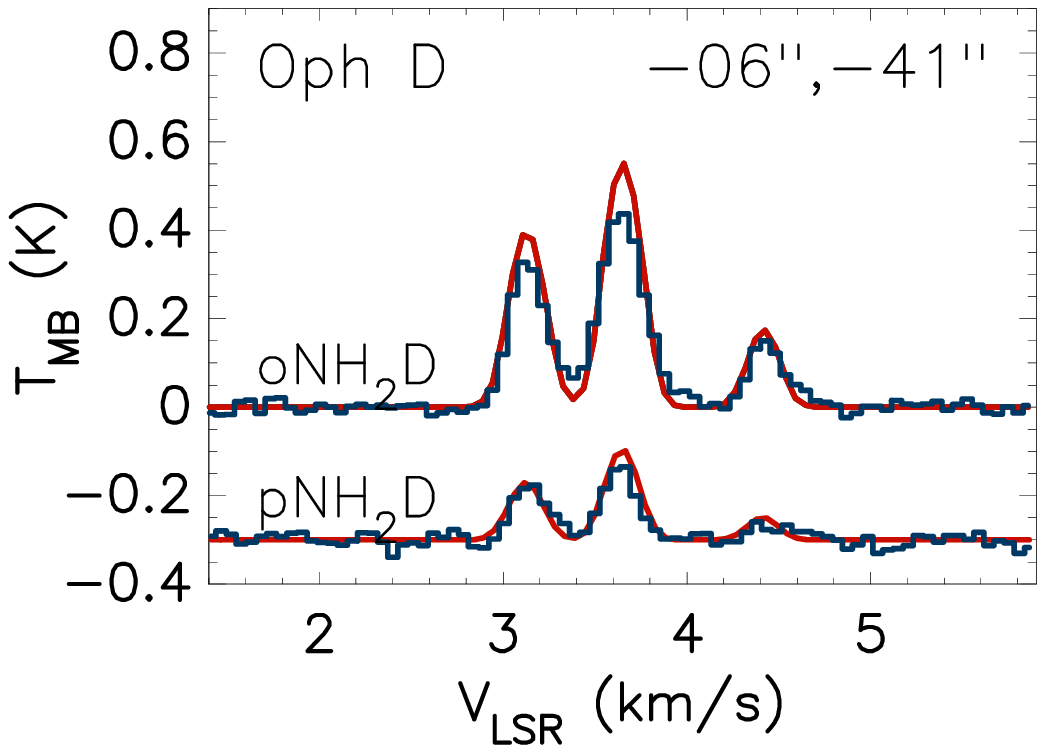}
\end{picture}}
\put(0,100){\large \bf Oph\,D \hspace{0.5cm} $\dammo$}
\end{picture}

\caption[]{$\dammo$ spectra observed towards Oph\,D in the Spring and Summer
2022. The model abundances $X(\odammo)=4.640\times10^{-9}$, 
$X(\pdammo)=1.469\times10^{-9}$.}
\label{ophd_best_fit_nh2d_spectra_summer}
\end{figure*}

\begin{figure*}
\centering
\unitlength=1.0mm
\begin{picture}(160,110)(0,0)
\put(60,40){
\begin{picture}(0,0) 
\includegraphics[width=3.75cm,angle=0]{ophd_nh2d_4.640E-09_1.469E-09_+00_+00.ps}
\end{picture}}
\put(39,4){
\begin{picture}(0,0) 
\includegraphics[width=3.75cm,angle=0]{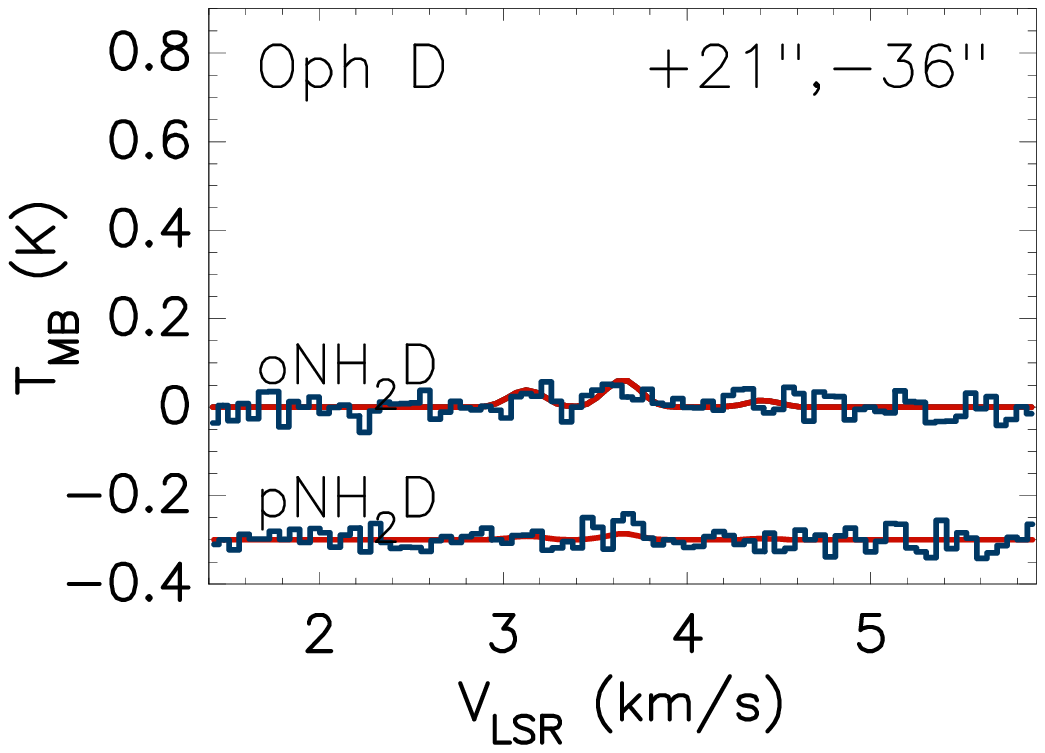}
\end{picture}}
\put(19,41){
\begin{picture}(0,0) 
\includegraphics[width=3.75cm,angle=0]{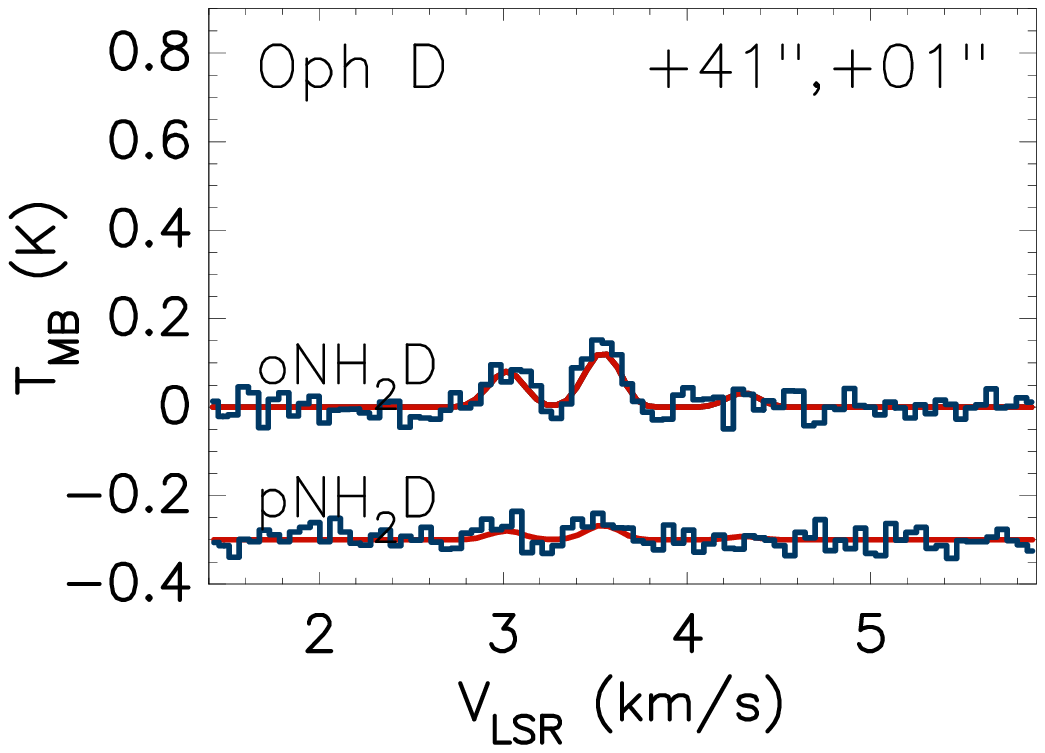}
\end{picture}}
\put(40,75){
\begin{picture}(0,0) 
\includegraphics[width=3.75cm,angle=0]{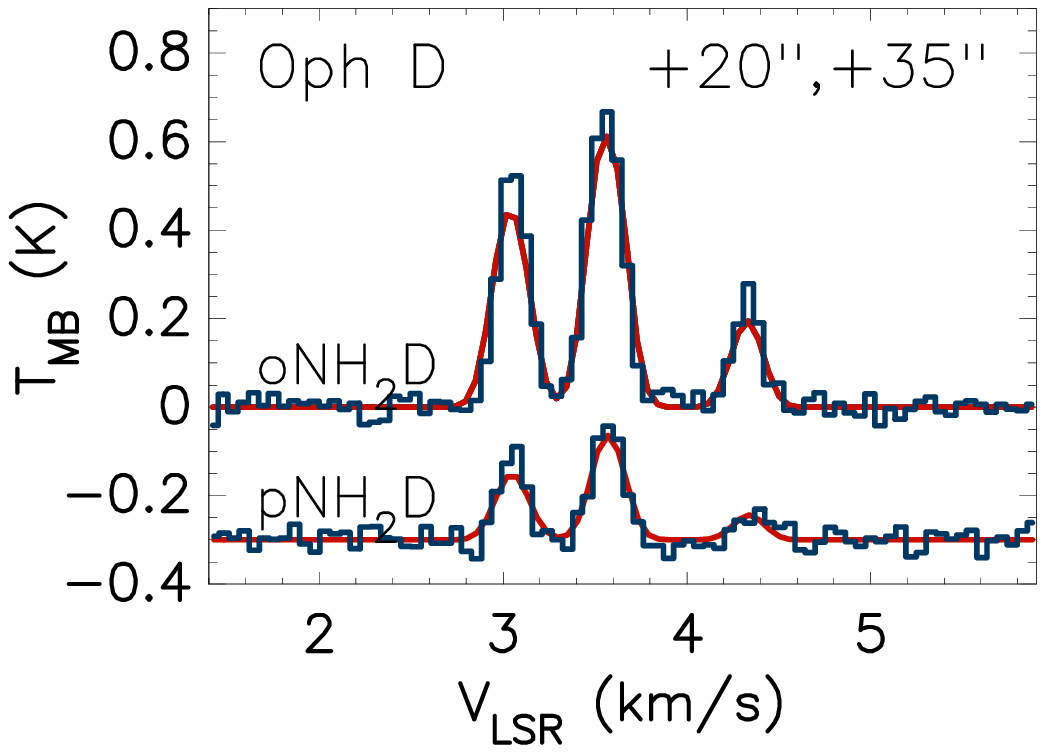}
\end{picture}}
\put(83,74){
\begin{picture}(0,0) 
\includegraphics[width=3.75cm,angle=0]{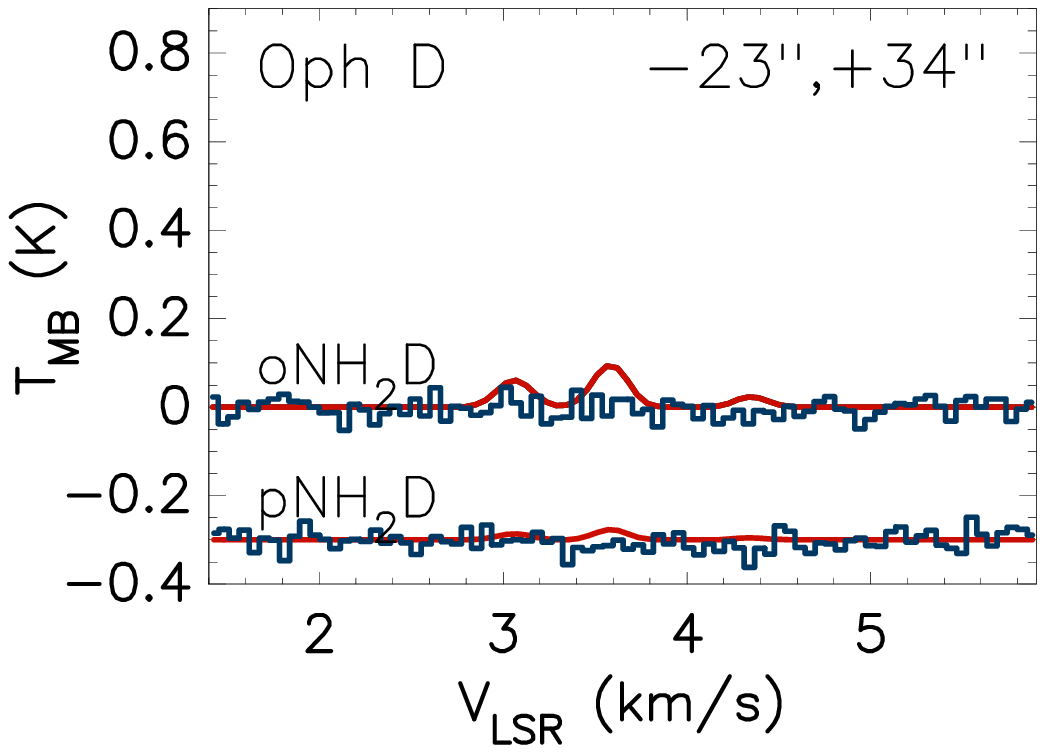}
\end{picture}}
\put(100,38){
\begin{picture}(0,0) 
\includegraphics[width=3.75cm,angle=0]{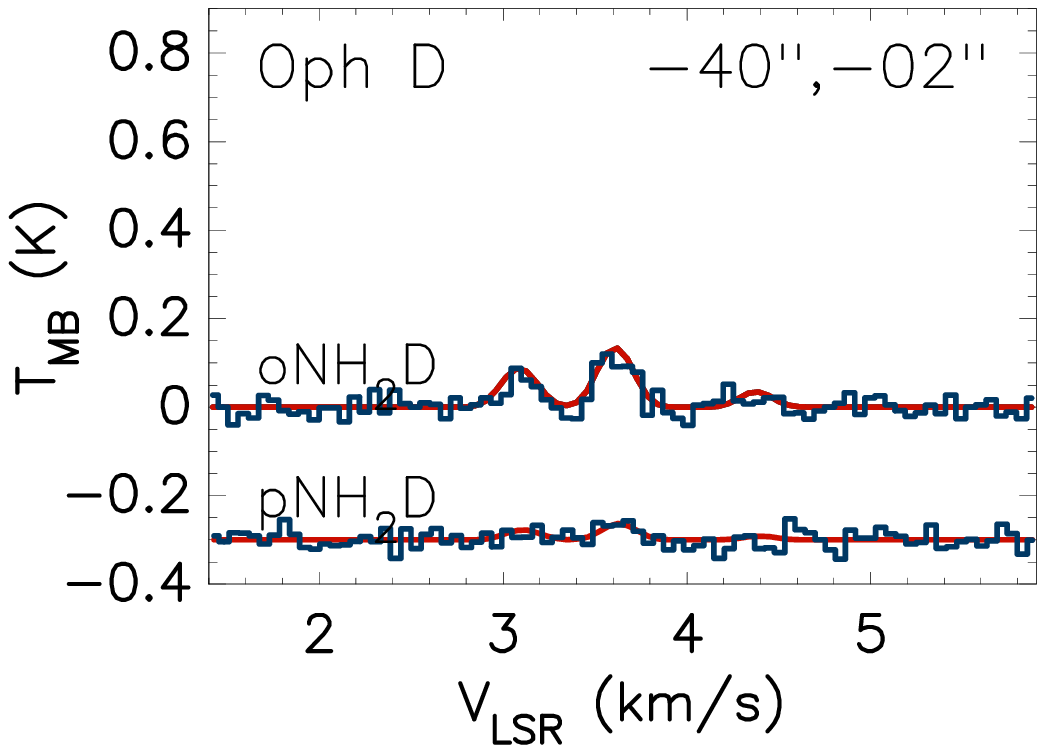}
\end{picture}}
\put(79,3){
\begin{picture}(0,0) 
\includegraphics[width=3.75cm,angle=0]{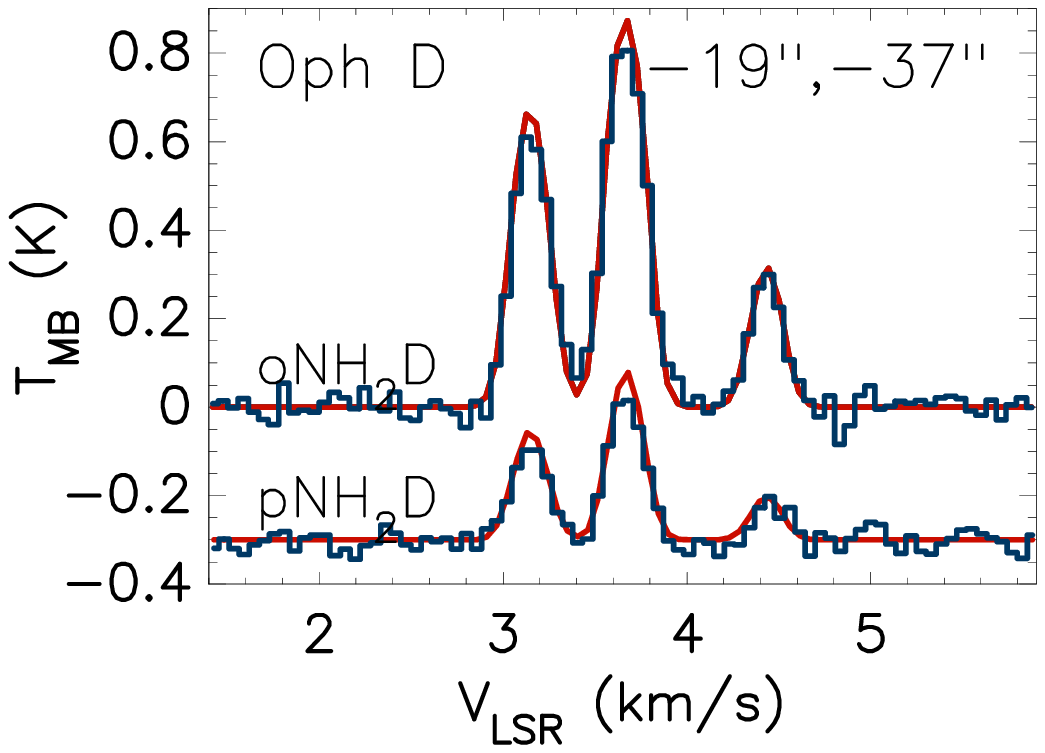}
\end{picture}}
\put(0,100){\large \bf Oph\,D \hspace{0.5cm} $\dammo$}
\end{picture}

\caption[]{$\dammo$ spectra towards Oph\,D observed in September 2022,
when the orientation of the LAsMA array was different from that used in the
Spring and Summer. The model abundances are the same as in
Fig.~\ref{ophd_best_fit_nh2d_spectra_summer}.}
\label{ophd_best_fit_nh2d_spectra_autumn}
\end{figure*}

\begin{figure*}
\centering
\unitlength=1.0mm
\begin{picture}(160,110)(0,0)
\put(60,40){
\begin{picture}(0,0) 
\includegraphics[width=3.75cm,angle=0]{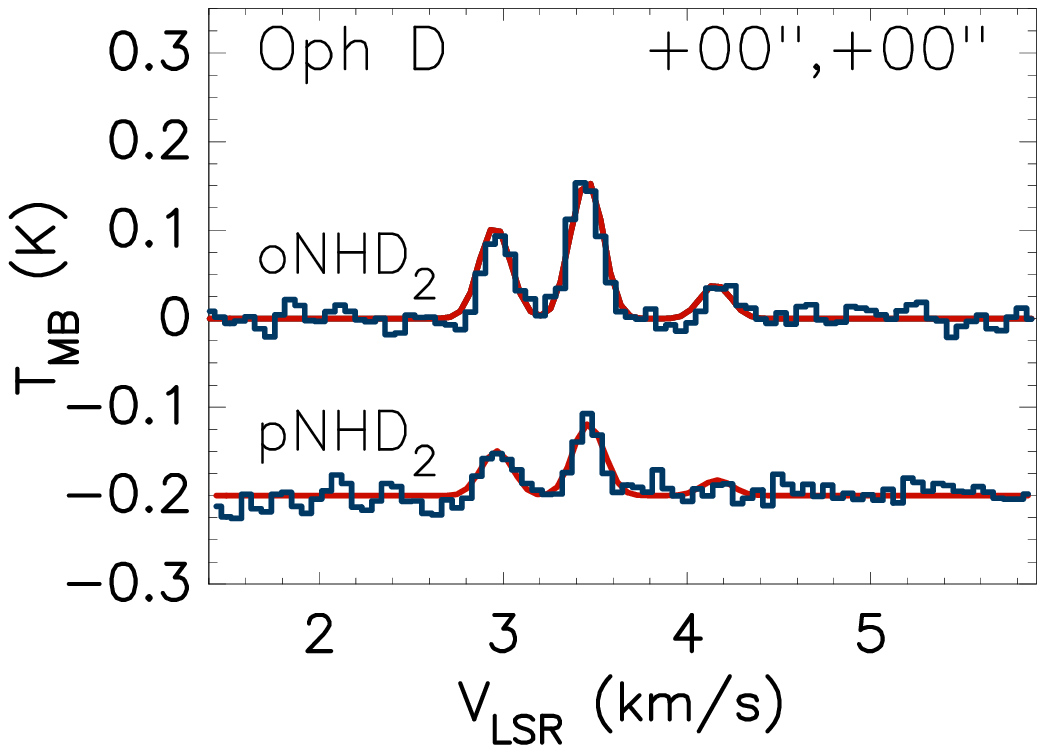}
\end{picture}}
\put(29,12){
\begin{picture}(0,0) 
\includegraphics[width=3.75cm,angle=0]{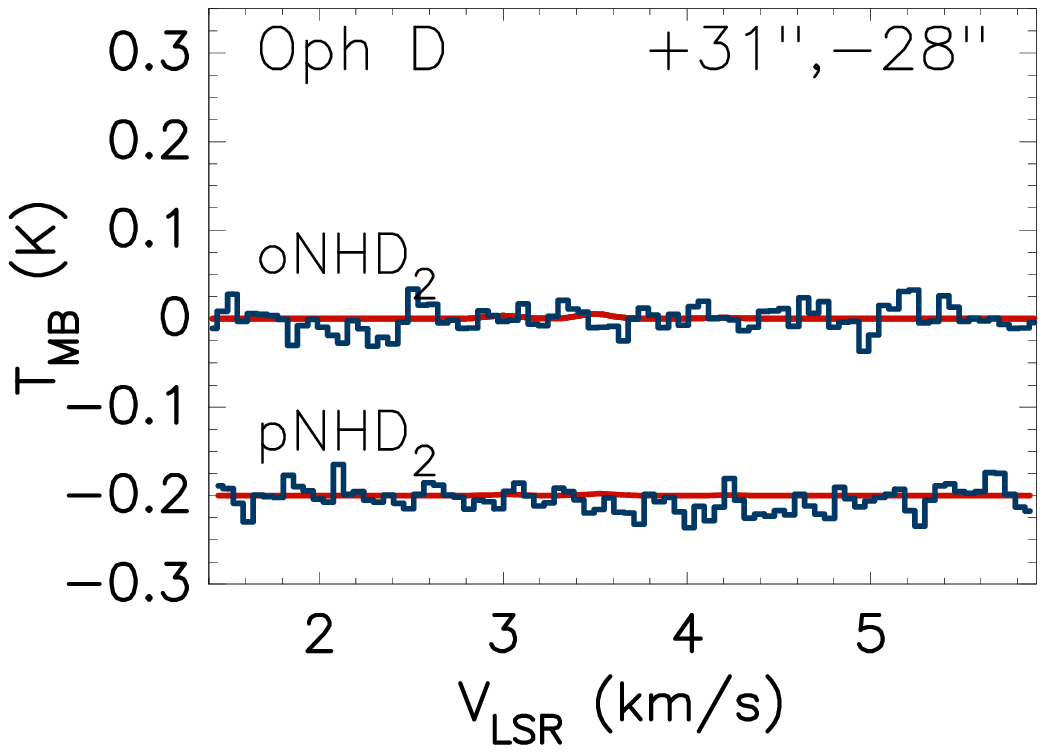}
\end{picture}}
\put(21,54){
\begin{picture}(0,0) 
\includegraphics[width=3.75cm,angle=0]{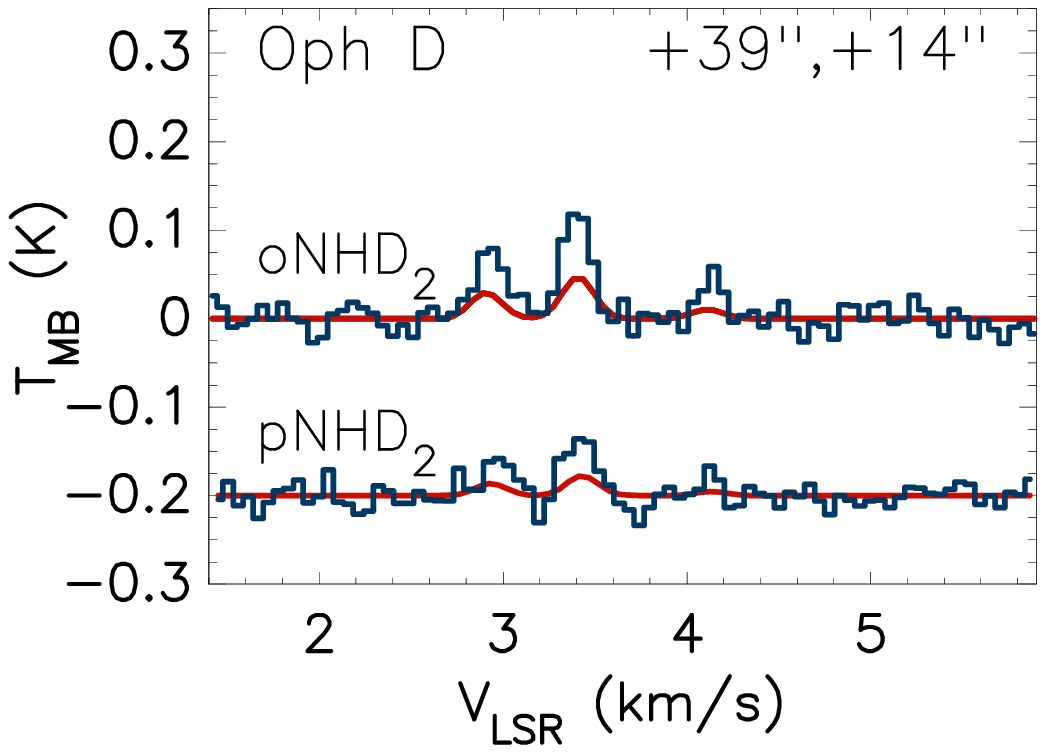}
\end{picture}}
\put(52,79){
\begin{picture}(0,0) 
\includegraphics[width=3.75cm,angle=0]{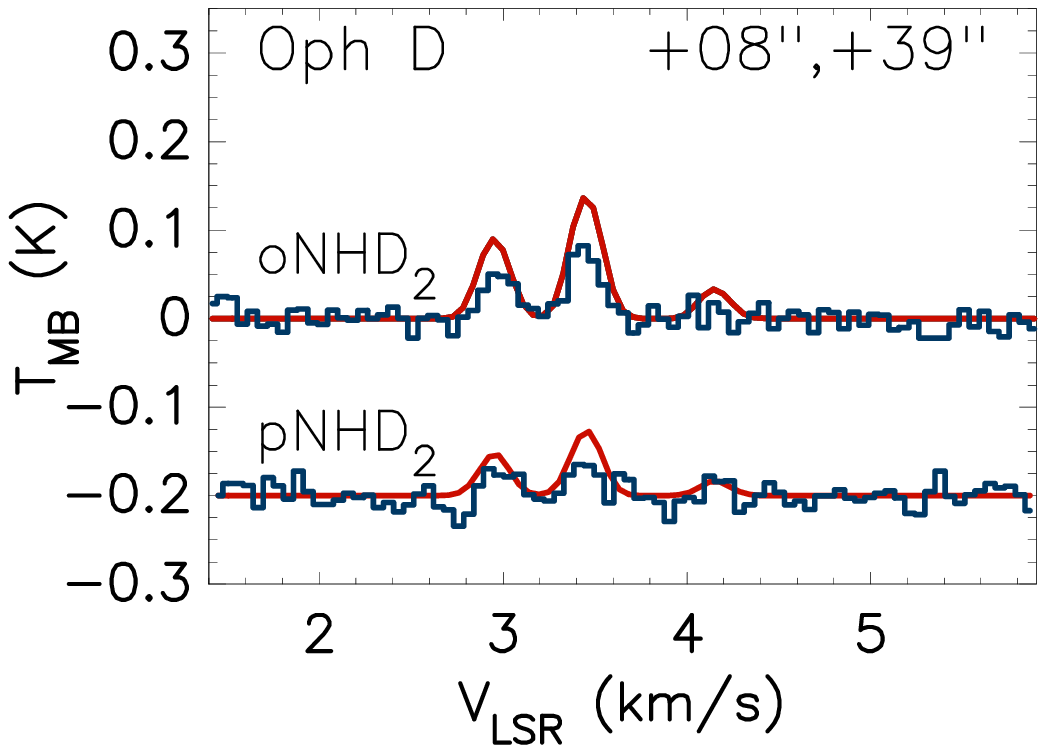}
\end{picture}}
\put(92,65){
\begin{picture}(0,0) 
\includegraphics[width=3.75cm,angle=0]{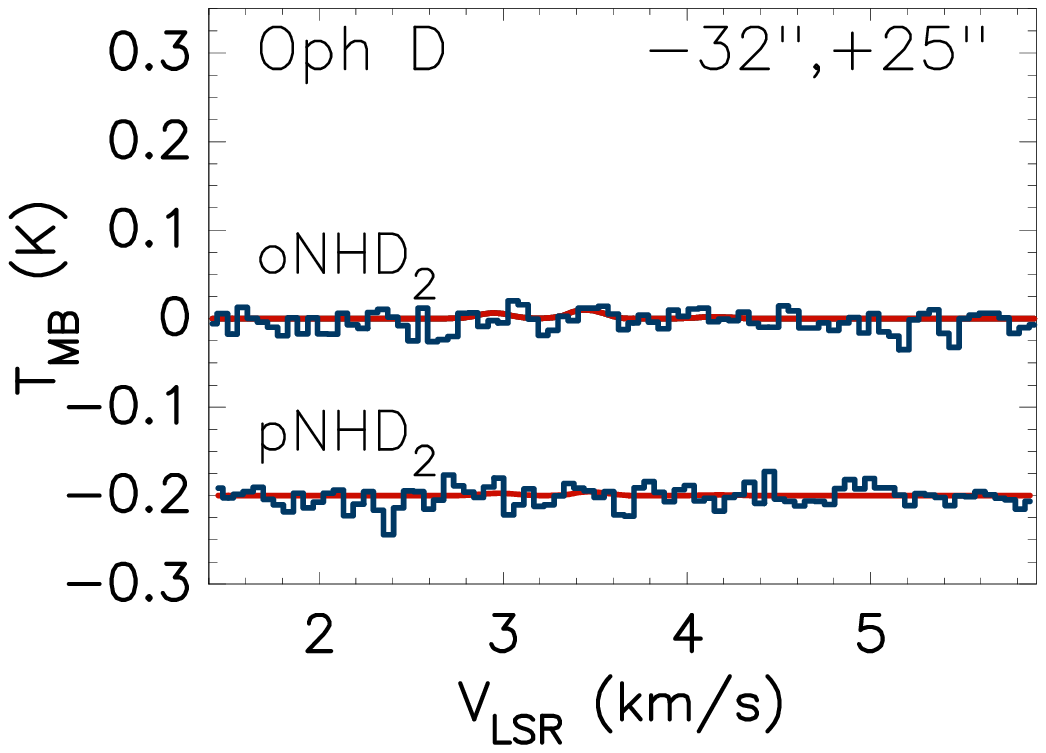}
\end{picture}}
\put(98,25){
\begin{picture}(0,0) 
\includegraphics[width=3.75cm,angle=0]{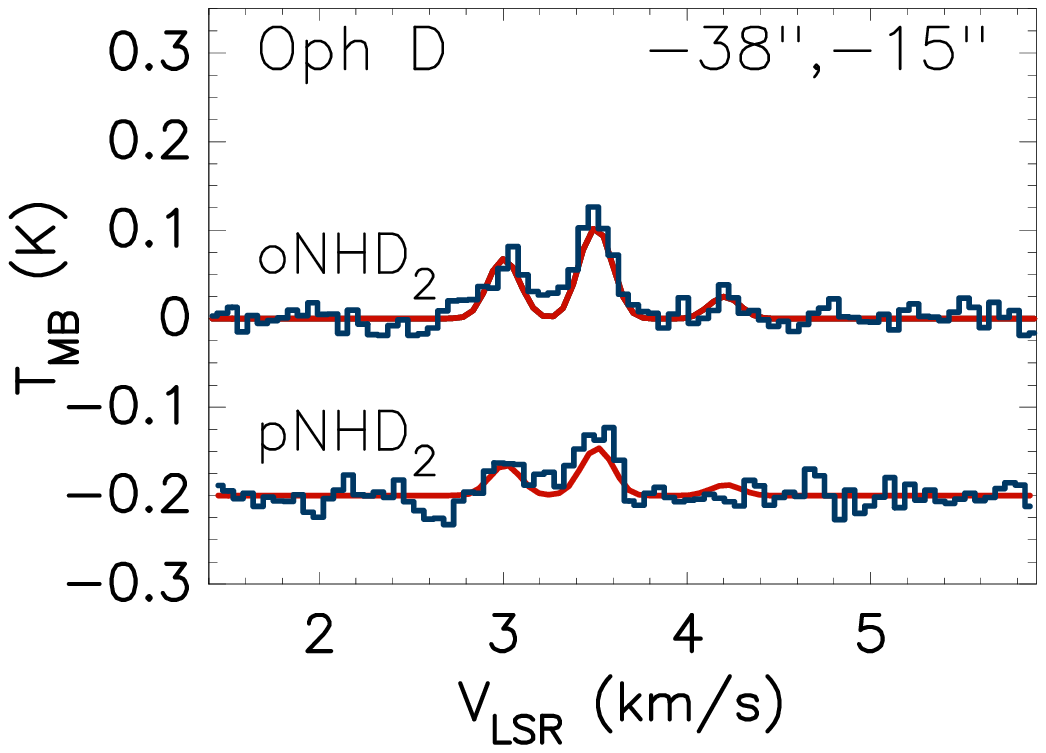}
\end{picture}}
\put(66,-1){
\begin{picture}(0,0) 
\includegraphics[width=3.75cm,angle=0]{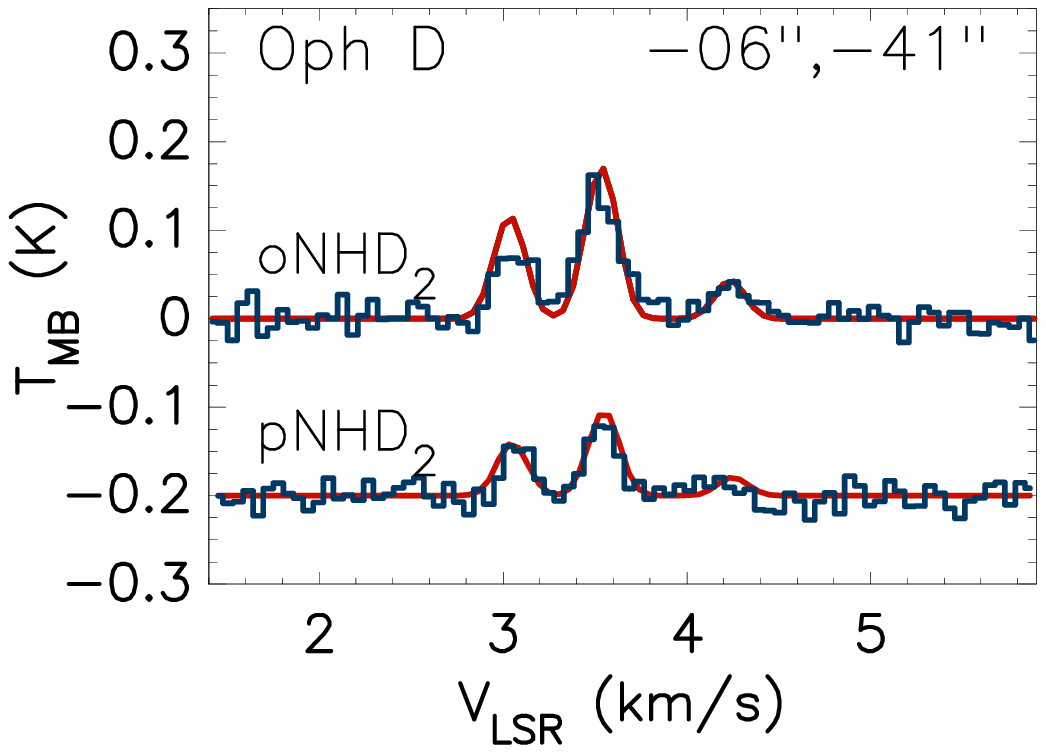}
\end{picture}}
\put(0,100){\large \bf Oph\,D \hspace{0.5cm} $\ddammo$}
\end{picture}

\caption[]{$\ddammo$ spectra observed towards Oph\,D in the Spring and Summer 2022. The model abundances $X(\oddammo)=1.122\times10^{-9}$, $X(\pdammo)=5.02\times10^{-10}$.}
\label{ophd_best_fit_nhd2_spectra_summer}
\end{figure*}

\begin{figure*}
\centering
\unitlength=1.0mm
\begin{picture}(160,110)(0,0)
\put(60,40){
\begin{picture}(0,0) 
\includegraphics[width=3.75cm,angle=0]{ophd_nhd2_1.122E-09_5.020E-10_+00_+00.ps}
\end{picture}}
\put(39,4){
\begin{picture}(0,0) 
\includegraphics[width=3.75cm,angle=0]{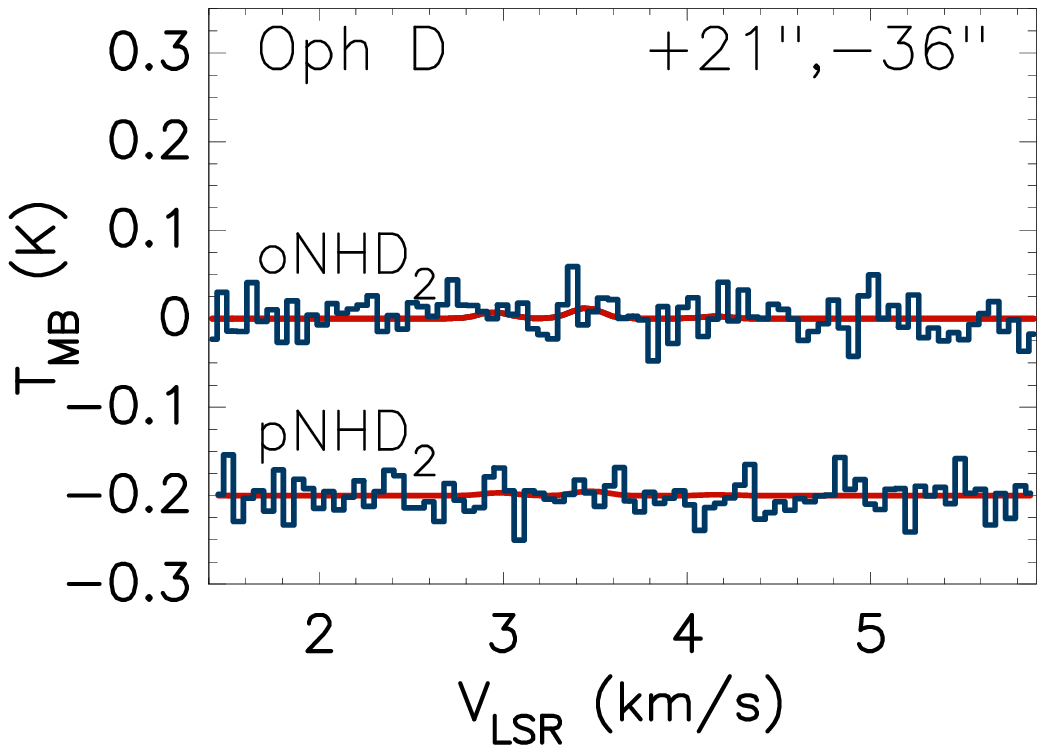}
\end{picture}}
\put(19,41){
\begin{picture}(0,0) 
\includegraphics[width=3.75cm,angle=0]{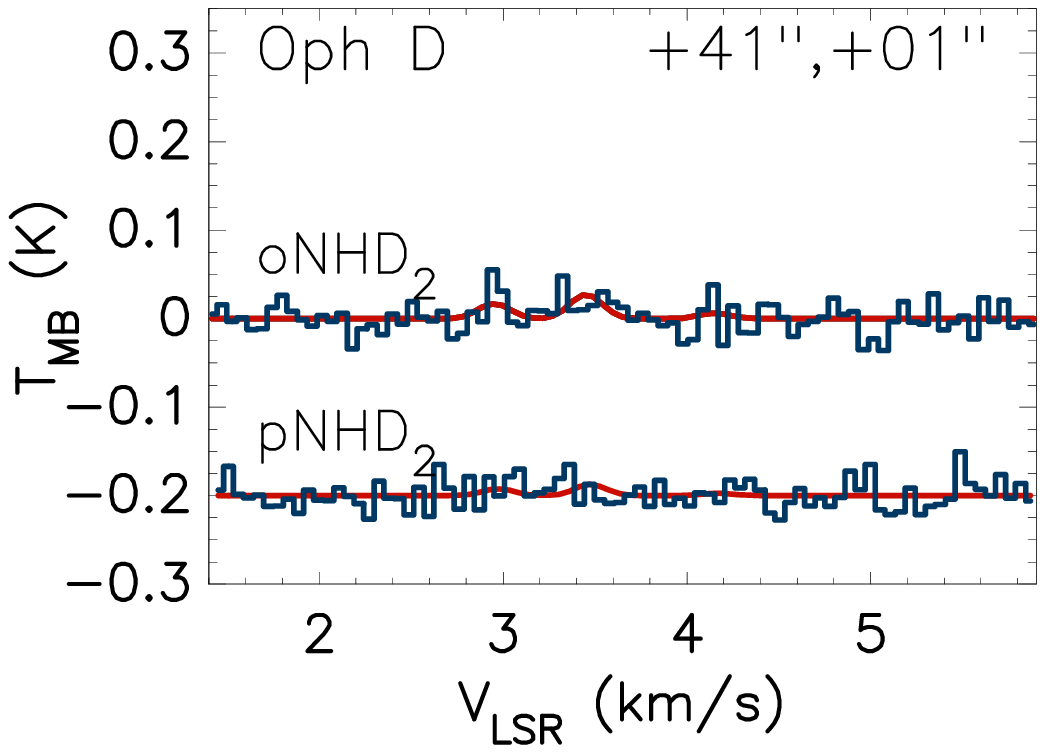}
\end{picture}}
\put(40,75){
\begin{picture}(0,0) 
\includegraphics[width=3.75cm,angle=0]{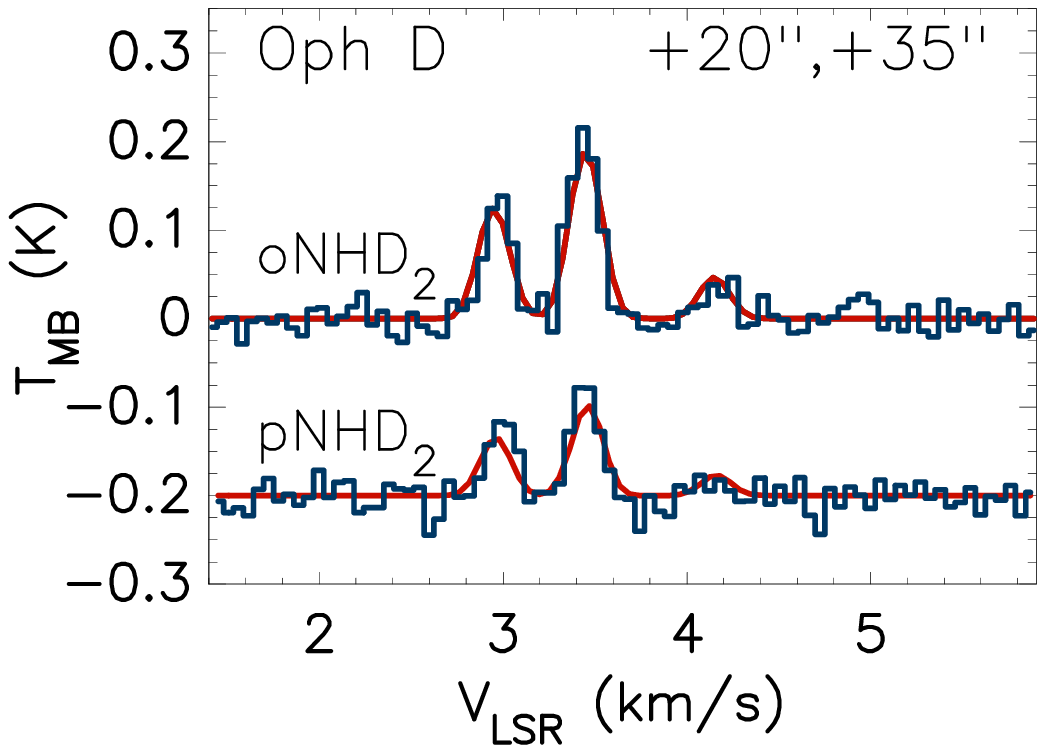}
\end{picture}}
\put(83,74){
\begin{picture}(0,0) 
\includegraphics[width=3.75cm,angle=0]{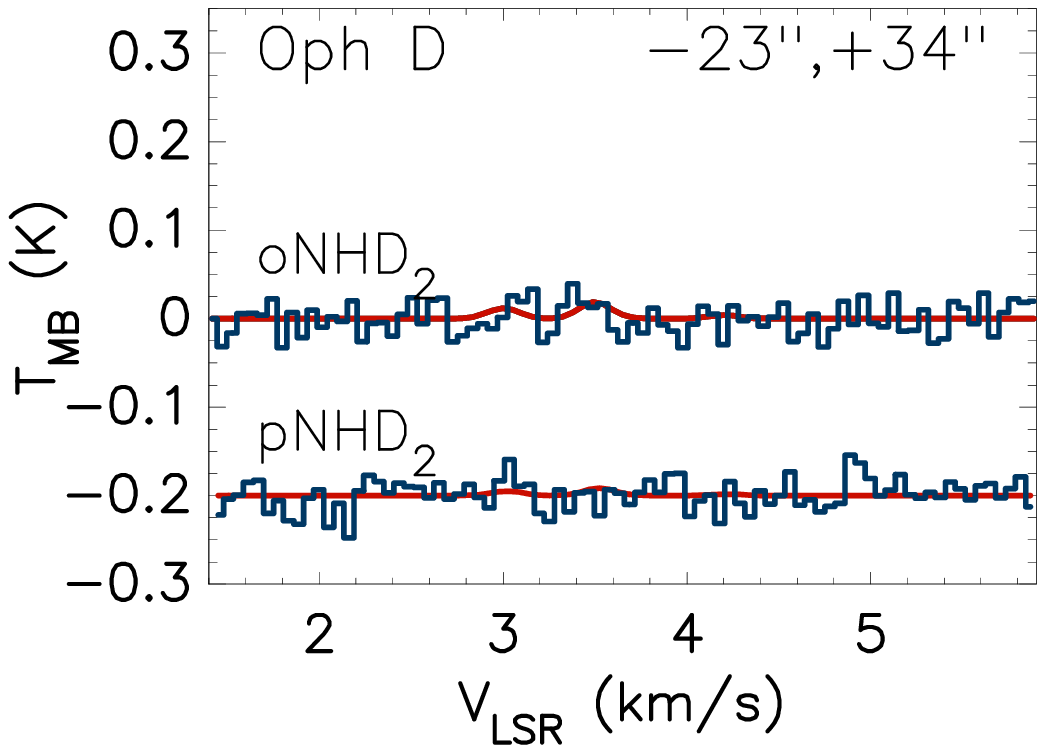}
\end{picture}}
\put(100,38){
\begin{picture}(0,0) 
\includegraphics[width=3.75cm,angle=0]{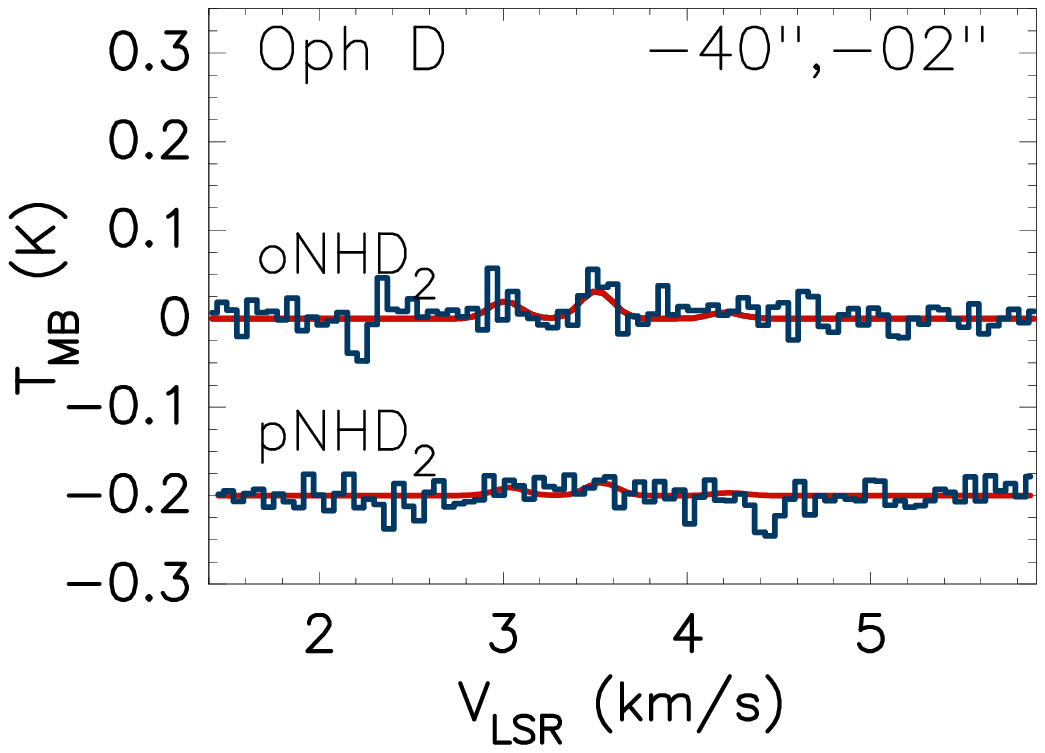}
\end{picture}}
\put(79,3){
\begin{picture}(0,0) 
\includegraphics[width=3.75cm,angle=0]{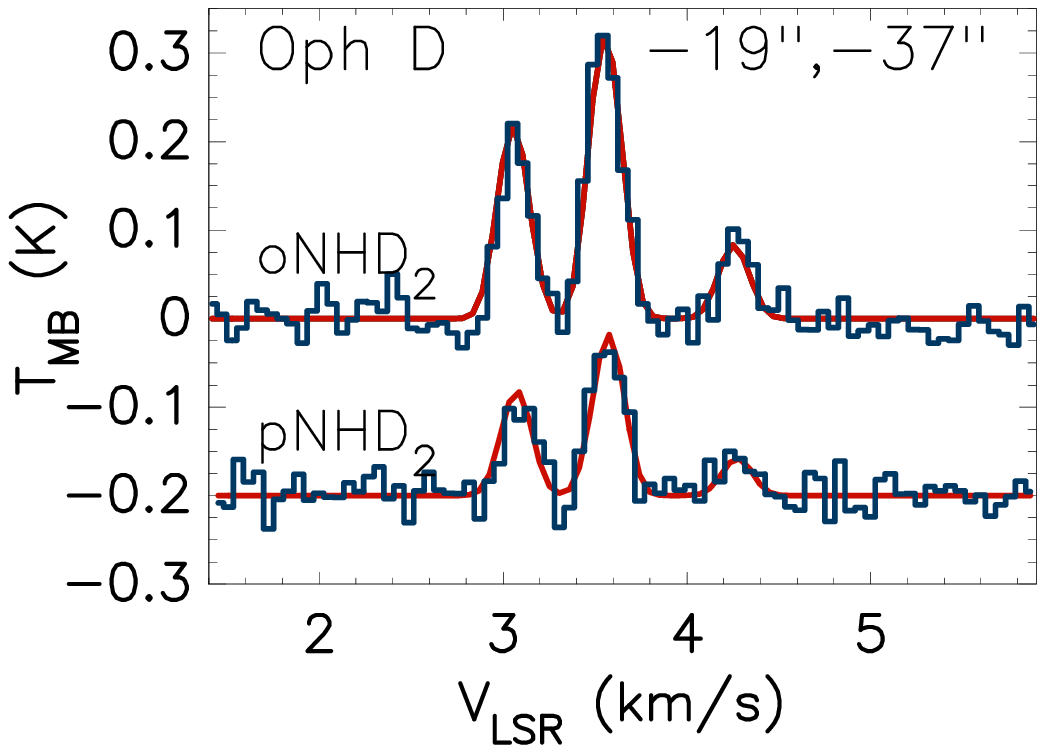}
\end{picture}}
\put(0,100){\large \bf Oph\,D \hspace{0.5cm} $\ddammo$}
\end{picture}

\caption[]{$\ddammo$ spectra observed towards Oph\,D in September 2022. The model abundances are the same as in Fig.~\ref{ophd_best_fit_nhd2_spectra_summer}.}
\label{ophd_best_fit_nhd2_spectra_autumn}
\end{figure*}

The $\dammo$ and $\ddammo$ spectra observed towards H-MM1 and Oph\,D with
LAsMA are shown in Figs.~\ref{hmm1_best_fit_nh2d_spectra} to 
\ref{ophd_best_fit_nhd2_spectra_autumn}. The simulated spectra (shown with
red) are calculated from the 3D models assuming that the fractional abundance
is constant. 

\section{Spin-state branching ratio diagrams for the reaction sequence $\ammo  {\rightarrow} {\rm NH_3D^+} \rightarrow \dammo$ }

{ Fig.~\ref{branching_diagram} illustrates the spin selection rules of the reaction sequence $\ammo \stackrel{\htwodplus}{\rightarrow} {\rm NH_3D^+} \stackrel{\rm e^-}{\rightarrow} \dammo$, assuming that this follows either the proton hop (PH) or the full scrambling (FS) scenario. This is one of the principal reaction sequences leading to $\dammo$. In the FS diagram, we have assumed that $\htwodplus$ is the more common para modification; in the PH mechanism, the spin state of the proton/deuteron donor does not matter.}

\begin{figure*}
\unitlength=1mm
\begin{picture}(180,80)(0,0)

  \put(6,60){\color{blue} o$\ammo$}
  \put(11,61){\color{blue} \circle{17}}
  \put(19,62){\vector(1,0){12}}
  \put(20.5,65){$\htwodplus$}
  \put(33.0,60){\color{blue} oNH$_3$D$^+$}
  \put(39,61){\color{blue} \circle{17}}
  \put(48.0,62){\vector(1,0){12}}
  \put(51,65){e$^-$}
  \put(62.5,60){\color{blue} o$\dammo$}
  \put(68,61){\color{blue} \circle{17}}
  \put(5,75){\bf PH}
  \put(6,17){\color{red} p$\ammo$}
  \put(11,18){\color{red} \circle{17}}
  \put(19,18){\vector(1,0){12}}
  \put(20.5,22){$\htwodplus$}
  \put(33.0,17){\color{red} pNH$_3$D$^+$}
  \put(39,18){\color{red} \circle{17}}
  \put(48,20){\vector(2,1){12}}
  \put(48,16){\vector(2,-1){12}} 
  \put(51,33){e$^-$}
  \put(48.5,24){1/2}
  \put(48.5,9){1/2}
  \put(62.5,27){\color{blue} o$\dammo$}
  \put(62.5,7){\color{red} p$\dammo$}
  \put(68,28){\color{blue} \circle{17}}
  \put(68,8){\color{red} \circle{17}}
  \put(101,60){\color{blue} o$\ammo$}
  \put(106,61){\color{blue} \circle{17}}
  \put(114,64){\vector(2,1){12}}
  \put(114,60){\vector(2,-1){12}}
  \put(115,69){1/2}
  \put(115,52){1/2}
  \put(113,75){p$\htwodplus$}
  \put(128.5,70){\color{blue} oNH$_3$D$^+$}
  \put(128.5,50){\color{red} pNH$_3$D$^+$}
  \put(135,71){\color{blue} \circle{17}}
  \put(135,51){\color{red} \circle{17}}
  \put(143.0,71){\vector(1,0){12}}
  \put(143.0,51){\vector(1,0){12}}
  \put(143.0,54){\vector(1,1){12}}
  \put(146,75){e$^-$}
  \put(157.5,70){\color{blue} o$\dammo$}
  \put(157.5,50){\color{red} p$\dammo$}
  \put(163,71){\color{blue} \circle{17}}
  \put(163,51){\color{red} \circle{17}}
  \put(143.5,60){1/2}
  \put(144.5,47){1/2}
  \put(100,75){\bf FS}
  \put(100,17){\color{red} p$\ammo$}
  \put(106,18){\color{red} \circle{17}}
  \put(114,20){\vector(2,1){12}}
  \put(114,16){\vector(2,-1){12}}
  \put(115,25){1/5}
  \put(115,9){4/5}
  \put(113,32){p$\htwodplus$}
  \put(127.5,27){\color{blue} oNH$_3$D$^+$}
  \put(127.5,7){\color{red} pNH$_3$D$^+$}
  \put(135,28){\color{blue} \circle{17}}
  \put(135,8){\color{red} \circle{17}}
  \put(143,28){\vector(1,0){12}}
  \put(143,8){\vector(1,0){12}}
  \put(143,10){\vector(1,1){12}} 
  \put(146,32){e$^-$}
  \put(143.5,17){1/2}
  \put(144.5,4){1/2}
  \put(156.5,27){\color{blue} o$\dammo$}
  \put(156.5,7){\color{red} p$\dammo$}
  \put(163,28){\color{blue} \circle{17}}
  \put(163,8){\color{red} \circle{17}}
  
\end{picture}
\caption{Spin-state branching ratio diagrams for the reaction sequence ${\ammo  \stackrel{\htwodplus}{\rightarrow} {\rm NH_3D^+} \stackrel{\rm e^-}{\rightarrow} \dammo}$ in the PH (left) and FS (right) models. Assuming that o/p-$\ammo=1$, the PH scenario leads to o/p-$\dammo=3$, whereas the FS scenario results in o/p-$\dammo=27/13$.}
      \label{branching_diagram}
\end{figure*}

\section{Simulated spectra using abundances from the chemistry model}
\label{appendix:simulated_spectra_pyRate}

\begin{figure*}
\unitlength=1mm
\begin{picture}(160,110)(0,0)

\put(115,50){
\begin{picture}(0,0) 
\includegraphics[width=6.5cm,angle=0]{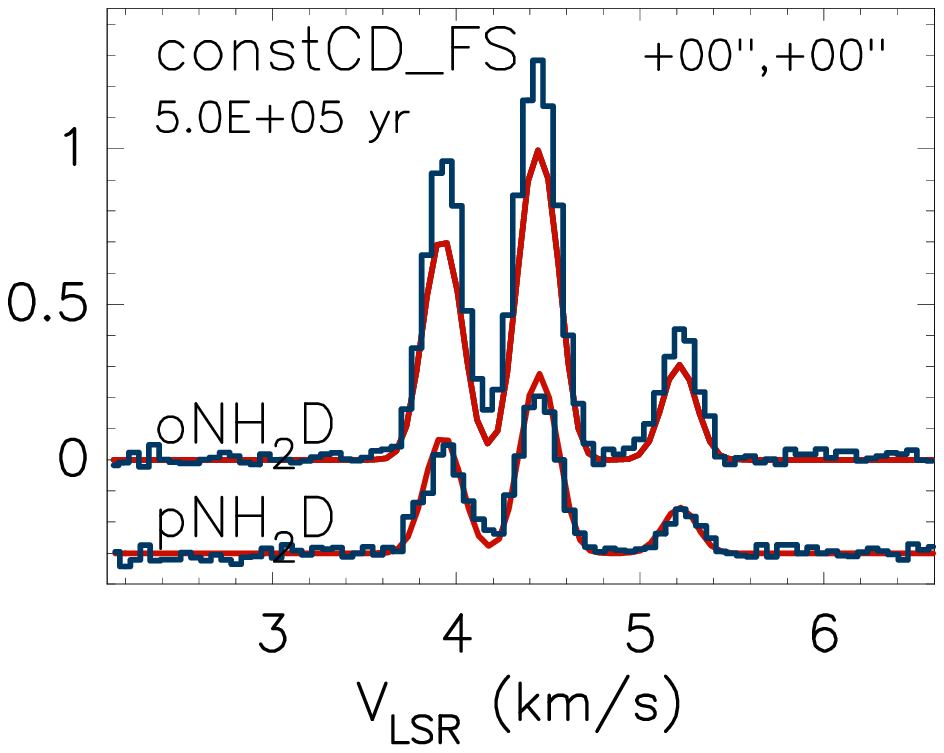}
\end{picture}}

\put(55,50){
\begin{picture}(0,0) 
\includegraphics[width=6.5cm,angle=0]{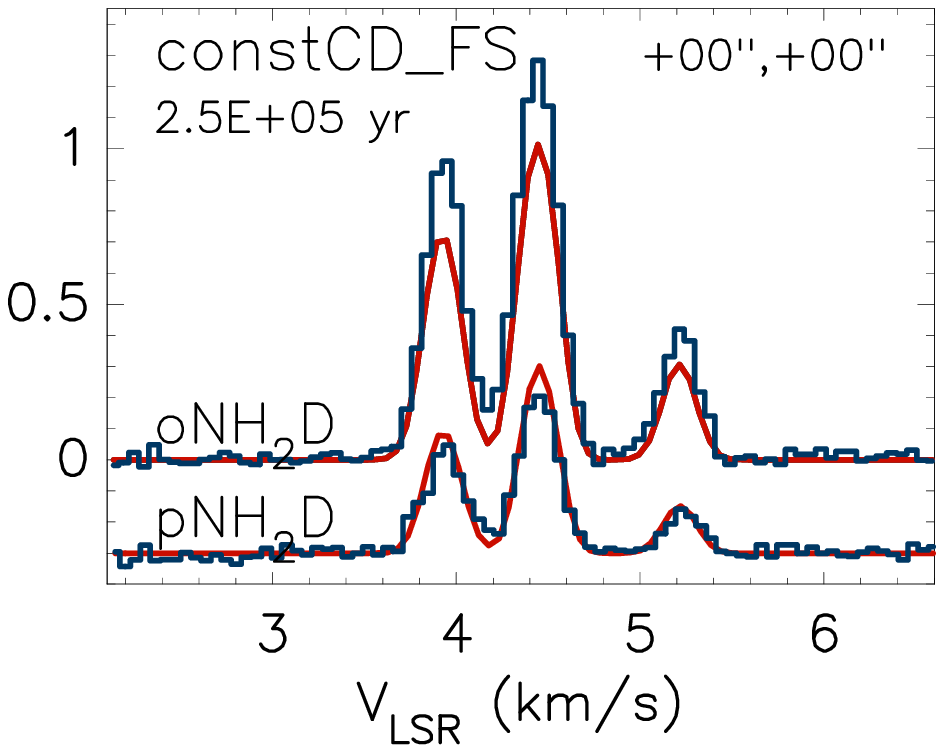}
\end{picture}}

\put(-6,50){
\begin{picture}(0,0) 
\includegraphics[width=6.5cm,angle=0]{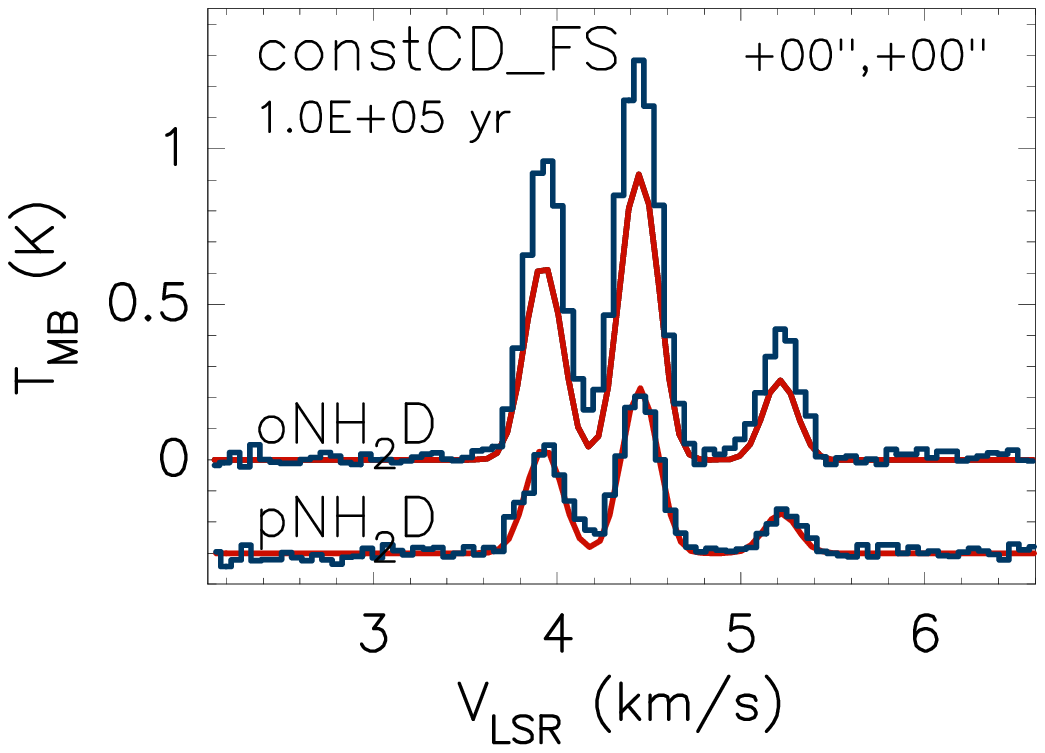}
\end{picture}}

\put(115,0){
\begin{picture}(0,0) 
\includegraphics[width=6.5cm,angle=0]{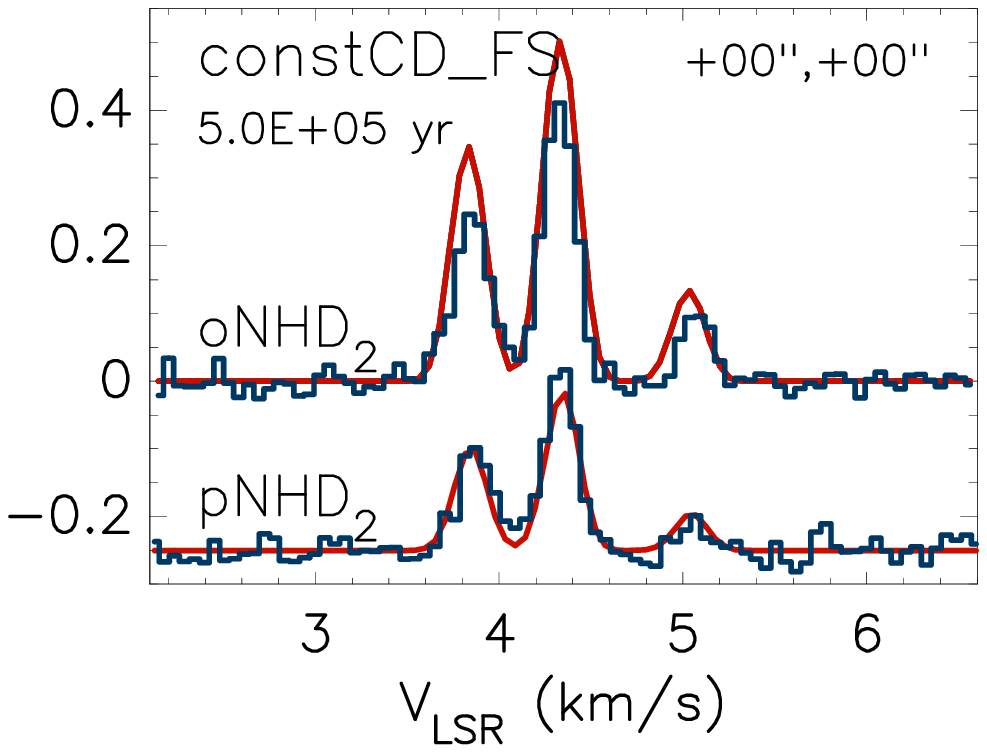}
\end{picture}}

\put(55,0){
\begin{picture}(0,0) 
\includegraphics[width=6.5cm,angle=0]{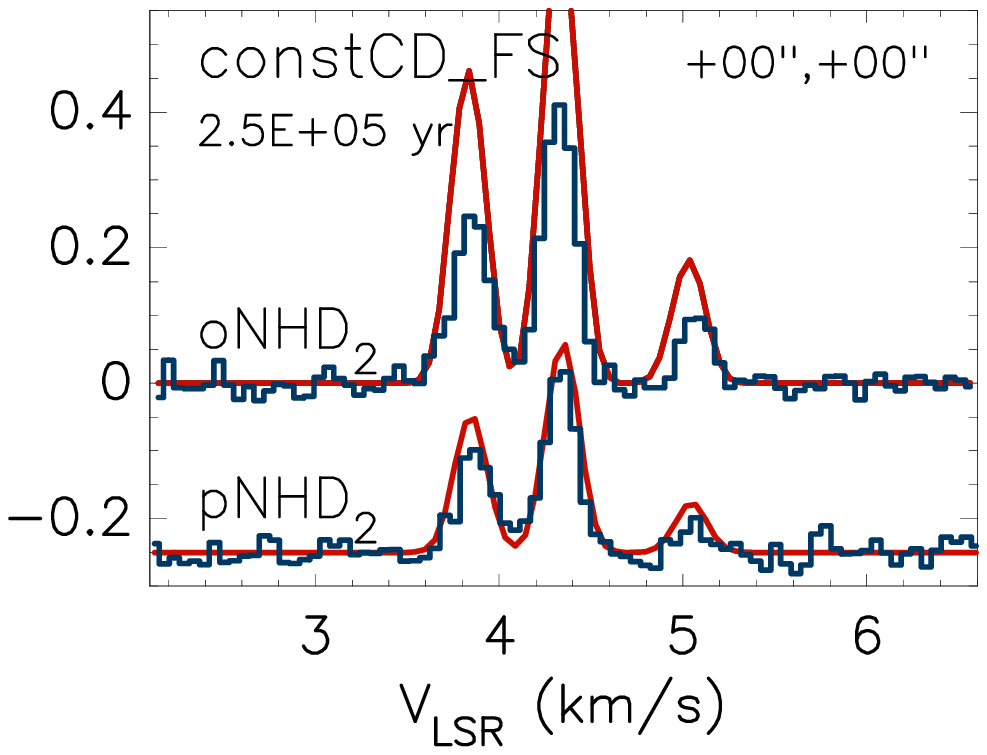}
\end{picture}}

\put(-6,0){
\begin{picture}(0,0) 
\includegraphics[width=6.5cm,angle=0]{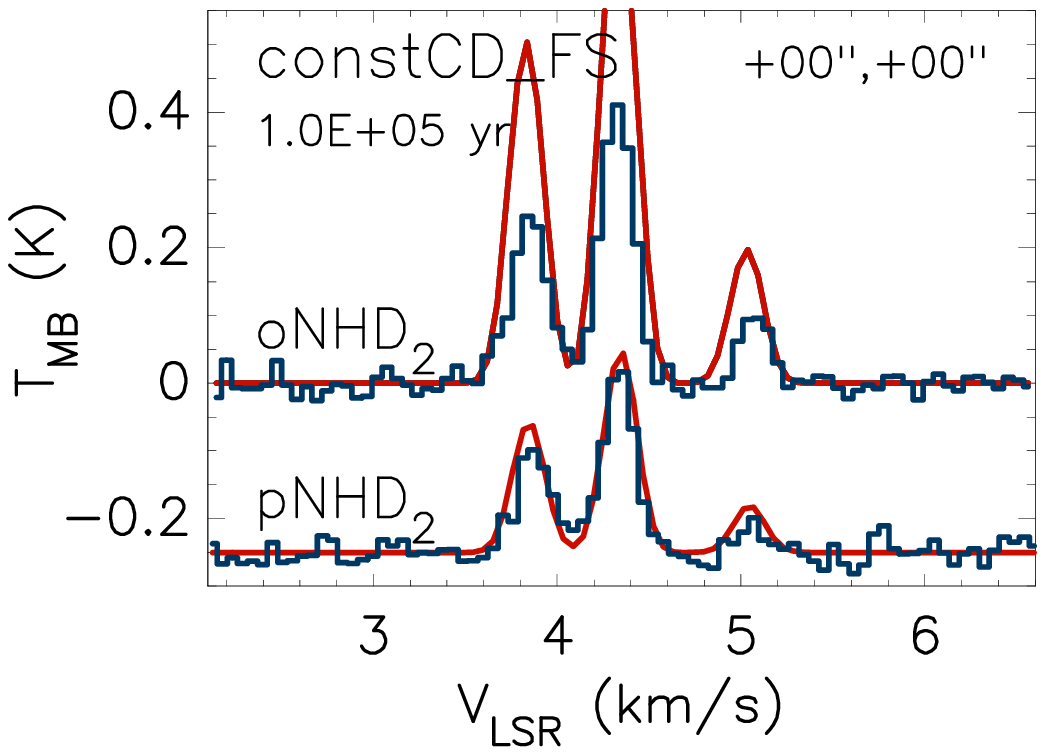}
\end{picture}}

\put(0,100){\bf Constant CD, FS}

\end{picture}
\caption{Simulated $\dammo$ and $\ddammo$ spectra {towards the centre of the 3D model for H-MM1 with abundance distributions interpolated from the chemistry model} assuming constant CD and full scrambling (FS, red lines). The spectra are convolved with the APEX beam. Spectra observed with APEX towards the centre of H-MM1 are shown with blue lines.}
\label{constCD_FS_spectra}
\end{figure*}

\begin{figure*}
\unitlength=1mm
\begin{picture}(160,110)(0,0)

\put(115,50){
\begin{picture}(0,0) 
\includegraphics[width=6.5cm,angle=0]{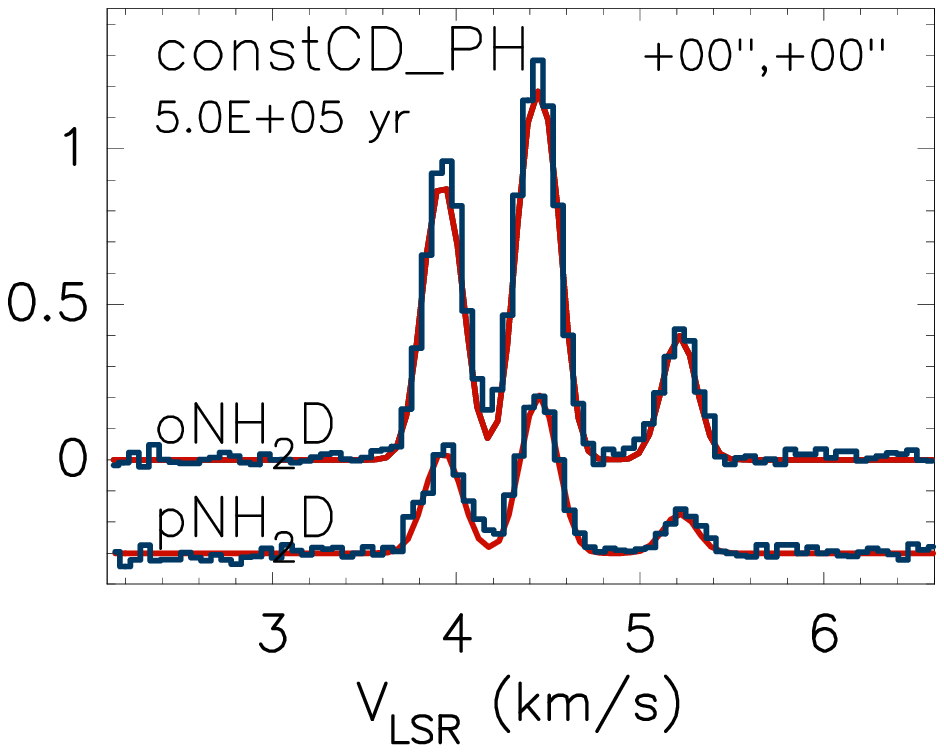}
\end{picture}}

\put(55,50){
\begin{picture}(0,0) 
\includegraphics[width=6.5cm,angle=0]{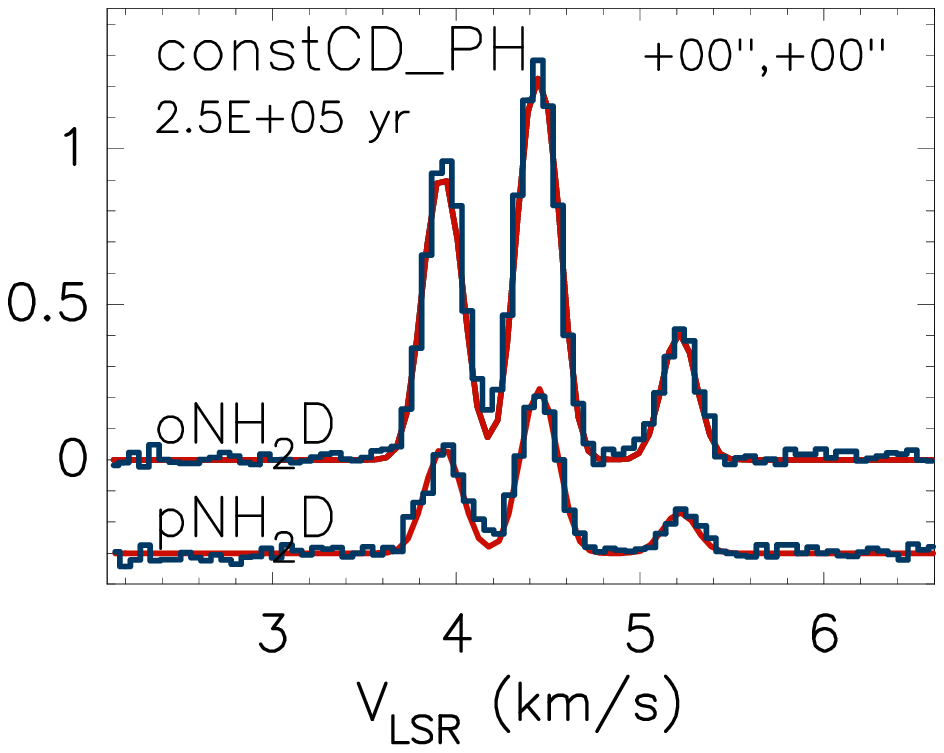}
\end{picture}}

\put(-6,50){
\begin{picture}(0,0) 
\includegraphics[width=6.5cm,angle=0]{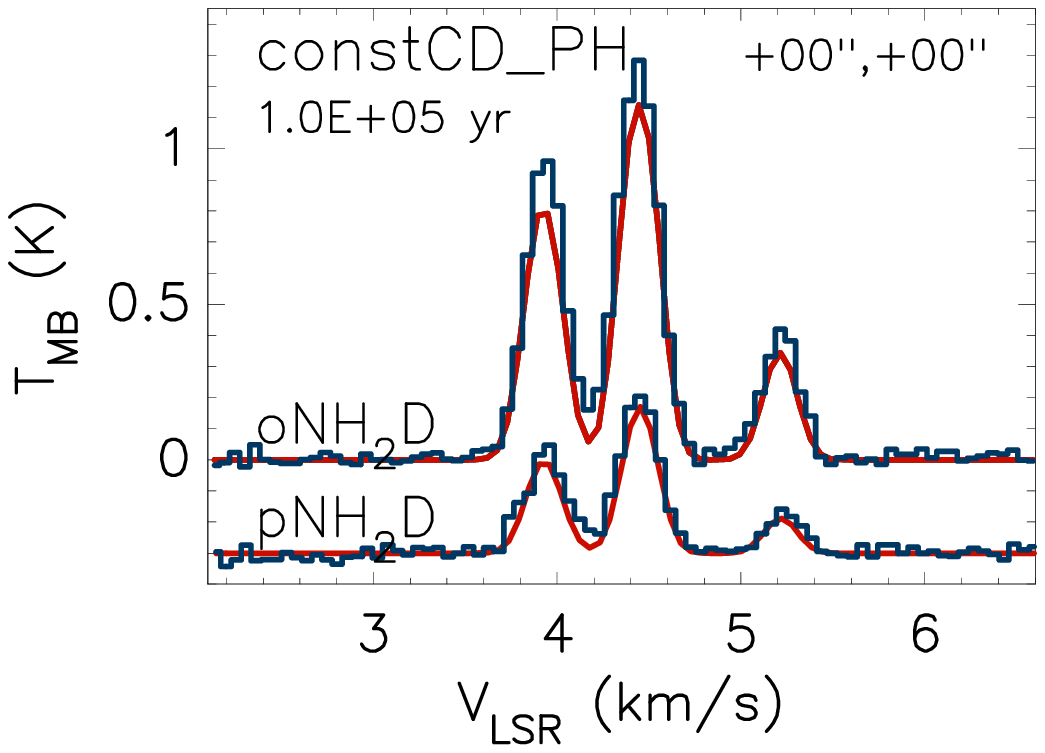}
\end{picture}}

\put(115,0){
\begin{picture}(0,0) 
\includegraphics[width=6.5cm,angle=0]{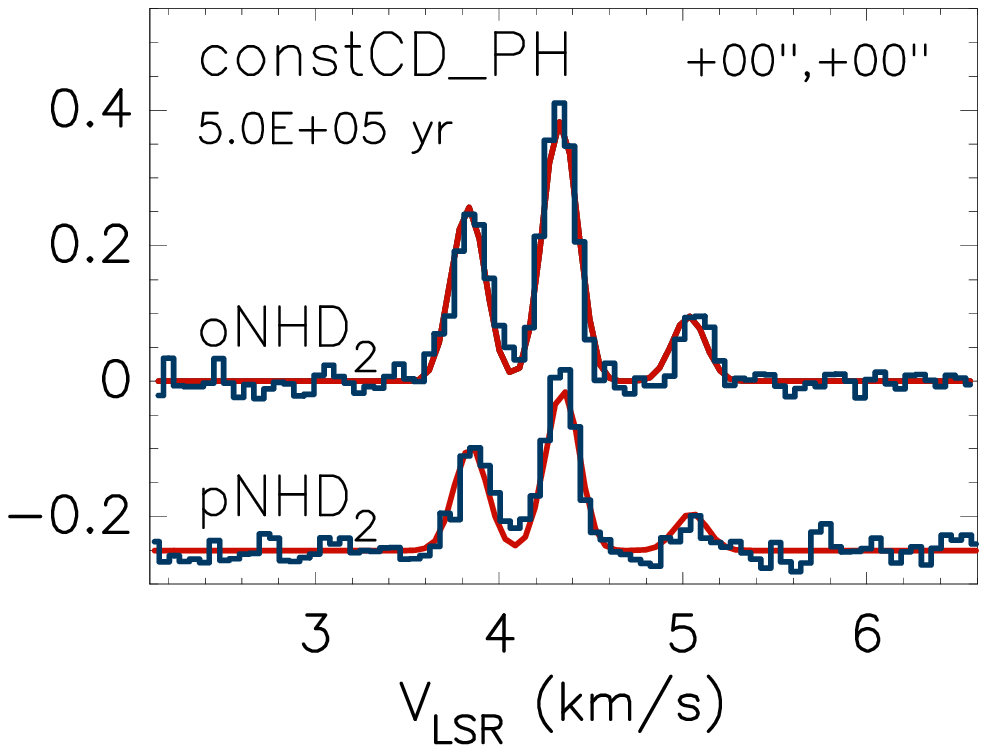}
\end{picture}}

\put(55,0){
\begin{picture}(0,0) 
\includegraphics[width=6.5cm,angle=0]{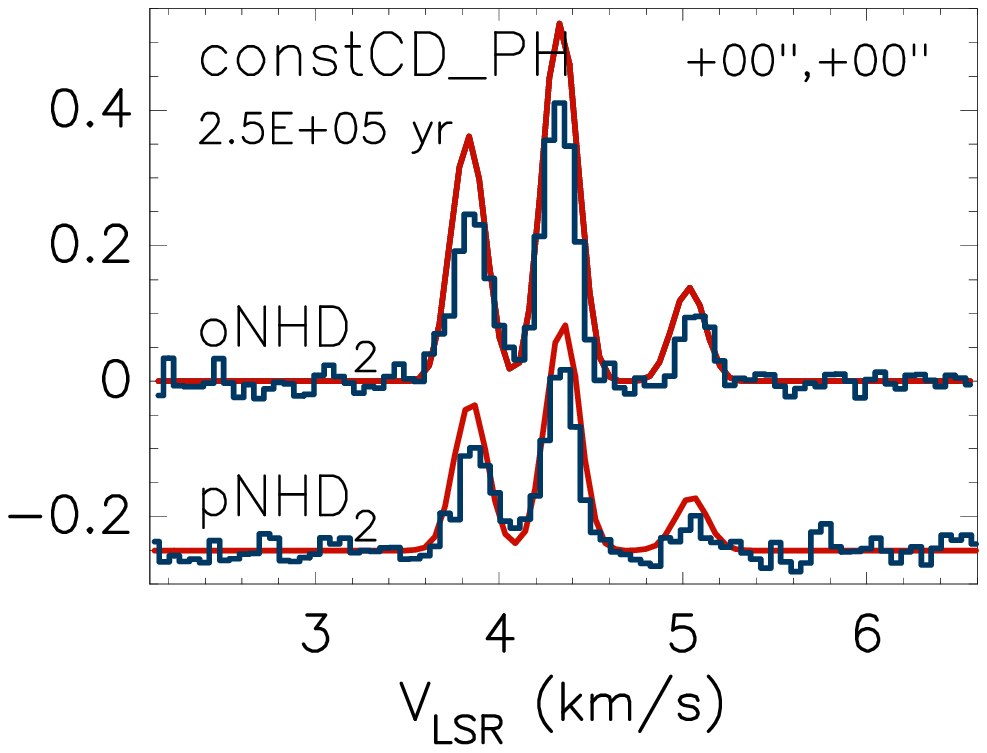}
\end{picture}}

\put(-6,0){
\begin{picture}(0,0) 
\includegraphics[width=6.5cm,angle=0]{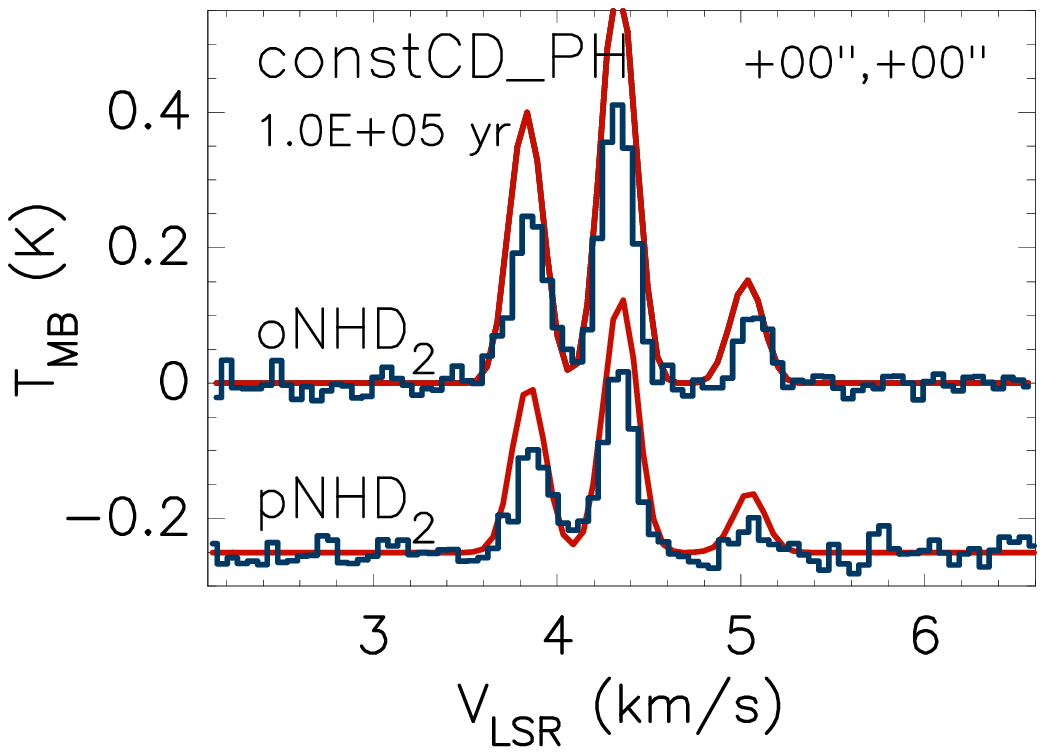}
\end{picture}}

\put(0,100){\bf Constant CD, PH}
\end{picture}
\caption{Same as Fig.~\ref{constCD_FS_spectra} but for a chemistry model assuming constant CD and proton hop (PH).}
\label{constCD_PH_spectra}
\end{figure*}

\begin{figure*}
\unitlength=1mm
\begin{picture}(160,110)(0,0)

\put(115,50){
\begin{picture}(0,0) 
\includegraphics[width=6.5cm,angle=0]{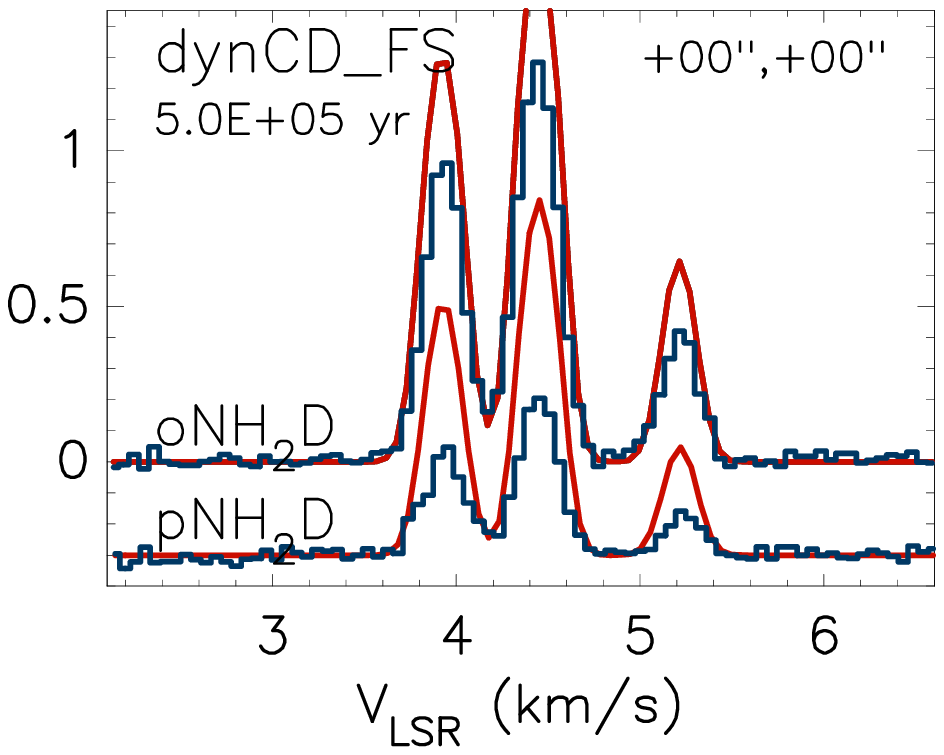}
\end{picture}}
 
\put(55,50){
\begin{picture}(0,0) 
\includegraphics[width=6.5cm,angle=0]{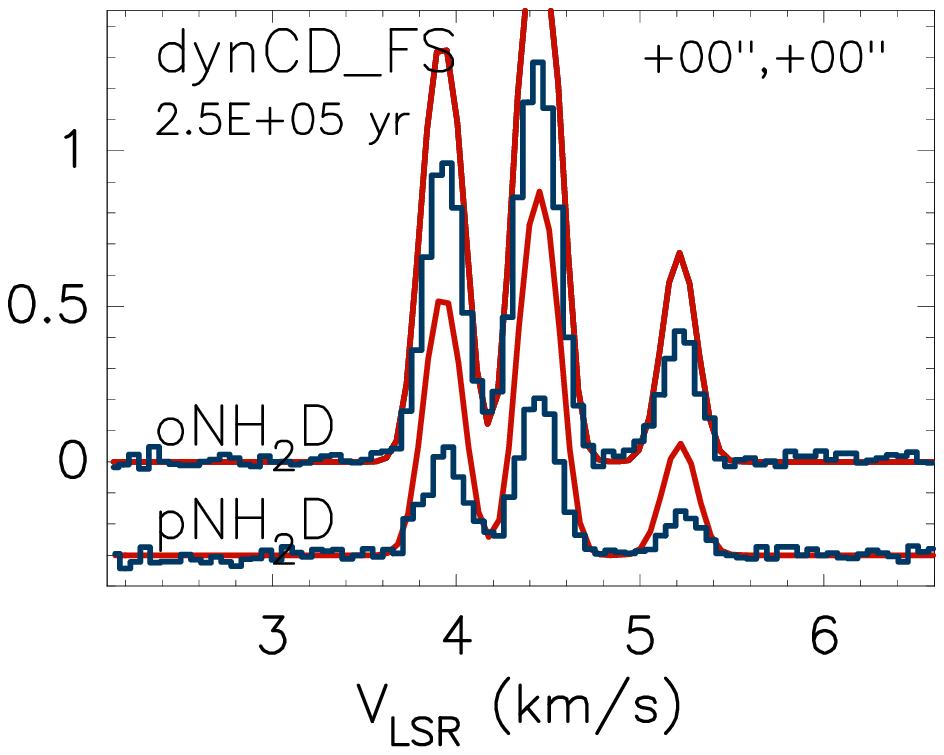}
\end{picture}}

\put(-6,50){
\begin{picture}(0,0) 
\includegraphics[width=6.5cm,angle=0]{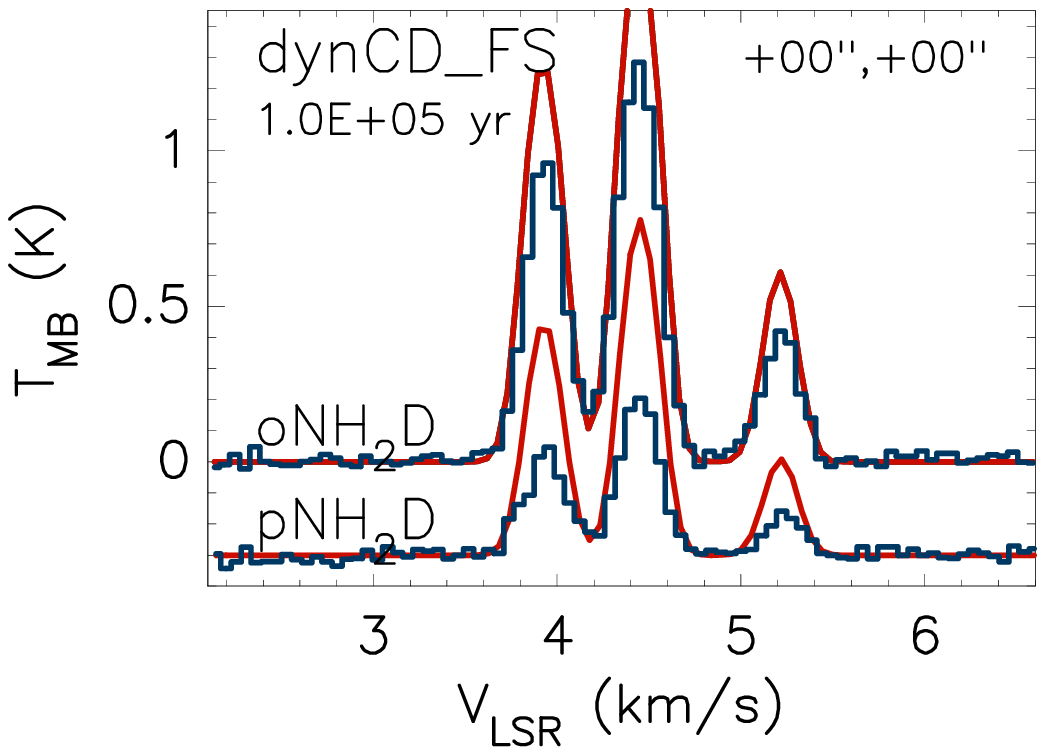}
\end{picture}}

\put(115,0){
\begin{picture}(0,0) 
\includegraphics[width=6.5cm,angle=0]{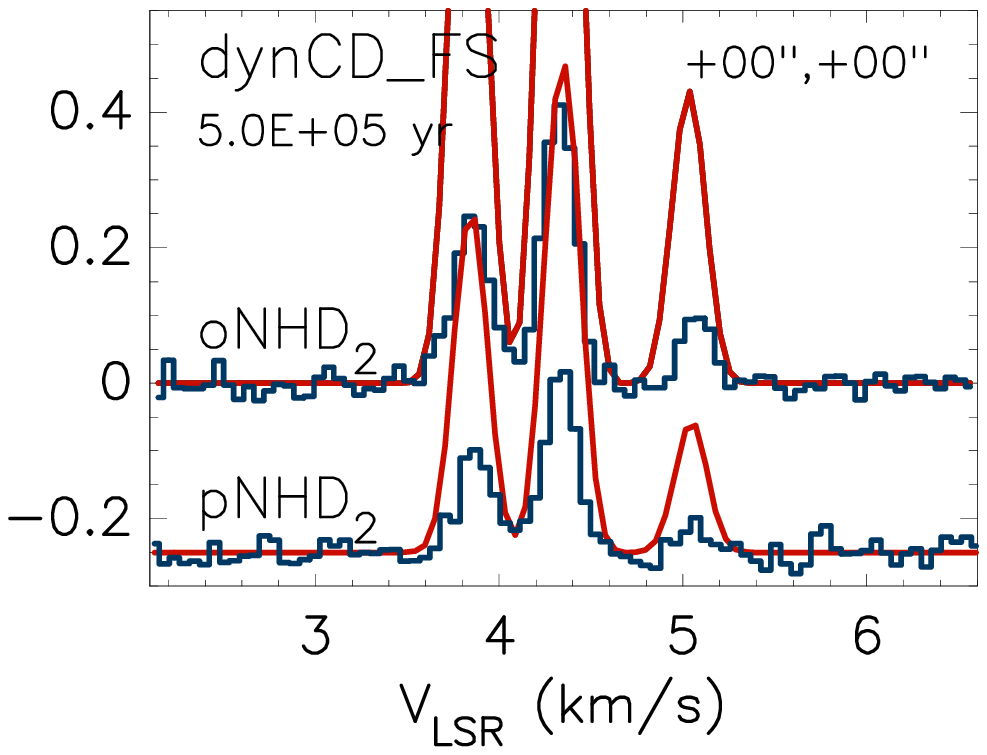}
\end{picture}}

\put(55,0){
\begin{picture}(0,0) 
\includegraphics[width=6.5cm,angle=0]{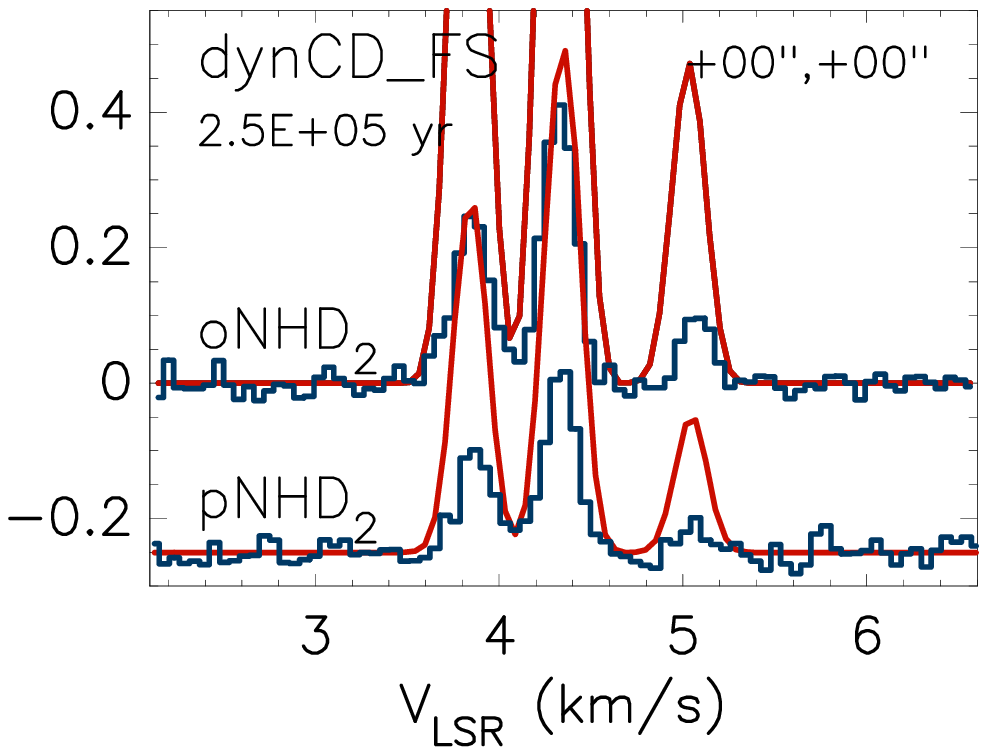}
\end{picture}}

\put(-6,0){
\begin{picture}(0,0) 
\includegraphics[width=6.5cm,angle=0]{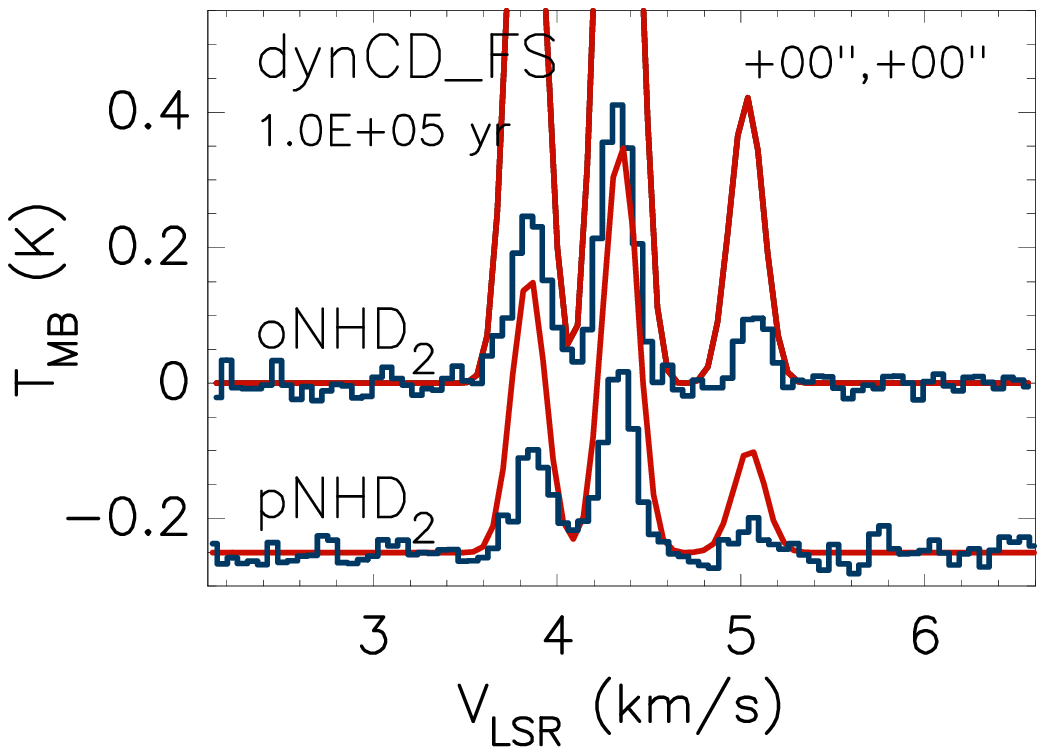}
\end{picture}}

\put(0,100){\bf Dynamic CD, FS}
\end{picture}
\caption{Same as Fig.~\ref{constCD_FS_spectra} but for a chemistry model assuming dynamic CD and full scrambling (FS).}
\label{dynCD_FS_spectra}
\end{figure*}

\begin{figure*}
\unitlength=1mm
\begin{picture}(160,110)(0,0)

\put(115,50){
\begin{picture}(0,0) 
\includegraphics[width=6.5cm,angle=0]{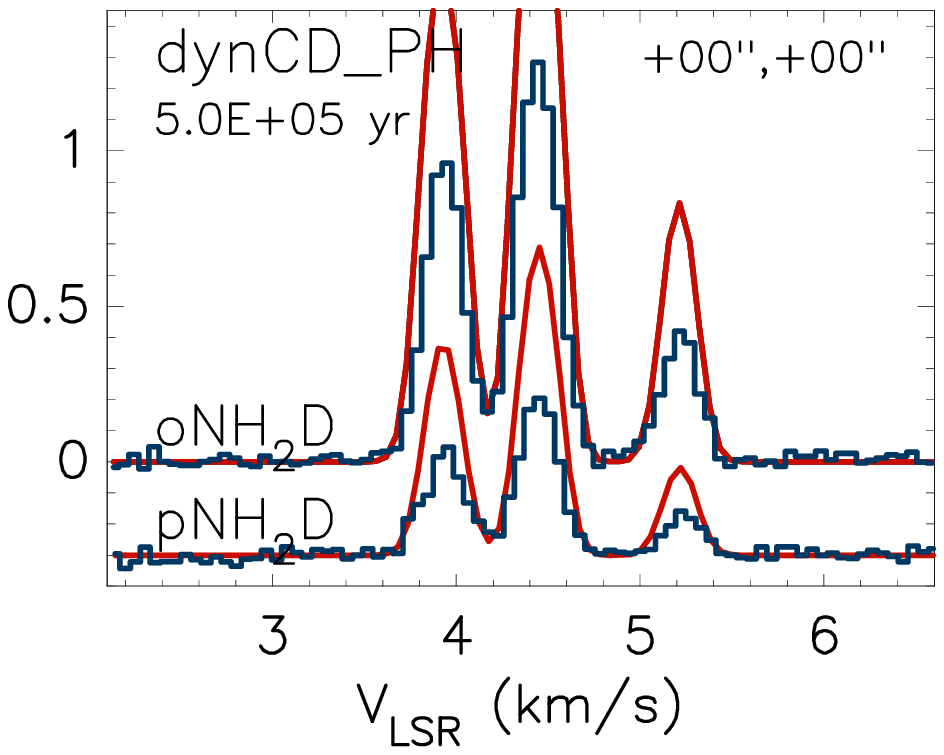}
\end{picture}}

\put(55,50){
\begin{picture}(0,0) 
\includegraphics[width=6.5cm,angle=0]{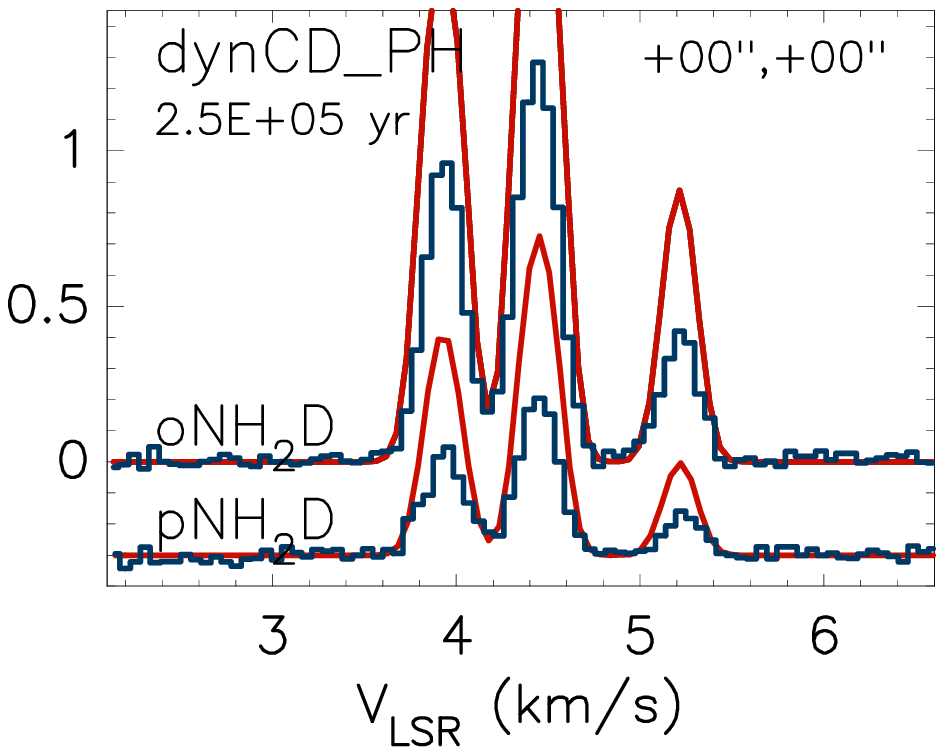}
\end{picture}}

\put(-6,50){
\begin{picture}(0,0) 
\includegraphics[width=6.5cm,angle=0]{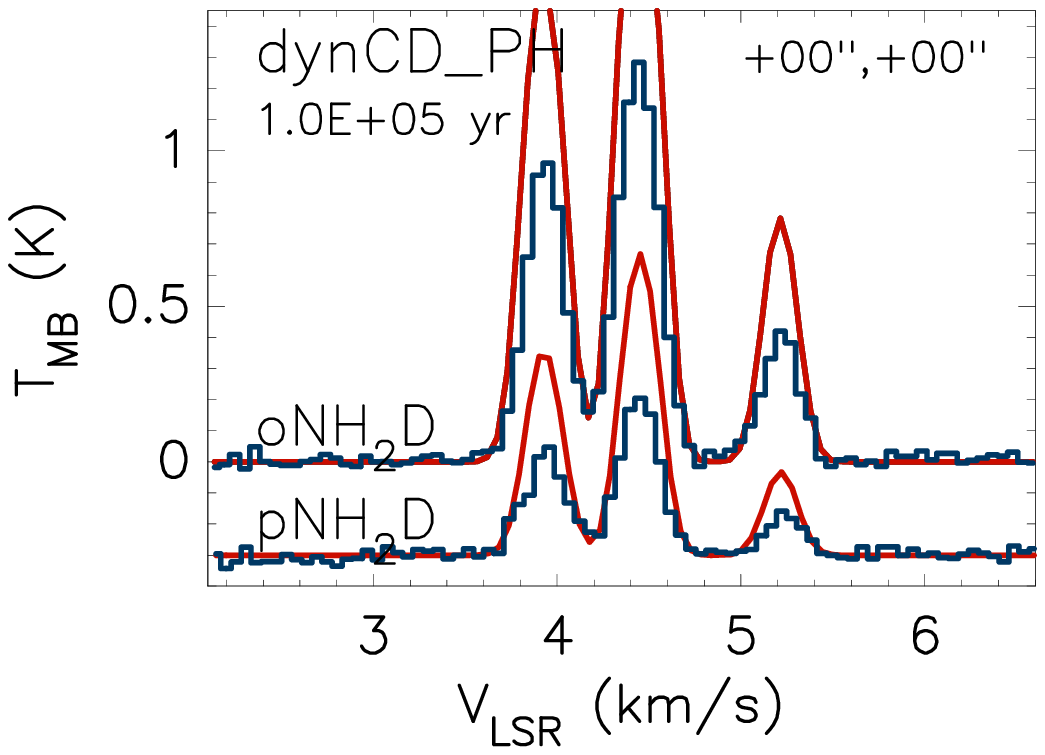}
\end{picture}}

\put(115,0){
\begin{picture}(0,0) 
\includegraphics[width=6.5cm,angle=0]{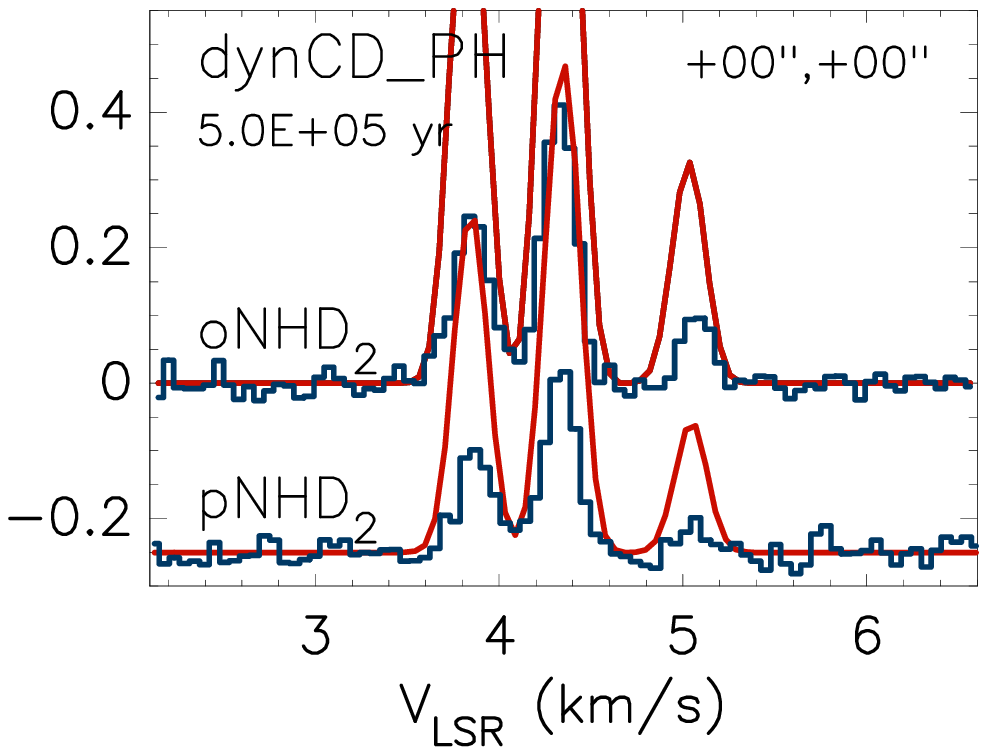}
\end{picture}}

\put(55,0){
\begin{picture}(0,0) 
\includegraphics[width=6.5cm,angle=0]{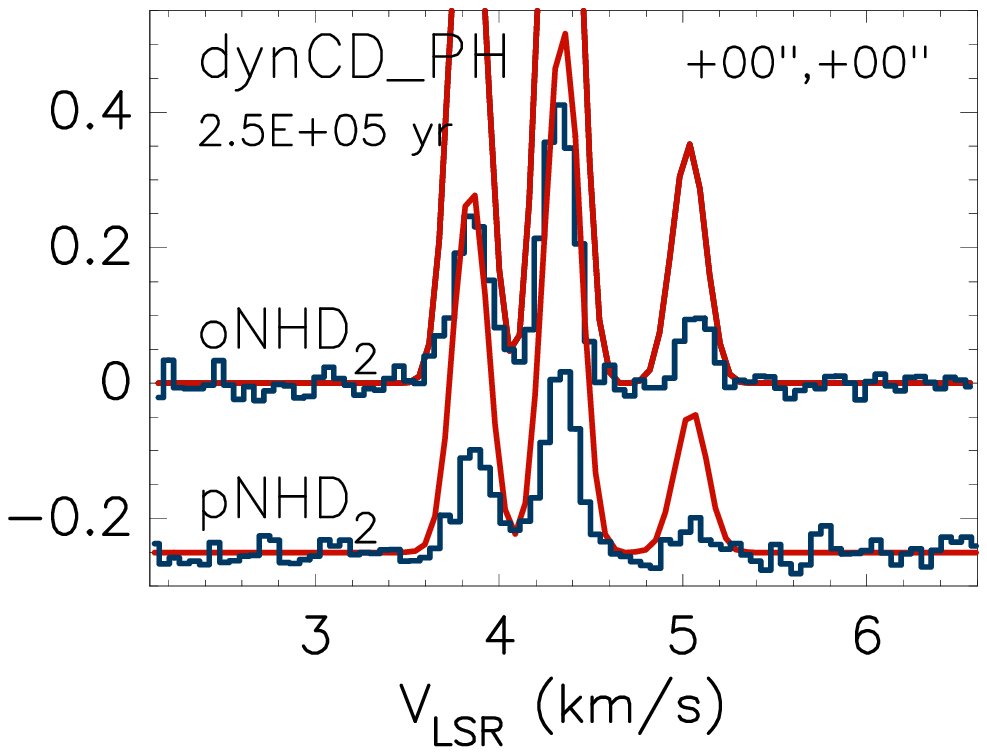}
\end{picture}}

\put(-6,0){
\begin{picture}(0,0) 
\includegraphics[width=6.5cm,angle=0]{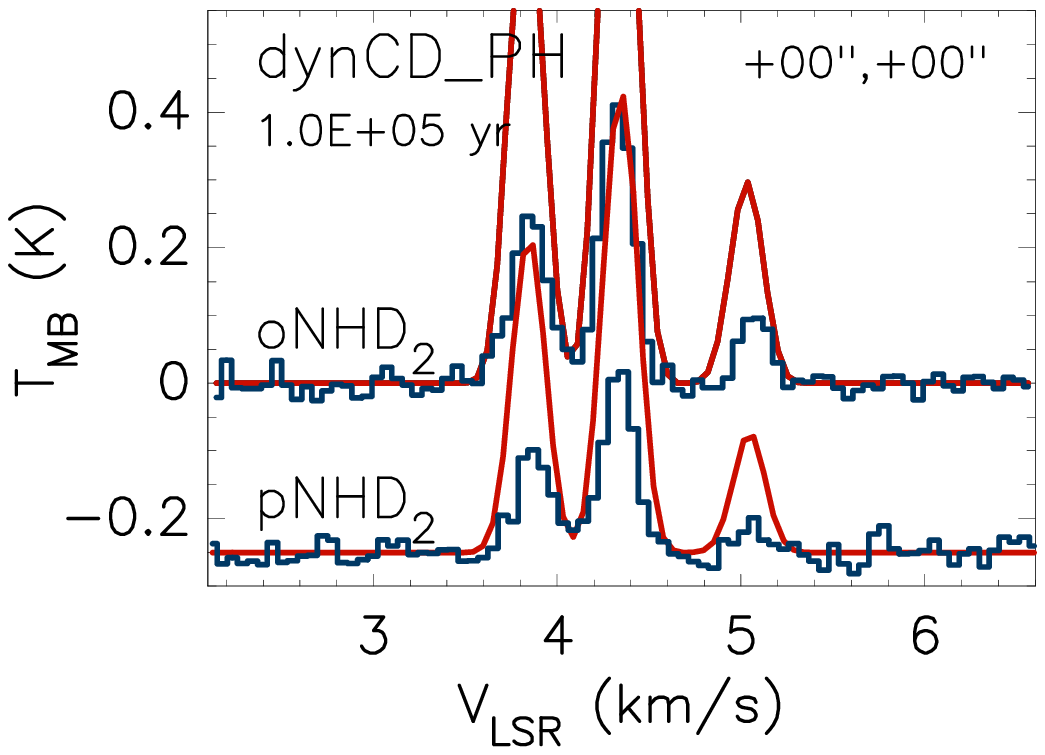}
\end{picture}}

\put(0,100){\bf Dynamic CD, PH}
\end{picture}

\caption{Same as Fig.~\ref{constCD_FS_spectra} but for a chemistry model assuming dynamic CD and proton hop (PH).}
\label{dynCD_PH_spectra}

\end{figure*}

{Figures~\ref{constCD_FS_spectra} to \ref{dynCD_PH_spectra} show $\dammo$ and $\ddammo$ spectra towards the centre of the 3D model of H-MM1 with abundances interpolated from the 1D core model used in the chemical model calculations. The spectra represent three simulation times} with different assumptions about the desorption efficiency and the viability of proton/deuteron exchanges in chemical reactions. Also shown are the observed spectra towards the centre of H-MM1.  

\vspace{5mm}

\newpage

\vspace{5mm}

\end{appendix}

\end{document}